\documentclass[a4paper,fleqn,usenatbib]{mnras}
\usepackage[T1]{fontenc}
\usepackage{ae,aecompl}

\usepackage{graphicx}	
\usepackage{amsmath}	
\usepackage{amssymb}	
\usepackage[utf8]{inputenc}
\usepackage{epsfig}
\usepackage{url}
\usepackage{capt-of}
\usepackage{multicol}
\usepackage{epstopdf}
\usepackage[singlelinecheck=false]{caption}
\usepackage{textcomp}
\usepackage[usenames,dvipsnames]{color}
\usepackage{float,lscape}
\usepackage{pdfpages}
\usepackage{pdflscape}
\usepackage{afterpage}
\usepackage{rotating}
\usepackage{arydshln}
\usepackage{soul}
\usepackage{threeparttable}

\usepackage{bm} 

\newcommand{\Msol}{\ensuremath{\mathrm{M}_{\odot}}}

\title[The K-CLASH survey]{K-CLASH: spatially-resolving star-forming galaxies in field and cluster environments at \textbf{$z$} $\mathbf{\approx 0.2}$--$\mathbf{0.6}$}

\author[Tiley et al.]{Alfred L.\ Tiley,$^{1,2,3\dagger}$ Sam P. Vaughan,$^{4,5,3}$ John P. Stott,$^{6}$ Roger L. Davies,$^{3}$
\newauthor Laura J. Prichard,$^{7}$ Andrew Bunker,$^{3}$ Martin Bureau,$^{3,8}$ Michele Cappellari,$^{3}$ 
\newauthor Matt Jarvis,$^{3,9}$ Aaron Robotham,$^{1}$ Luca Cortese,$^{1,5}$  Sabine Bellstedt$^{1}$ and
\newauthor Behzad Ansarinejad$^{2}$\\
$^{1}$International Centre for Radio Astronomy Research, The University of Western Australia, 35 Stirling Highway, Crawley \\ WA 6009, Australia\\
$^{2}$Centre for Extragalactic Astronomy, Department of Physics, Durham University, South Road, Durham, DH1 3LE, UK\\
$^{3}$Sub-department of Astrophysics, Department of Physics, University of Oxford, Denys Wilkinson Building, Keble Road, \\ Oxford, OX1 4RH, UK\\
$^{4}$Sydney Institute for Astronomy, School of Physics, Building A28, The University of Sydney, NSW 2006, Australia\\
$^{5}$ARC Centre of Excellence for All Sky Astrophysics in 3 Dimensions (ASTRO3D), Australia\\
$^{6}$Department of Physics, Lancaster University, Lancaster LA1 4YB, UK\\
$^{7}$Space Telescope Science Institute, 3700 San Martin Drive, Baltimore MD 21218, USA\\
$^{8}$Yonsei Frontier Lab and Department of Astronomy, Yonsei University, 50 Yonsei-ro, Seodaemun-gu, Seoul 03722, Republic\\
 of Korea\\
$^{9}$Department of Physics, University of the Western Cape, Bellville 7535, South Africa\\
$^{\dagger}$ E-mail: alfred.tiley@uwa.edu.au
}

\date{Accepted XXX. Received YYY; in original form ZZZ}

\pubyear{2020}

\begin{document}
\label{firstpage}
\pagerange{\pageref{firstpage}--\pageref{lastpage}}
\maketitle

\begin{abstract}
We present the KMOS-CLASH (K-CLASH) survey, a K-band Multi-Object Spectrograph (KMOS) survey of the spatially-resolved gas properties and kinematics of 191 (predominantly blue) H$\alpha$-detected galaxies at $0.2 \lesssim z \lesssim 0.6$ in field and cluster environments. K-CLASH targets galaxies in four Cluster Lensing And Supernova survey with Hubble (CLASH) fields in the KMOS $IZ$-band, over $7'$ radius ($\approx2$--$3$\,Mpc) fields-of-view. K-CLASH aims to study the transition of star-forming galaxies from turbulent, highly star-forming disc-like and peculiar systems at $z\approx1$--$3$, to the comparatively quiescent, ordered late-type galaxies at $z\approx0$, and to examine the role of clusters in the build-up of the red sequence since $z\approx1$. In this paper, we describe the K-CLASH survey, present the sample, and provide an overview of the K-CLASH galaxy properties. We demonstrate that our sample comprises star-forming galaxies typical of their stellar masses and epochs, residing both in field and cluster environments. We conclude K-CLASH provides an ideal sample to bridge the gap between existing large integral-field spectroscopy surveys at higher and lower redshifts. We find that star-forming K-CLASH cluster galaxies at intermediate redshifts have systematically lower stellar masses than their star-forming counterparts in the field, hinting at possible ``downsizing'' scenarios of galaxy growth in clusters at these epochs. We measure no difference between the star-formation rates of H$\alpha$-detected, star-forming galaxies in either environment after accounting for stellar mass, suggesting that cluster quenching occurs very rapidly during the epochs probed by K-CLASH, or that star-forming K-CLASH galaxies in clusters have only recently arrived there, with insufficient time elapsed for quenching to have occured. 
\end{abstract}




\section{Introduction}
\label{sec:intro}

Galaxy growth has occurred predominantly over the last 10 Gyr since $z\approx2$, during which $\approx$80 per cent of the stellar mass in the cosmos was assembled \citep[e.g.][]{PerezGonzalez:2008}. Over the same period, we also observe drastic changes to galaxy morphologies, with the emergence of the Hubble Sequence \citep[][]{Hubble:1926,Hubble:1936}, including the build-up of the ``red sequence'' of passive S0s and elliptical galaxies in clusters \citep[e.g.][]{Bell:2004,Stott:2007}, and a corresponding decline in star-forming disc-like systems in the same environments \citep[e.g.][]{Butcher:1978,Hilton:2010,PintosCastro:2013}. This coincides with the parallel transition of the star-forming population from rapidly star-forming, gas-rich, disc-like and peculiar systems with high levels of chaotic motions at $z\approx1$--$3$ \citep[e.g.][]{Wisnioski:2015,Stott:2016}, to the comparatively quiescent spiral galaxies common in the local Universe, with dynamics dominated by ordered circular motion. The same 10 Gyr also bear witness to fundamental changes in the balance of luminous and dark matter in galaxies, evidenced by evolution in the slopes, normalisations and scatters of key dynamical galaxy scaling relations relating their baryonic and dark mass contents, for example the Tully-Fisher relation \citep[TFR; e.g.][]{Tully:1977aa,Gnerucci:2011,Tiley:2016,Ubler:2017,Turner:2017,Tiley:2019a}, and the specific angular momentum-stellar mass relation \citep[e.g.][]{Swinbank:2017,Harrison:2017}. 

Understanding how and why galaxies have changed so dramatically during this period is vital for a complete theory of galaxy formation and evolution over the history of our Universe. Large spectroscopic and imaging campaigns over the past few decades, including the Two Degree Field (2dF) Galaxy Redshift Survey \citep{Folkes:1999}, the Sloan Digital Sky Survey \citep{York:2000}, the Visible Multi-Object Spectrograph \citep[VIMOS;][]{LeFevre:2000} Very Large Telescope (VLT) Deep Survey \citep{LeFevre:2004}, the Cluster Lensing And Supernova survey with Hubble \citep[CLASH;][]{Postman:2012}-VLT survey \citep[CLASH-VLT;][]{Rosati:2014}, the Canadian Network for Observational Cosmology (CNOC) Cluster Redshift Survey \citep{Yee:1996}, and the CNOC2 Field Galaxy Redshift Survey \citep{Yee:2000}, have already facilitated great advances in our understanding of galaxy evolution over this period. Together, these studies provide statistically large censuses of the properties and distributions of galaxies out to $z\approx3$ and beyond, over large areas of the sky and a wide variety of galaxy environments. However, the recent advent of integral-field spectroscopy (IFS), that allows the simultaneous capture of images and spectra in a single observation, has opened a new window on the spatially-resolved properties of galaxies in our Universe \citep[for a comprehensive review, see][]{Cappellari:2016}. Large IFS surveys such as the ATLAS$^{3\rm{D}}$ project \citep{Cappellari:2011}, the Calar Alto Legacy Integral Field Area \citep[CALIFA;][]{Sanchez:2012} Survey, the Sydney-Australian-Astronomical Observatory Multi-object Integral-field Spectrograph \citep[SAMI;][]{Croom:2012} Galaxy Survey \citep{Bryant:2015}, and the Mapping Nearby Galaxies at Apache Point Observatory \citep[MaNGA;][]{Bundy:2015} survey have now mapped the spatially-resolved visible properties of many thousands of galaxies at $z\approx0$. 

Recent complementary advances in near-infrared IFS technologies have allowed a parallel unveiling of the spatially-resolved rest-frame visible properties of star-forming galaxies at $z\approx1$--$3$ via their bright nebular line emission (predominantly H$\alpha$ and [N {\sc ii}]). IFS surveys such as the Spectroscopic Imaging survey in the Near-infrared with SINFONI \citep[SINS;][]{ForsterSchreiber:2009}, the  $K$-band Multi-Object Spectrograph \citep[KROSS;][]{Sharples:2013} Redshift One Spectroscopic Survey \citep[KROSS;][]{Stott:2016,Harrison:2017}, the KMOS$^{3\rm{D}}$ survey \citep{Wisnioski:2015}, the KMOS Galaxy Evolution Survey (KGES; Tiley et al., in preparation), the KMOS Clusters Survey \citep[KCS;][]{Beifiori:2017}, and the KMOS Deep Survey \citep[KDS;][]{Turner:2017} have also now surveyed several thousands of star-forming galaxies at $1 \lesssim z \lesssim 3$. Comparisons between these IFS galaxy samples and those in the local Universe have revealed stark differences between the galaxy populations at each epoch. But, whilst the spatially-resolved properties of galaxies at the two extreme ends of this key period for galaxy evolution are well studied, there is an absence of equivalent IFS studies at intermediate redshifts; the origins of the changes to galaxies between $z\approx1$--$3$ and $z\approx0$ are in many cases still unclear, as are the physical processes that underpin them and the timescales on which they occur.

\subsection{The K-CLASH survey}

In this work, we present the KMOS-CLASH (K-CLASH) survey, that aims to bridge this gap in redshift and understanding by studying the spatially-resolved gas kinematics and emission line properties of star-forming galaxies, in both field and cluster environments at $z \approx 0.2$--$0.6$, to investigate how galaxies in both environments transition into their present-day populations. The K-CLASH survey is a University of Oxford guaranteed time survey with the European Southern Observatory (ESO) KMOS on the VLT, Chile. Using 8.5 nights of KMOS time, we targeted the H$\alpha$ and [N {\sc ii}] line emission from 282 galaxies in four CLASH cluster fields. We carried out KMOS observations in the $IZ$-band to detect H$\alpha$ emission from star-forming field galaxies along cluster sight lines in the redshift range $0.2 \lesssim z \lesssim 0.6$. Similarly, we selected cluster fields to observe with KMOS such that the cluster redshift was in the range $0.3 \lesssim z \lesssim 0.6$. In this way, in the same observations, we were able to simultaneously construct both field and cluster samples of star-forming galaxies. 

K-CLASH provides unique samples for comparison with similar larger IFS samples of galaxies in both the field and clusters at higher and lower redshifts. Unlike existing IFS samples at similar redshifts \citep[e.g.][]{Swinbank:2017} that typically target bluer nebular emission lines (e.g. [O {\sc ii}] or [O {\sc iii}]) with the VLT Multi-Unit Spectroscopic Explorer \citep{Bacon:2010}, K-CLASH benefits from its use of H$\alpha$ and [N {\sc ii}] to trace the galaxies' ongoing star formation, kinematics and gas properties. These lines are redder and thus less affected by dust extinction than the bluer indicators. More generally, K-CLASH allows for {\it direct} comparisons between its galaxies' gas kinematics and properties and those of galaxies in other major IFS surveys at other redshifts that share common tracers and probe the same gas phases. 

As K-CLASH galaxies are located within CLASH cluster fields, we also benefit from a wealth of ancillary data associated with CLASH itself, including optical and near-infrared imaging from the Subaru Telescope (Subaru) and the {\it Hubble Space Telescope} ({\it HST}),\footnote{Due to the spatial distribution of the K-CLASH galaxies in the cluster fields, only a small subset benefit from {\it HST} coverage.} infrared imaging from {\it Spitzer}, partial coverage in submillimeter Bolocam \citep{Glenn:1998} imaging, and X-ray imaging from {\it Chandra}. This allows for a comprehensive, multi-wavelength perspective on star-forming galaxies across a range of environments and spanning $\approx3$ Gyr of cosmic history.

The broad goals of K-CLASH are to investigate the transition of star-forming field galaxies from turbulent, highly star-forming discs at $z\approx1$--$3$, to the comparatively quiescent late-types at $z\approx0$ with more ordered dynamics, as well as to examine the role of environment in the build-up of the red sequence in galaxy clusters since $z\approx1$. In this paper, we describe the K-CLASH survey, including the survey design and observing strategy, as well as the general physical properties of the K-CLASH galaxies.  A detailed examination of the properties of the K-CLASH galaxies in cluster environments is undertaken by \citet{vaughan:2020}, who present evidence for star-formation quenching in clusters using the K-CLASH sample. In \S~\ref{sec:clusterselect}, we detail the criteria used to select CLASH clusters for follow up observations with KMOS, as well as the properties of each selected cluster. We describe the KMOS target selection, observations, and observing strategy in \S~\ref{sec:obsandsample}, and in \S~\ref{sec:kmosmeasurements} we discuss the measurements we make using the KMOS data. In \S~\ref{sec:sampleoverview}, we provide an overview of the K-CLASH sample. In \S~\ref{sec:galaxyproperties}, we present measurements of the key properties of the K-CLASH galaxies. We then discuss K-CLASH in the context of exisiting works in \S~\ref{sec:discussion}, drawing simple comparisons between the properties of the K-CLASH field and cluster galaxies, as well as between the properties of K-CLASH galaxies and those observed at higher and lower redshifts as part of other large IFS surveys. Finally, in \S~\ref{sec:conclusions} we provide concluding remarks and discuss planned future work with K-CLASH.

A Nine-Year {\it Wilkinson Microwave Anisotropy Probe} \citep[{\it WMAP9};][]{Hinshaw:2013} cosmology is used throughout this work (Hubble constant at $z = 0$, $H_{0} = 69.3 $ km s$^{-1}$ Mpc$^{-1}$; non-relativistic matter density at $z=0$, $\Omega_{0} = 0.287$; dark energy density at $z=0$, $\Omega_{\Lambda} = 0.713$). All magnitudes are quoted in the AB system \citep{Oke:1983}. All stellar masses ($\rm{M}_{*}$) assume a \citet{Chabrier:2003} initial mass function (IMF). 

\section{Field Selection and Cluster Properties}
\label{sec:clusterselect}

\begin{table*}
\caption{Basic properties of the CLASH clusters observed with KMOS.} \label{tab:clashclusters}\smallskip
\begin{tabular}{ lcccccccc}
\hline
Cluster & RA$_{\rm{BCG}}$ & Dec$_{\rm{BCG}}$ & Redshift & $k_{\rm{B}}T_{\rm{X}}$ & $L_{\rm{X}}$ & $\rm{R}_{200,\rm{cluster}}$ & $\rm{M}_{200}$ & Angular  \\
 &  & & $\rm{(a)}$  & $\rm{(b)}$  & $\rm{(b)}$  & $\rm{(c)}$  & $\rm{(d)}$  & scale  \\
 &  (J2000) &  (J2000) &  &  (keV) & ($10^{44}$ erg s$^{-1}$)  & (Mpc) & ($10^{14}\ \Msol h^{-1}$) & ($\frac{\rm{Mpc}}{\rm{arcmin}}$)\\
\hline
{\bf MACS2129}.4$-$0741 & 21:29:26.12 & $-$07:41:27.8 & 0.589$^{\rm{(e)}}$ & $9.0\pm1.2$ & 22.6 $\pm$ 1.5 & $1.90^{+0.05}_{-0.04}$  & $10.0^{+0.8}_{-0.6}$ & 0.40  \\
{\bf MACS1311}.0$-$0310 & 13:11:01.80 & $-$03:10:39.7  & 0.494\phantom{$^{\rm{(e)}}$} & $5.9\pm0.4$ & \phantom{0}9.4 $\pm$ 0.4  & $1.39^{+0.37}_{-0.28}$ & $\phantom{1}3.5^{+3.6}_{-1.7}$ & 0.37 \\
{\bf MACS1931}.8$-$2635 & 19:31:49.63 & $-$26:34:32.5 & 0.352\phantom{$^{\rm{(e)}}$} & $6.7\pm0.4$ & 20.9 $\pm$ 0.6 & $1.87^{+0.08}_{-0.12}$ & $\phantom{1}7.3^{+0.9}_{-1.3}$ & 0.30 \\
{\bf MS2137}$-$2353\phantom{CC.0} & 21:40:15.17 & $-$23:39:40.3 & 0.313$^{\rm{(f)}}$ & $5.9\pm0.3$ & \phantom{0}9.9 $\pm$ 0.3 & $1.26^{+0.13}_{-0.12}$ & $\phantom{1}2.1^{+0.7}_{-0.5}$ & 0.28 \\
\hline
\multicolumn{9}{l}{%
\begin{minipage}{17.5cm}%
\smallskip
Notes. $^{\rm{(a)}}$Published cluster redshifts from \citet{Postman:2012} (except for MACS2129). Unless otherwise stated, cluster redshifts were calculated by the Massive Cluster Survey \citep[][]{Ebeling:2007,Ebeling:2010} based on {\it Chandra} X-ray spectra. $^{\rm{(b)}}$Published X-ray temperatures from \citet{Postman:2012}.  $^{\rm{(c)}}$Cluster radius (within which the mean density is 200 times the critical density at the cluster's redshift), calculated as $\rm{R}_{200,\rm{cluster}} = c_{200}\rm{r}_{s}$ from lensing measurements of each cluster according to the methods of \citet{Zitrin:2009,Zitrin:2013,Zitrin:2015}, where r$_{s}$ and $c_{200}$ are respectively the scale radius and the concentration of the Navarro-Frenk-White \citep{Navarro:1997} dark matter profile. The parameters are an updated version of those available from the Mikulski Archive for Space Telescopes (\url{https://archive.stsci.edu/prepds/clash/}), provided by Zitrin et al. (private communication). $^{\rm{(d)}}$Cluster masses, calculated as $\rm{M}_{200} = (4\pi/3) 200 \rho_{\rm{crit}}(z) \rm{R}_{200,\rm{cluster}}^{3}$, where $\rho_{\rm{crit}}(z)$ is the Universe critical density. $^{\rm{(e)}}$MACS2129 redshift determined by \citet{Stern:2010} via spectroscopic redshift measurements of cluster members. $^{\rm{(f)}}$MS2137 redshift determined by \citet{Stocke:1991} via optical spectroscopic follow up of individual galaxy redshifts.
\end{minipage}%
}\\
\end{tabular}
\end{table*}

The CLASH sample comprises 25 clusters, the majority of which (20 out of the 25) are X-ray selected (X-ray temperatures $T_{\rm{X}} \geq 5$ keV), massive, dynamically-relaxed systems. The remainder were selected on the basis of their strong lensing properties. For a detailed analysis of the properties of the entire sample of CLASH clusters, see \citet{Postman:2012} (and \citealt{Zitrin:2015} for an analysis of their strong and weak lensing properties). 

\subsection{Cluster field selection}
\label{subsec:fieldselect}

K-CLASH cluster fields were selected for KMOS observations if they satisfied the following criteria: 

\begin{itemize}
\item the cluster field has associated wide-field photometry (in visible bandpasses) and photometric redshift catalogues that are publicly available;
\item the cluster redshift is in the range $0.2 \lesssim z \lesssim 0.6$, such that the H$\alpha$ and [N {\sc ii}] emission lines from cluster members falls within the $IZ$-band of KMOS;
\item the redshift of the cluster is such that the observed  H$\alpha$ and [N {\sc ii}] emission lines from cluster members do not fall within strong atmospheric absorption features and do not overlap with bright sky emission features.
\end{itemize}

Combining these selection criteria with scheduling constraints resulted in complete observations of four CLASH cluster fields within the 8.5 night K-CLASH observing programme, in the direction of the MS 2137.3-2353, MACSJ1931-26, MACSJ1311-03, and MACSJ2129-07 clusters (hereafter referred to respectively as MS2137, MACS1931, MACS1311, and MACS2129). The properties of each of these clusters are briefly discussed in \S~\ref{subsec:clusterprops}.

\subsection{Cluster properties}
\label{subsec:clusterprops}

\begin{figure}
\begin{minipage}[]{.5\textwidth}
\includegraphics[width=.95\textwidth,trim= 15 15 10 10,clip=True]{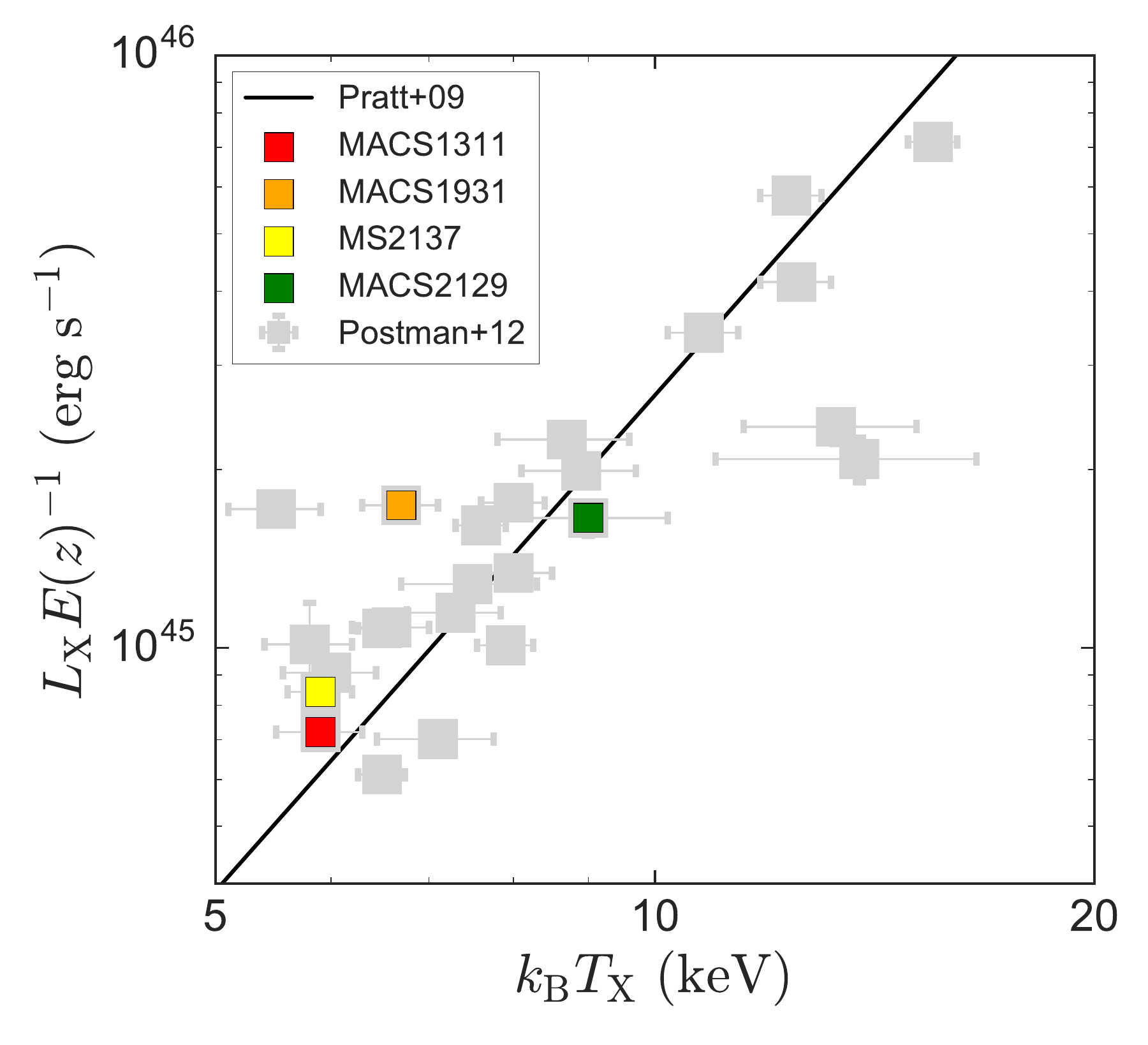}
\end{minipage}
\caption{%
X-ray luminosities of the CLASH clusters (grey squares) as a function of their X-ray temperatures. The former are scaled assuming self-similar evolution ($E(z) = \sqrt{\Omega_{0}(1+z)^{3} + \Omega_{\Lambda}}$) for direct comparison with the ``L2-T2'' relation of \citet{Pratt:2009}, for 31 X-ray-selected nearby clusters (black line). The K-CLASH clusters (coloured squares) are situated in the lower half of the temperature range of the CLASH sample clusters. Three of the four (MACS1311, MS2137, and MACS2129) closely follow the general trend between luminosity and temperature of the total sample of CLASH clusters, suggesting they are each dynamically relaxed. MACS1931 is a moderate outlier from this trend, suggesting it is not fully relaxed or the cluster X-ray luminosity is contaminated by the presence of an AGN.  %
     }%
\label{fig:scalingrelationcontext}
\end{figure}

In this section we consider the individual properties of the four clusters observed with KMOS as part of K-CLASH, providing a short summary for each. The basic properties of each of the observed clusters are listed in Table~\ref{tab:clashclusters}. In Figure~\ref{fig:scalingrelationcontext}, we show the position of each of the K-CLASH clusters in the X-ray luminosity ($L_{\rm{X}}$)--temperature ($T_{\rm{X}}$) plane, in comparison to all other clusters in the CLASH sample.   
 
\subsubsection{MACS2129}
\label{subsubsec:MACS2129}

MACS2129 is the highest redshift cluster ($z=0.589$) we observed with KMOS. The most massive cluster in K-CLASH, it is one of five clusters originally targeted by the CLASH survey on the basis of its significant lensing strength. Several previous studies have examined background lensed sources with multiple images in the cluster field \citep[e.g.][]{Zitrin:2015,Monna:2017}. Unlike the other CLASH clusters that are typically along lines-of-sight with negligible Galactic extinction, MACS2129 suffers from significant Galactic cirrus obscuration in its north-eastern quadrant. Accordingly, we preferentially targeted galaxies outside of this region of extinction, in the remaining three quadrants of the cluster. 

\subsubsection{MACS1311}
\label{subsubsec:MACSJ1311}

MACS1311 is the second highest redshift cluster in our sample ($z=0.494$). X-ray-selected for the CLASH survey, and of intermediate mass, MACS1311 appears dynamically-relaxed according to its X-ray surface brightness symmetry \citep{Postman:2012}.

\subsubsection{MACS1931}
\label{subsubsec:MACS1931}

MACS1931 is the second most massive cluster of the four observed with KMOS, and is at relatively low redshift ($z=0.352$). X-ray-selected by CLASH, MACS1931 has one of the most X-ray luminous cores of any known cluster. The brightest cluster galaxy (BCG) hosts an active galactic nucleus (AGN), evident in spatially-extended radio emission surrounding the galaxy, with corresponding X-ray cavities or ``bubbles" \citep{Ehlert:2011}. The BCG appears elongated in optical imaging along the same axis, with long extended filaments of nebular emission. The cluster has a high cooling rate but without the corresponding central metallicity peak normally associated with cool core clusters, suggesting a bulk flow of cool gas from the central region of the cluster to its outer regions \citep[e.g.][]{Ehlert:2011}.

\subsubsection{MS2137}
\label{subsubsec:MS2137}

MS2137 is another X-ray-selected cluster, with the lowest total mass of our sample. It is also the lowest redshift cluster in our sample ($z=0.313$). MS2137 appears dynamically relaxed, with a prominent lensed arc north of its BCG \citep[e.g.][]{Donnarumma:2009}.

\section{Galaxy Target Selection and KMOS Observations}
\label{sec:obsandsample}

In this section we give details of the initial K-CLASH galaxy sample selection criteria, as well as the observing strategy employed to target galaxies with KMOS. 

\subsection{Target selection criteria}
\label{subsec:targetselection}

Individual galaxies were selected for observation with KMOS using photometry and photometric redshifts measured by \citet{Umetsu:2014} from deep, spatially-extended Subaru Suprime-Cam \citep{Miyazaki:2002} imaging in the $B$, $V$, $R_{\rm{C}}$, $I_{\rm{C}}$, and $Z$ bands (available for MS2137, MACS1931, and MACS2129), as well as $B$- and $V$-band imaging from the ESO Wide Field Imager \citep{Baade:1999} and $z'$-band imaging from the Inamori-Magellan Areal Camera and Spectrograph\footnote{\url{http://www.lco.cl/telescopes-information/magellan/instruments/imacs/}} \citep[IMACS;][]{Dressler:2011} used as supplementary photometry for MACS1311 (for which only $R_{\rm{C}}$-band imaging is available from Suprime-Cam). All images were reduced using techniques described in \citet{Nonino:2009} and \citet{Medezinski:2013}. \citet{Umetsu:2014} extracted magnitudes by running {\sc SExtractor} in dual-image mode using the {\sc Colorpro} code \citep{Coe:2006}. Photometric redshifts were calculated using the Bayesian photometric analysis and spectral energy distribution (SED) fitting package, {\sc bpz} \citep{Benitez:2000} in {\sc Python}.\footnote{\url{http://www.stsci.edu/~dcoe/BPZ/}} The full catalogues of photometry and photometric redshifts for each cluster used in this work are publicly available online at the Mikulski Archive for Space Telescopes.\footnote{\url{https://archive.stsci.edu/prepds/clash/}} For a detailed summary of the ancillary data available for the full CLASH cluster sample, see \citet{Postman:2012} and tables therein.

\begin{figure*}
\centering
\begin{minipage}[]{1.\textwidth}
\centering
\includegraphics[width=.68\textwidth,trim= 70 5 70 5,clip=True]{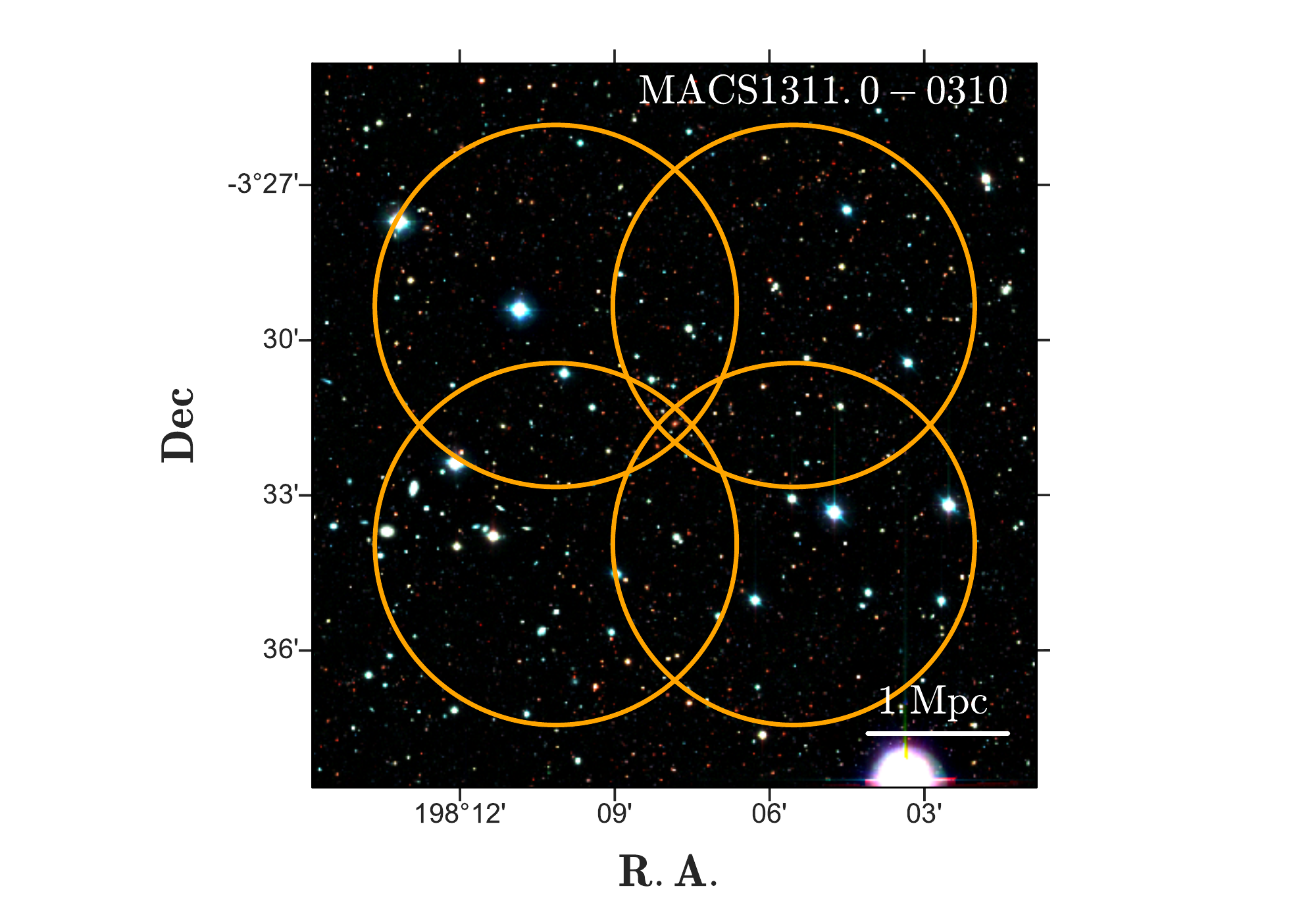}
\end{minipage}
\caption{%
Three-colour-composite (Subaru) image of the MACS1311 cluster, overlaid with a schematic of the KMOS daisy wheel observing strategy. Each cluster was observed with four separate but overlapping KMOS pointings (the $7'$ KMOS patrol fields are represented by orange circles). Each KMOS pointing includes the BCG, visible here for MACS1311 as the red elliptical galaxy in the centre of the image (where all four pointings overlap). This observing strategy allowed us to targets galaxies out to a projected distance of $2$--$3$ Mpc (depending on the cluster redshift) from each cluster center, providing a sample of galaxies with projected cluster-centric radii ranging from the core to the outskirts. %
     }%
\label{fig:kmospointings}
\end{figure*}

Galaxies were initially selected to be bright ($V<22$ for MACS1931 and MS2137, and $V<23$ otherwise), with a good {\sc bpz} SED fit (as defined by a {\sc bpz} modified chi-squared parameter, {\it chisq2} $ < 1$), and with a sharply-peaked unimodal redshift likelihood distribution (as defined by the {\sc bpz} parameter {\it ODDS} $ > 0.7$ for MS2137, MACS1311, and MACS2129, and {\it ODDS} $ > 0.8$ for MACS1931). We also required that the maximum likelihood {\sc bpz} redshift of each galaxy, $z_{\rm{b}}$, fell within the range $0.2 \leq z_{\rm{b}} \leq 0.6$, so that the H$\alpha$ and [N {\sc ii}] emission lines from each galaxy can be detected in the $IZ$-band of KMOS. These criteria were each chosen to maximise the likelihood of a nebular line detection with KMOS, whilst also maintaining a reasonable target sample size in the direction of each cluster. To observe galaxies in lines-of-sight toward both the core and the outskirts of each cluster, we adopted a ``daisy wheel" observing pattern combining four separate KMOS pointings per cluster, with each pointing allocating one KMOS arm to the BCG (Figure~\ref{fig:kmospointings}). This effectively imposed a further selection criterion, that targeted galaxies must fall within a roughly $7'$ radius of each cluster BCG.

Following this, priority was given first to those galaxies that were blue ($B-V \leq 0.9$ for $z_{\rm{b}} \leq 0.4$, and $V-R_{\rm{C}} \leq 0.9$ for $z_{\rm{b}} > 0.4$, such that each filter pair straddles the 4000~\AA\  break in the restframe of the galaxy) and with a $z_{\rm{b}}$ close to the cluster peak in the photometric redshift distribution. The latter was decided by visual inspection of the $z_{\rm{b}}$ distribution to manually identify a redshift window. The ``cluster peak'' was chosen as the peak of the $z_{\rm{b}}$ distribution closest to the $z_{\rm{b}}$ of the BCG ($z_{\rm{BCG}}$). The width of the window in redshift space was determined on a cluster-by-cluster basis, and was chosen to incorporate the full width of the peak down to a number count approximately equal to the background (field) population (typically $z_{\rm{peak}}\pm0.07$). This was designed to provide a sufficiently large, but nevertheless robust, sample of star-forming cluster members. We note that whilst selecting for a blue colour should increase the chance of detecting H$\alpha$ emission in the target galaxies, it will likely also preferentially exclude from our sample those galaxies in the cluster environment with the lowest star-formation rates \citep[SFRs; e.g.][]{Koopmann:2004a,Koopmann:2004b}. A full discussion of the SFRs of K-CLASH galaxies in cluster and field environments is presented in \S~\ref{subsubsec:comparingSFRs}.

To provide a sample of star-forming field galaxies, second priority was given to those galaxies that were blue and more likely to reside in the field i.e.\ those that were blue and did not fall within each cluster's redshift window defined above. We note that, as a compromise to maximise the number of blue targets for our KMOS observations, we impose no further constraint on the redshifts of these field candidates, beyond that their $z_{\rm{b}}$ places their H$\alpha$ emission within the KMOS $IZ$-band (unlike the {\it cluster} selection criteria listed in \S~\ref{subsec:fieldselect}, that also avoid strong telluric and sky emission line spectral regions). Third priority was given to red cluster galaxies, i.e.\ those galaxies within each cluster's redshift window with either $B-V > 0.9$ for clusters at $z \leq 0.4$ or $V-R_{\rm{C}} > 0.9$ for clusters at $z > 0.4$.

In summary, we target blue galaxies in each cluster field, giving preference to those most likely to reside in the cluster itself. In the absence of enough blue targets, we allocate the remaining KMOS arms to the red targets in each cluster field that are most likely to reside in the cluster. Since we rely on photometric redshifts to prioritise observations of likely cluster members, and since the fraction of star-forming systems should be lower in cluster environments compared to the field, we inevitably expect the number of star-forming cluster members we detect to be considerably smaller than the number of star-forming field galaxies detected.

\subsection{KMOS observations and data reduction}
\label{subsec:kmosobs}

All K-CLASH observations were carried out with KMOS on Unit Telescope 1 of the ESO VLT, Cerro Paranal, Chile. The K-CLASH observations were undertaken over two years, during ESO observing periods P97--P100\footnote{Programme IDs 097.A-0397, 098.A-0224, 099.A-0207, and 0100.A-0296.}. 

KMOS consists of 24 individual integral-field units (IFUs), each with a $2\farcs8 \times 2\farcs8$ field-of-view (FOV), deployable in a $7'$ diameter circular patrol field. The resolving power of KMOS in the $IZ$-band ranges from $R\approx2800$ to $\approx3800$, corresponding to a velocity resolution of $\sigma \approx 34$ -- $46$ km s$^{-1}$ (depending on the wavelength). The mean and standard deviation of the seeing (i.e.\ the full-width-at-half-maximum (FWHM) of the point spread function (PSF)) in the $IZ$-band for K-CLASH observations were $0\farcs78$ and $0\farcs15$, respectively.

As stated in \S~\ref{subsec:targetselection}, each of the four CLASH clusters was observed in a daisy wheel configuration of four KMOS pointings, as demonstrated for the MACS1311 cluster field in Figure~\ref{fig:kmospointings}. In each pointing we allocated one KMOS IFU to the BCG, meaning the BCG exposure time was, on average, four times longer than that of a typical galaxy in the K-CLASH sample. This aimed to maximise the chance of detecting low level nebular emission in these massive elliptical galaxies from residual star formation or AGN activity, or even detect stellar absorption lines. At least one additional IFU was allocated to a reference star in each pointing to monitor the PSF throughout the observations, and for improved centering when reconstructing data cubes. The remaining IFUs were allocated to various target galaxies according to the selection criteria and priorities laid out in \S~\ref{subsec:targetselection}.

Each KMOS pointing was observed for a total of 3--4.5 hours with KMOS (in 2--3 observing blocks lasting 1.5 hours each), in an ``OSOOSOOS'' nod-to-sky observing pattern, where ``O'' and ``S'' are on-source (i.e.\ science) and sky frames, respectively. The MS2137, MACS1931, and MACS2129 clusters were observed for a total of 16.5 hours each. MACS1311 was observed for a total of 12 hours.    

A single sky-subtracted (O-S) data cube was reconstructed for each science frame (i.e.\ each OS pair) using the standard ESO {\sc esorex} pipeline,\footnote{\url{http://www.eso.org/sci/software/cpl/download.html}} with the {\it skytweak} routine employed for further residual sky subtraction. The pipeline performs standard flat, dark, and arc calibrations during the reconstruction process. Each reconstructed cube was flux calibrated using corresponding observations of standard stars taken on the same night as the science data. Calibrated cubes were centred using the positioning of the reference star(s) observed in each science frame, and then stacked using the {\sc esorex} {\it ksigma} iterative ($3\sigma$) clipping routine. 

\section{KMOS Measurements}
\label{sec:kmosmeasurements}

In this section we describe the basic measurements taken from the KMOS data, including H$\alpha$ and [N {\sc ii}] fluxes for each galaxy (integrated within circular apertures) and the construction of two-dimensional maps of line properties and galaxy kinematics. We use some of these basic measurements to provide a summary of the K-CLASH sample in \S~\ref{sec:sampleoverview}. The galaxy properties we derive from these basic measurements are discussed in \S~\ref{sec:galaxyproperties} (along with additional galaxies properties we measure from ancillary data). 

\subsection{Aperture line fluxes and spectroscopic redshifts}
\label{subsec:measuringlinefluxes}

We measure the H$\alpha$ and [N {\sc ii}] fluxes of each galaxy by calculating the integrals of the H$\alpha$ and [N {\sc ii}] components of the best-fitting triple Gaussian triplet model to the observed H$\alpha$, [N {\sc ii}]6548 and [N {\sc ii}]6583 emission lines in the integrated galaxy spectrum. 

We extract galaxy spectra from the data cube of each K-CLASH galaxy in three circular apertures of increasing diameter. Each aperture is centred on the position of the peak of the continuum emission (or the peak of the H$\alpha$ flux if no continuum is detected) with a diameter of $0\farcs6$, $1\farcs2$, and $2\farcs4$, respectively. We consider multiple spectra extracted from different sized apertures (within the bounds of the KMOS IFUs' $2\farcs8 \times 2\farcs8$ fields-of-view) to account for differences in the spatial extent and distribution of H$\alpha$ flux between galaxies and to find the best compromise between maximising the source signal and minimising the noise for each galaxy we observe. 

Before fitting the emission lines, we first model and subtract any continuum emission in the extracted spectra. To account for the presence of (sometimes substantial) residual sky line emission in the cube, we adopt an iterative ``clip and fit" approach to accurately model the continuum emission whilst avoiding bias due to contaminating sky emission in iterative steps. An iterative approach is required since we see a distribution of amplitudes in the residual sky line emission across each spectrum. To model the continuum we divide each spectrum into segments of 50 pixels. We perform a 2$\sigma$ iterative clip on each segment to roughly remove residual sky line emission (resulting from over- or under-subtraction during the sky removal process). We then fit and subtract a 6$^{\rm{th}}$ order polynomial from the remaining flux. Following this, we repeat the sigma-clipping process to remove any remaining sky line emission, and fit and subtract a 3$^{\rm{rd}}$ order polynomial. Lastly, to account for non-perfect subtraction from the polynomial fitting, we calculate the median value of the resulting spectrum. We then construct our continuum model for the {\it original} spectrum as the sum of this median, and the two best-fitting polynomials. We subtract this continuum model from the original spectrum to produce our final, continuum-subtracted spectrum ready for emission line fitting.

\begin{figure*}
\centering
\begin{minipage}[]{1.\textwidth}
\centering
\includegraphics[width=.66\textwidth,trim= 10 0 0 0,clip=True]{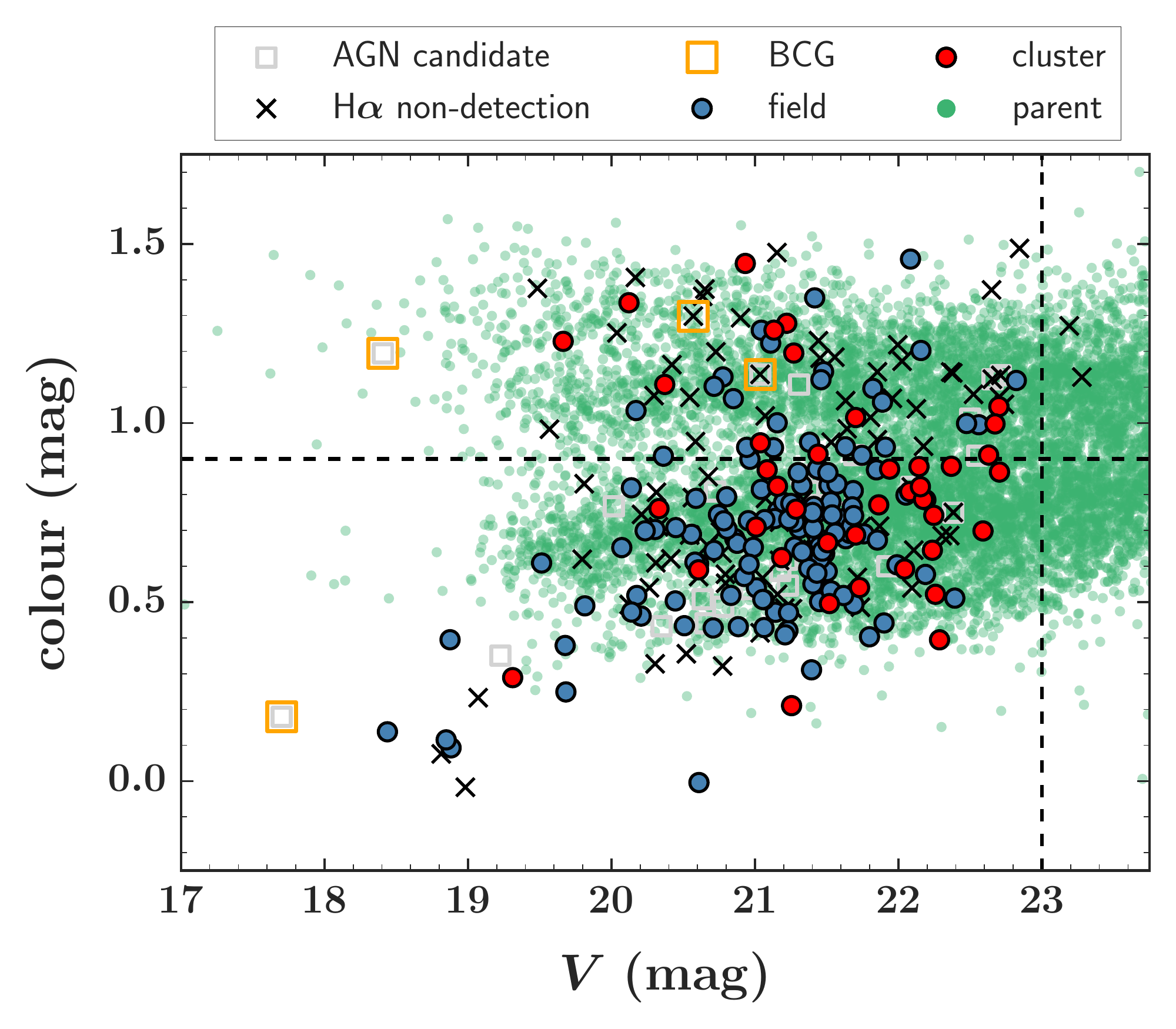}
\end{minipage}
\caption{%
The colour ($B-V$ for $z_{\rm{b}} \leq 0.4$, and $V-R_{\rm{C}}$ for $z_{\rm{b}} > 0.4$) of the K-CLASH galaxies as a function of their $V$-band magnitude. Our main selection preference is for galaxies that are blue (colour $\leq 0.9$ mag; horizontal dashed line), and bright ($V < 23$; vertical dashed line. We note that we only consider galaxies with $V<22$ for MACS1931 and MS2137. For clarity, we indicate whether each K-CLASH galaxy is a non-detection in H$\alpha$ (black crosses; see \S~\ref{subsec:measuringlinefluxes} and \S~\ref{subsec:detectionstats}), whether it is a BCG (yellow squares), or a candidate AGN host (grey squares; see \S~\ref{subsec:AGNselection}), and whether it is a member of either our {\it field} or {\it cluster} sub-samples (respectively blue and red filled circles; see \S~\ref{subsec:clustermembership}). For reference we also include the parent sample (green filled circles), comprising all galaxies in the K-CLASH fields, from which the K-CLASH targets were selected. K-CLASH galaxies detected in H$\alpha$ are predominantly blue in colour and are dominated by field galaxies. %
     }%
\label{fig:colourmagnitude}
\end{figure*}

To perform the emission line fit, we use a model comprising three Gaussians, representing the H$\alpha$ line and the [N {\sc ii}] doublet, that are forced to share a common width and redshift. The width and redshift themselves are free parameters during the fitting process. The intensities of the H$\alpha$ line and the [N {\sc ii}] doublet are also free to vary independently of one another, but the amplitudes of the [N {\sc ii}] lines are coupled according to theory, whereby the intensity of the bluer of the two lines is a factor of 2.95 less than that of the redder line \citep[e.g.][]{Acker:1989}. The best-fitting model is found via a $\chi^{2}$ minimisation process using {\sc mpfit} \citep{Markwardt:2009} in {\sc Python} that employs the Levenberg–Marquardt least-squares fitting algorithm. 

We deem a galaxy to be detected in H$\alpha$ emission if it has a signal-to-noise ratio in the H$\alpha$ line, S/N$_{\rm{H}\alpha} \geq 5$ in at least one of the spectra extracted from different apertures. We adopt the methods of the KROSS survey \citep[e.g.][]{Stott:2016,Harrison:2017,Tiley:2019a} to calculate this as 

\begin{equation}
\rm{S/N}_{\rm{H}\alpha} \equiv \sqrt{\chi^{2}_{\rm{H}\alpha}-\chi^{2}_{\rm{base}}}\,\,,
\label{eq:S2N}
\end{equation}

\noindent where $\chi^{2}_{\rm{H}\alpha}$ is the chi-squared of the H$\alpha$ component of the best-fitting triplet, and $\chi^{2}_{\rm{base}}$ is the chi-squared of a horizontal line with zero-point equal to the median of the baseline in a region near to the line emission
\citep[but excluding the emission itself; e.g.][]{Neyman:1933,Bollen:1989,Labatie:2012}. At this stage we also visually inspect the fit to each spectrum to ensure the best fit is not biased by remaining bright sky lines. We take the spectroscopic redshift of the galaxy as the best-fitting redshift from the spectrum with the highest S/N$_{\rm{H}\alpha}$ (out of the three apertures).

\subsection{Constructing KMOS maps}
\label{subsec:constructingmaps}

To map the spatially-resolved line properties of the K-CLASH galaxies, we first model and subtract any stellar continuum from the data cube of each (where the data cube is re-sampled from the native $0\farcs2$ spaxels to $0\farcs1$ spaxels), before modelling the H$\alpha$ and [N {\sc ii}] nebular line emission from small groups of spaxels at each spatial position in the cube. We adopt the same method described in \S~\ref{subsec:measuringlinefluxes} to subtract the continuum and model the emission lines. To construct the maps, we employ an adaptive spatial smoothing process similar to that used in the KROSS survey \citep[e.g. in][]{Stott:2016} and equivalent to an adaptive convolution with a square top-hat kernel, whereby for each spaxel we sum the flux from spectra in an increasing number of surrounding spaxels until we achieve $\rm{S/N}_{\rm{H}\alpha} \geq 5$. We start by summing the flux within a $0\farcs3 \times 0\farcs3$ bin (i.e.\ $3 \times 3$ $0\farcs1$ spaxels - approximately equivalent to half the FWHM of the typical PSF) centred on the spaxel in question. If $\rm{S/N}_{\rm{H}\alpha} < 5$, we then consider a $0\farcs5 \times 0\farcs5$ bin, and finally a $0\farcs7 \times 0\farcs7$ bin. If we have still not achieved $\rm{S/N}_{\rm{H}\alpha} \geq 5$ with a $0\farcs7 \times 0\farcs7$ bin, we mask this spaxel in the resulting maps. This process is repeated for every spaxel in the cube, to create maps of the best-fitting emission line properties as a function of spatial position across each galaxy. 

In this work, we construct maps of the H$\alpha$ intensity, mean line-of-sight velocity ($v_{\rm{obs}}$), and line-of-sight velocity dispersion ($\sigma_{\rm{obs}}$) by considering respectively the integral, central position, and (sigma-)width of the best-fitting Gaussian to the H$\alpha$ emission. The velocities and velocity dispersions are calculated in the rest frame of each galaxy. We apply an iterative masking process to the maps, described in Appendix~\ref{sec:mapmasking}, to remove ``bad'' pixels where the fit to the emission lines is adversely affected by residual sky contamination and edge effects, as well as (spatially) non-resolved features. 

The spatially-resolved gas properties and kinematic measurements of the K-CLASH galaxies will be the subject of detailed studies, the results of which will be presented in subsequent papers. In this paper, we therefore limit our analysis of the spatially-resolved properties to providing an overview of the numbers of galaxies we spatially-resolve in H$\alpha$ emission in \S~\ref{subsec:resolvednumbers}, as well as presenting maps for a sub-set of the H$\alpha$-resolved systems (\S~\ref{subsubsec:maps}).

\section{K-CLASH Sample Overview}
\label{sec:sampleoverview}

In this section, we provide an overview of the final targeted K-CLASH sample. In \S~\ref{subsec:detectionstats} we provide an outline of the KMOS detection statistics. In \S~\ref{subsec:AGNselection} we describe our methods to flag candidate AGN hosts within the sample, and in \S~\ref{subsec:clustermembership} we explain how we determine whether a K-CLASH galaxy resides in a cluster or field environment. Finally, in \S~\ref{subsec:resolvednumbers}, we provide a summary of the number of K-CLASH galaxies that we spatially-resolve in H$\alpha$ emission.

The positions of the K-CLASH galaxies in the visible colour-magnitude plane (used for initial target selection) and the normalised cluster-centric radius-velocity plane (used to determine cluster membership) are shown in respectively Figure~\ref{fig:colourmagnitude} and \ref{fig:R_V}. In each case, we highlight whether or not a galaxy is detected in H$\alpha$ with KMOS, whether it is a candidate AGN host, and whether it resides in a cluster or the field. In Table~\ref{tab:kclashvals} we provide flags for each K-CLASH galaxy detailing these same classifications. We also include each galaxy's coordinates on the sky, along with other key derived properties discussed in \S~\ref{sec:galaxyproperties}.

\subsection{Detection statistics}
\label{subsec:detectionstats}

In total we observed 282 galaxies with KMOS across the four CLASH cluster fields. Visual inspection of the data reveals either stellar continuum or nebular (H$\alpha$ or [N {\sc ii}]) line emission in 243 (86 per cent of the) galaxies in our sample. However, we only robustly detect H$\alpha$ emission in 191 (68 per cent of the) galaxies in our sample, in the galaxy's integrated spectrum (see \S~\ref{subsec:measuringlinefluxes}). We note that the H$\alpha$-detected galaxies include the BCGs of MACS1931 and MS2137, but not those of MACS1311 and MACS2129.

Of the 191 H$\alpha$-detected galaxies, 149 (78 per cent) were targeted as blue systems in the initial selection (\S~\ref{subsec:targetselection}; including the BCG of MACS1931). The remaining 42 (22 per cent of the) detected galaxies were red systems (including the BCG of MS2137). Conversely, of the 91 galaxies not detected in H$\alpha$, 50 (55 per cent) are blue and 41 (45 per cent) are red (the latter including the BCGs of MACS1311 and MACS2129). In other words, we targeted 199 blue galaxies and 83 red galaxies, and detected H$\alpha$ emission in 149 (75 per cent) and 42 (51 per cent) of them, respectively. 

The lower H$\alpha$ detection rate of the lower priority (i.e.\ red) galaxies is perhaps unsurprising given that their colours, ignoring the effects of any dust in the galaxies, suggest these are systems less likely to exhibit substantial ongoing star formation (and thus to have large H$\alpha$ fluxes). Our H$\alpha$ detection rate of 75 per cent for our primary targets (i.e. blue and bright systems) implies that our target selection cuts were effective in selecting star-forming (or at least H$\alpha$-emitting) galaxies. There are two possible explanations for  why we do not detect H$\alpha$ emission from the remaining 25 per cent of blue targets. Firstly, the wavelength of the H$\alpha$ line emission coincides with that of strong telluric features or bright sky line emission or, secondly, the H$\alpha$ emission is weak and/or the depth of our KMOS observations is insufficient to robustly detect it (or a combination of both). 

The latter explanation is the least likely to account for the blue non-detections given that the distribution of $V$-band luminosities (that should correlate with the galaxies' stellar masses and star-formation rates and thus their H$\alpha$ luminosities) for the detected and non-detected blue targets are very similar. Here we calculate the extinction corrected $V$-band luminosity as

\begin{equation}
L_{V} = 4 \pi D_{L}^{2}\ 10^{0.4A_{V}}\ F_{V}\,\,, 
\label{eq:L_V}
\end{equation}

\noindent where $D_{L}$ is the luminosity distance calculated from the galaxy redshift (spectroscopic for galaxies detected in H$\alpha$, {\sc bpz} otherwise) assuming a {\it WMAP9} cosmology, $F_{V}$ is the $V$-band flux (\S~\ref{subsec:stellarmasses}) and $A_{V}$ is the (inferred) $V$-band extinction (\S~\ref{subsubsec:halphaSFRs}). A Kolmogorov-Smirnov (K-S) two-sample test between the two luminosity distributions, assuming a null hypothesis that the two are drawn from the same underlying distribution, returns a $p$-value of $p = 0.40$ (2 s.f.). Adopting a critical value of $p = 0.05$, we cannot reject the null hypothesis. We conclude that the blue H$\alpha$ non-detections are not intrinsically fainter than the blue detections in the $V$-band. Assuming a correlation with the H$\alpha$ luminosity, we find no evidence to suggest that the blue non-detections are intrinsically fainter in their H$\alpha$ emission than the blue detections. However, we still cannot definitively rule out that, despite their colours and $V$-band luminosity, the non-detected blue targets (or a sub-set of them) may simply have H$\alpha$ fluxes below the K-CLASH detection limit. 

Considering the photometric redshifts of the 50 non-detected blue targets, 36 suggest that the galaxy's H$\alpha$ emission may fall within a strong telluric absorption feature in the $IZ$-band, implying we are unlikely to detect it. The remaining 14 photometric redshifts suggest that the H$\alpha$ emission falls within strong sky line emission features. However, given the limited accuracy of the photometric redshift estimates, we cannot say for certain that this is the case. We can only conclude that it is possible that a large number of the blue non-detections are the result of a conspiracy between their redshifts and the positions of strong telluric absorption and sky line emission features, implying their H$\alpha$ emission is either strongly absorbed by the atmosphere or is strongly confused with residual sky line emission (resulting from non-perfect sky subtraction). 

The red non-detections in K-CLASH follow a similar story, 21 of 41 galaxies having a photometric redshift suggesting the redshifted H$\alpha$ emission should fall within strong telluric features in the $IZ$-band, and a further 15 indicating that H$\alpha$ may be observed at a wavelength corresponding to bright sky line emission features. The remaining 5 red non-detections are most likely due to intrinsically low SFRs (and thus correspondingly low H$\alpha$ luminosities). We note that, given their colours and the large inaccuracies of their photometric redshifts, this is probably also the case for {\it any} of the 41 red non-detections.

\subsection{AGN candidate identification}
\label{subsec:AGNselection}

Since for the K-CLASH survey we are interested in ``normal'' star-forming systems, before undertaking any analysis it is important to first determine whether the nebular emission detected from any given K-CLASH galaxy is driven by heating from ionising photons from ongoing star formation (i.e. young, massive stars) or rather from an AGN. The limited wavelength range of our KMOS observations does not encompass the H$\beta$ and [O{\sc iii}] emission lines for all of our targets, so we are unable to place our objects on many of the common emission line diagnostic diagrams used to identify AGN contamination \citep[e.g. the BPT diagram;][]{BPT, Kauffmann:2003, Kewley:2006}. Instead, we turn to ancillary data as well as the [N{\sc ii}]/H$\alpha$ flux ratio to identify candidate AGN hosts within our sample galaxies.

We first cross-match our sample with publicly available X-ray imaging from the {\it Chandra} Advanced CCD Imaging Spectrometer \citep[ACIS; e.g.][]{Grant:2014} Survey of X-ray Point Sources \citep{Wang:2016}, verifying that {\it Chandra} pointings overlap fully with our observations. In total, 6 of our target galaxies have detected X-ray emission, including 3 out of the 4 BCGs. Of these 6 galaxies, we detect 5 in H$\alpha$ emission. We infer X-ray luminosities in the range $10^{42}-10^{44}$ erg s$^{-1}$ for these objects. These luminosities are not corrected for absorption by Galactic hydrogen along the line-of-sight, and as such should be treated as lower limits. It should also be noted that these X-ray observations do not have a uniform exposure time, with exposures varying from 10,847 seconds for MACS2129 to 98,922 seconds for MACS1931, implying that we may be missing weak X-ray AGN in the shallower fields. 

\begin{figure*}
\centering
\begin{minipage}[]{1.\textwidth}
\centering
\includegraphics[width=0.63\textwidth, trim = 50 0 50 0]{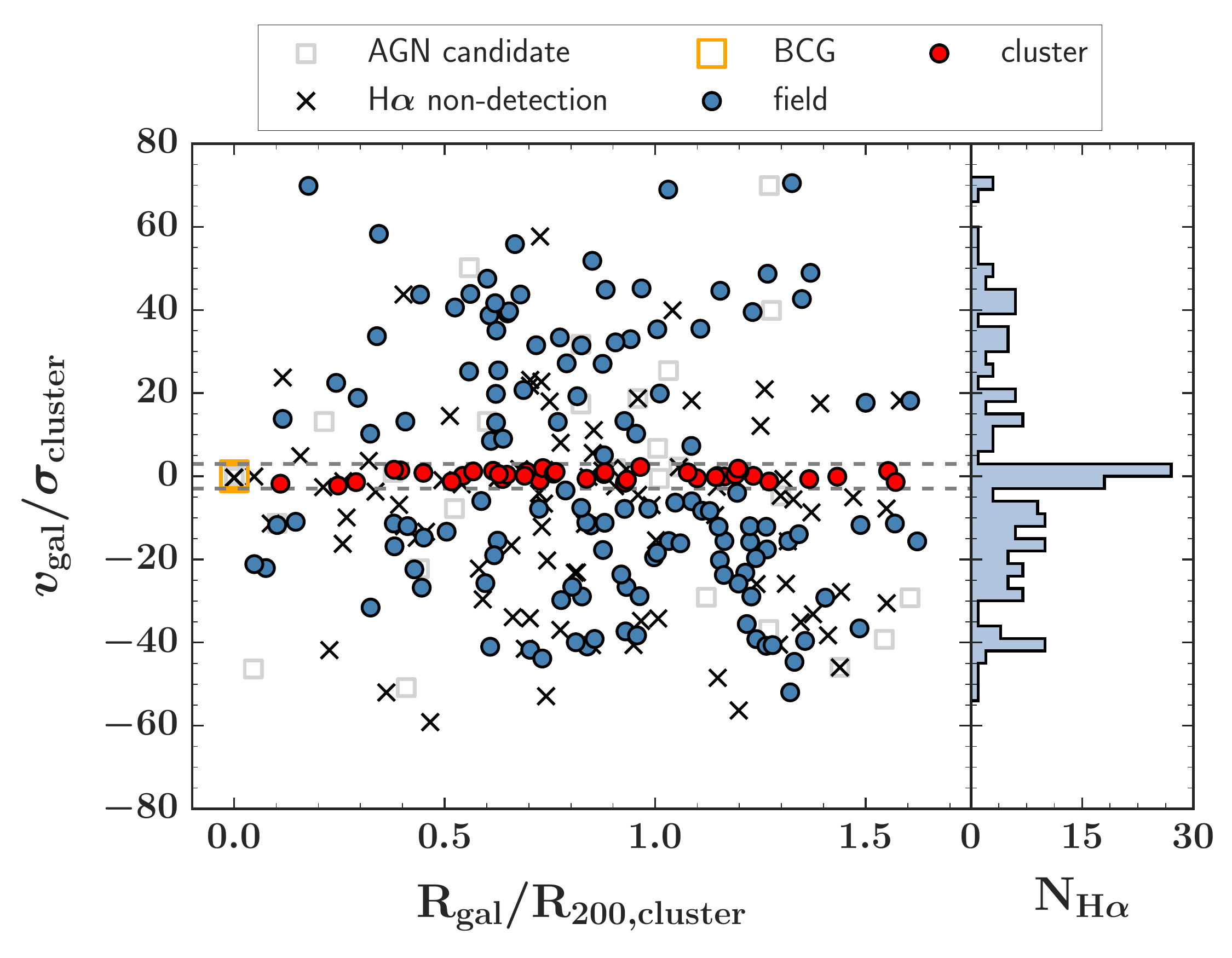}
\end{minipage}
\caption{{\bf Left:} The positions of the K-CLASH galaxies in the normalised, cluster-centric radius-velocity plane. For the non-H$\alpha$-detected MACS1311 and MACS2129 BCGs, here we adopt the spectroscopic redshifts tabulated in the \citet{Molino:2017} catalogues, available for each cluster at \url{https://archive.stsci.edu/missions/hlsp/clash/}. Symbols are as described in Figure~\ref{fig:colourmagnitude}. We include dashed horizontal lines to highlight the $v_{\rm{gal}} = \pm3\sigma_{\rm{cluster}}$ cluster boundaries in the normalised cluster-centric radius-velocity plane. The majority of H$\alpha$-detected K-CLASH galaxies are located outside of the four clusters targeted with KMOS (i.e.\ they are field galaxies). {\bf Right:} The number of H$\alpha$-detected galaxies ($\rm{N}_{\rm{H}\alpha}$) in bins of $v_{\rm{gal}}/\sigma_{\rm{cluster}}$, which peaks in the range $-3 \leq v_{\rm{gal}}/\sigma_{\rm{cluster}} < 3$, corresponding to the cluster boundaries in cluster-centric normalised velocity space. } 
\label{fig:R_V}
\end{figure*}

Next, we cross-match our observations with the {\it Wide-field Infrared Survey Explorer} \citep[{\it WISE};][]{Wright:2010,NeoWISE} All{\it WISE} source catalogue\footnote{\url{http://wise2.ipac.caltech.edu/docs/release/allwise/}}, requiring spatial offsets of less than $10''$, corresponding to $\approx1.5$ times the FWHM of the {\it WISE} PSF. A total of 231 K-CLASH targets have a corresponding entry in the catalogue. We use the {\it WISE} observations in the $W1$ and $W2$ bands to select candidate AGN hosts, adopting a criterion $W1 - W2 > 0.8$ following the method of \citet{Stern:2012aa}. A total of 13 objects satisfy this criterion, of which we detect 10 in H$\alpha$. Since not all the K-CLASH targets are detected in {\it WISE}, we also apply an additional (weak) cut to our sample, selecting AGN hosts based on their colours in the {\it Spitzer} Infrared Array Camera \citep[IRAC;][]{Fazio:2004} $3.6$~$\mu$m ($[3.6]$) and $4.5$~$\mu$m ($[4.5]$) channels. Unfortunately, we are unable to use the full ($[3.6]-[4.5]$ versus $[5.8]-[8.0]$) colour-colour cuts from \citet{Donley:2012} since no $5.8$~$\mu$m nor $8.0$~$\mu$m {\it Spitzer} IRAC observation is available for the K-CLASH sample galaxies. Nevertheless, we are able to identify a single candidate AGN host with a conservative cut of $[3.6]-[4.5]>1.0$ (based on the full \citealt{Donley:2012} selection), for which we do not detect H$\alpha$ emission. We also note that four K-CLASH targets were detected neither in {\it WISE} imaging nor mid-infrared (MIR) {\it Spitzer} observations, meaning we were unable to establish whether they are candidate AGN hosts or not from their MIR colour.

As a final step, we directly identify galaxies that are likely to suffer from a strong contribution to their nebular line emission from non-stellar excitation sources by examining for each object the ratio of the [N {\sc ii}] to H$\alpha$ integrated line flux in a $1\farcs2$ diameter circular aperture. We classify those galaxies with $\log_{10}$([N {\sc ii}]$/\rm{H}\alpha)> -0.1$ as likely AGN hosts \citep[similarly to][]{Wisnioski:2018}. A total of 13 objects satisfy this criterion, all of which we robustly detect H$\alpha$ emission from. We note that whilst these objects are strong candidates to each host an AGN, we can still fail to identify galaxies hosting weak AGN surrounded by regions of strong star formation.

In summary, we identify the following candidate AGN hosts within the K-CLASH sample:

\begin{itemize}
\item 6 galaxies detected in X-rays, of which 5 are detected in H$\alpha$;
\item 13 galaxies with a {\it WISE} colour $W_1-W_2>0.8$, of which 10 are detected in H$\alpha$;
\item 1 galaxy with a \textit{Spitzer} colour $[3.6]-[4.5]>1.0$, that is undetected in H$\alpha$;
\item 13 galaxies with $\log_{10}($[N {\sc ii}]/H$\alpha{})>-0.1$, all of which are detected in H$\alpha$. 
\end{itemize}

In total, we identify 28 unique K-CLASH galaxies that each potentially host an AGN, detecting 23 of these in H$\alpha$ with KMOS. Only 5 of these are classified as a candidate host using two or more of the diagnostics described above. In total, we thus find that $\approx10$ per cent of the K-CLASH targets are likely to host an AGN. This fraction is lower than that found in the KMOS$^{3\rm{D}}$ survey (25 per cent of ``normal" galaxies at $0.6<z<2.7$ with $9.0<\log(\rm{M}_{*}/\rm{M}_{\odot})<11.7$ are candidate hosts; \citealt{ForsterSchreiber:2018}), but is consistent with the findings of \cite{Kartaltepe:2010}, who use multi-wavelength observations of $0.3<z<1$ star-forming galaxies to establish that 10--20 per cent of them host an AGN. This is perhaps unsurprising, since we expect a steep decline in AGN activity from $z\approx1$ to the present day \citep[e.g.][]{Fanidakis:2012}.

Finally we note that each of the BCGs, except that of MACS1311, is a candidate AGN host. This includes the two BCGs that we detect in H$\alpha$ emission (belonging to MACS1931 and MS2137).

\subsection{Cluster member identification}
\label{subsec:clustermembership}

In this sub-section, we describe our method to determine cluster membership for the K-CLASH galaxies. We employ a simple set of criteria, based on sky position (i.e.\ right ascension and declination) and redshift, to define whether a galaxy is associated with one of the four targeted CLASH clusters, and to separate the sample into two crude categories: galaxy cluster members and galaxies in the field (i.e.\ not in a cluster environment). We note here that we also considered a more sophisticated, probabilistic mixture model to determine cluster membership, that gives a continuous measure of the probability that each galaxy resides in the cluster environment. However, given that our results remain unchanged whether we employ this more complex approach or the more basic one, we prefer to adopt the simpler of the two methods, which we outline here. 

Due to the large scatter between the photometric and spectroscopic redshifts (the median difference between the two measurements for H$\alpha$-detected galaxies is only $-0.011 \pm 0.007$, but the standard deviation of the same distribution is $0.18 \pm 0.08$), we only attempt to determine cluster membership for galaxies with a robust measurement of the latter, i.e.\ those detected in H$\alpha$ (S/N$_{\rm{H}\alpha} \geq 5$). The number of H$\alpha$-detected K-CLASH galaxies is not sufficiently large nor complete to define each cluster's radial escape velocity profile based on the cluster members themselves, as is commonly done in larger spectroscopic surveys \citep[e.g.][]{Owers:2017}. We therefore instead turn to previously published properties of the CLASH clusters to infer the membership of the galaxies in our sample.  Specifically, we use the Navarro-Frenk-White \citep[NFW;][]{Navarro:1997} scale radius ($r_{\rm{s}}$) and concentration ($c_{200}$) determined from combined strong and weak lensing analyses of the CLASH clusters \citep{Zitrin:2015}\footnote{We adopt the best-fitting parameters from the most recent version of the cluster lens models, provided by Zitrin et al. (private communication).}, as well as the cluster X-ray properties from \cite{Postman:2012}. 

We calculate the size of each cluster as 

\begin{equation}
\frac{\rm{R}_{200,\rm{cluster}}}{\rm{kpc}} = c_{200} \times \frac{r_{\rm{s}}}{\rm{kpc}}\,\,,
\label{eq:R200}
\end{equation}

\noindent where $\rm{R}_{200,\rm{cluster}}$ is the radius within which the mean density is 200 times the critical density at the cluster's redshift \citep[which corresponds to the virial radius, e.g.][]{Navarro:1996,White:2001}. We calculate the predicted cluster velocity dispersion, $\sigma_{\rm{cluster}}$, from the cluster X-ray temperature assuming the $\sigma_{\rm{cluster}}$--$T_{\rm{X}}$ relation for clusters of \citet{Girardi:1996}. We note that we also verified that our cluster selection remains unchanged if we instead adopt the $\sigma_{\rm{cluster}}$--$T_{\rm{X}}$ relation from \citet{Wu:1999}, or if we assume a hydrostatic isothermal model with a perfect galaxy-gas energy equipartition ($\sigma_{\rm{cluster}}^{2} = k_{\rm{B}}T_{\rm{X}}/\mu m_{\rm{p}}$, where $k_{\rm{B}}$, $m_{\rm{p}}$, and $\mu$ are respectively the Boltzmann constant, the mass of the proton, and the mean molecular weight). Unless otherwise stated in Table~\ref{tab:clashclusters}, we take the redshift of the cluster ($z_{\rm{cluster}}$) as the published value from \citet{Postman:2012}. For each galaxy, we then use its spectroscopic redshift ($z_{\rm{spec}}$) to calculate its projected velocity with respect to the rest-frame of the cluster as 

\begin{equation}
v_{\rm{gal}}=c \times \frac{z_{\rm{spec}}-z_{\rm{cluster}}}{1+z_{\rm{cluster}}}\,\,,
\end{equation}

\vspace{0.25cm}

\noindent where $c$ is the speed of light.

Our approach is then to simply classify those galaxies with a BCG-centric projected distance $R_{\rm{gal}} < 2R_{200,\rm{cluster}}$ and a projected cluster-centric velocity $|v_{\rm{gal}}| < 3\sigma_{\rm{cluster}}$ as cluster members. Correspondingly, those galaxies that do not satisfy these conditions are deemed to reside in the field. The ranges in radius and velocity used to define cluster membership were a conservative compromise, to account for the possibility that the clusters may not be completely relaxed whilst also minimising contamination from non-cluster members (i.e. field galaxies). In practice, all the galaxies targeted by K-CLASH comfortably satisfy the projected radius criterion (with projected radii $\lesssim 1.6\ \rm{R}_{200,\rm{cluster}}$), making this criterion redundant in this case. Whilst basic, our approach is straightforward and, as mentioned previously, provides similar results to more sophisticated methods of selecting cluster members. 

Excluding BCGs, we deem 45 of the 191 H$\alpha$-detected K-CLASH galaxies to be cluster members and the remaining 146 to reside in the field, i.e. the vast majority ($76$ per cent) of K-CLASH galaxies detected in H$\alpha$ reside in a field environment, with only $24$ per cent made up of cluster members. As discussed in \S~\ref{subsec:targetselection}, this is as expected given our simple photometric redshift and colour selection criteria. Using the criteria outlined in \S~\ref{subsec:AGNselection}, we find that 5 out of the 45 H$\alpha$-detected cluster members (including two BCGs) and 18 out of the 146 field galaxies are candidate AGN hosts. Since for K-CLASH we are predominantly interested in ``normal'' star-forming systems, we exclude these AGN candidates from any further analysis, leaving 40 cluster members and 128 field galaxies (respectively 24 and 76 per cent of H$\alpha$ detected targets, excluding AGN). We hereafter refer to these two sub-samples as respectively the {\it field} sub-sample (or {\it field} galaxies) and the {\it cluster} sub-sample (or {\it cluster} galaxies). The positions of the two sub-samples in the cluster-centric radius-velocity plane are shown in Figure~\ref{fig:R_V}. For context, we also show H$\alpha$ non-detections (for which the cluster-centric velocity is calculated from the photometric redshift) and candidate AGN hosts.

\subsection{Spatially-resolved H$\alpha$ emission}
\label{subsec:resolvednumbers}

In \S~\ref{subsec:constructingmaps}, we described our method to construct spatially-resolved maps of the line properties and kinematics of the K-CLASH galaxies from their KMOS data cubes. In this sub-section, we provide a summary of the number of K-CLASH targets that are spatially resolved in H$\alpha$ emission. 

We consider a galaxy to be spatially resolved in H$\alpha$ if its maps, following bad pixel masking, contain at least one contiguous region of pixels with an area larger than one resolution element (as defined by the FWHM of the KMOS PSF). We include a 10 per cent margin of error to consider only those galaxies that are robustly resolved and to exclude marginal cases. In total, we spatially resolve H$\alpha$ emission in 146 K-CLASH galaxies, corresponding to 76 per cent of those detected in H$\alpha$ (but only 52 per cent of targeted galaxies). These 146 spatially-resolved systems comprise 34 (85 per cent of) {\it cluster} galaxies and 94 (73 per cent of) {\it field} galaxies. The remaining 18 are candidate AGN hosts, which we ignore in this work (see \S~\ref{subsec:AGNselection}). As discussed in \S~\ref{subsec:constructingmaps}, the spatially-resolved properties of the K-CLASH sample will be the subject of future papers, so we refrain from a full analysis in this work. However, we do provide a brief discussion of these properties in \S~\ref{subsubsec:maps}, where we also present example KMOS maps of a subset of spatially-resolved targets. 

\section{K-CLASH Galaxy Properties}
\label{sec:galaxyproperties}

In this section, we present the key properties of the K-CLASH galaxies. In \S~\ref{subsec:stellarmasses} we outline our method to calculate the galaxy stellar masses by modelling their spectral energy distributions (SEDs) using several different SED-fitting routines. We describe our measurements of the stellar sizes of the K-CLASH galaxies in \S~\ref{subsec:sizes}. In \S~\ref{subsec:fluxandsfr} we explain how we calculate the total H$\alpha$ line flux of each galaxy in our sample, and detail our prescription for estimating the galaxies' H$\alpha$-derived SFRs. Finally, in \S~\ref{subsubsec:maps} we present example KMOS maps of the emission line properties and kinematics of a subset of those K-CLASH galaxies that are spatially-resolved in H$\alpha$ emission.

The (integrated) galaxy properties presented in this section are tabulated for each K-CLASH galaxy in Table~\ref{tab:kclashvals}.

\subsection{Stellar masses}
\label{subsec:stellarmasses}

\begin{figure*}
\centering
\begin{minipage}[]{1.\textwidth}
\centering
\includegraphics[width=0.67\textwidth,trim = 10 0 0 0]{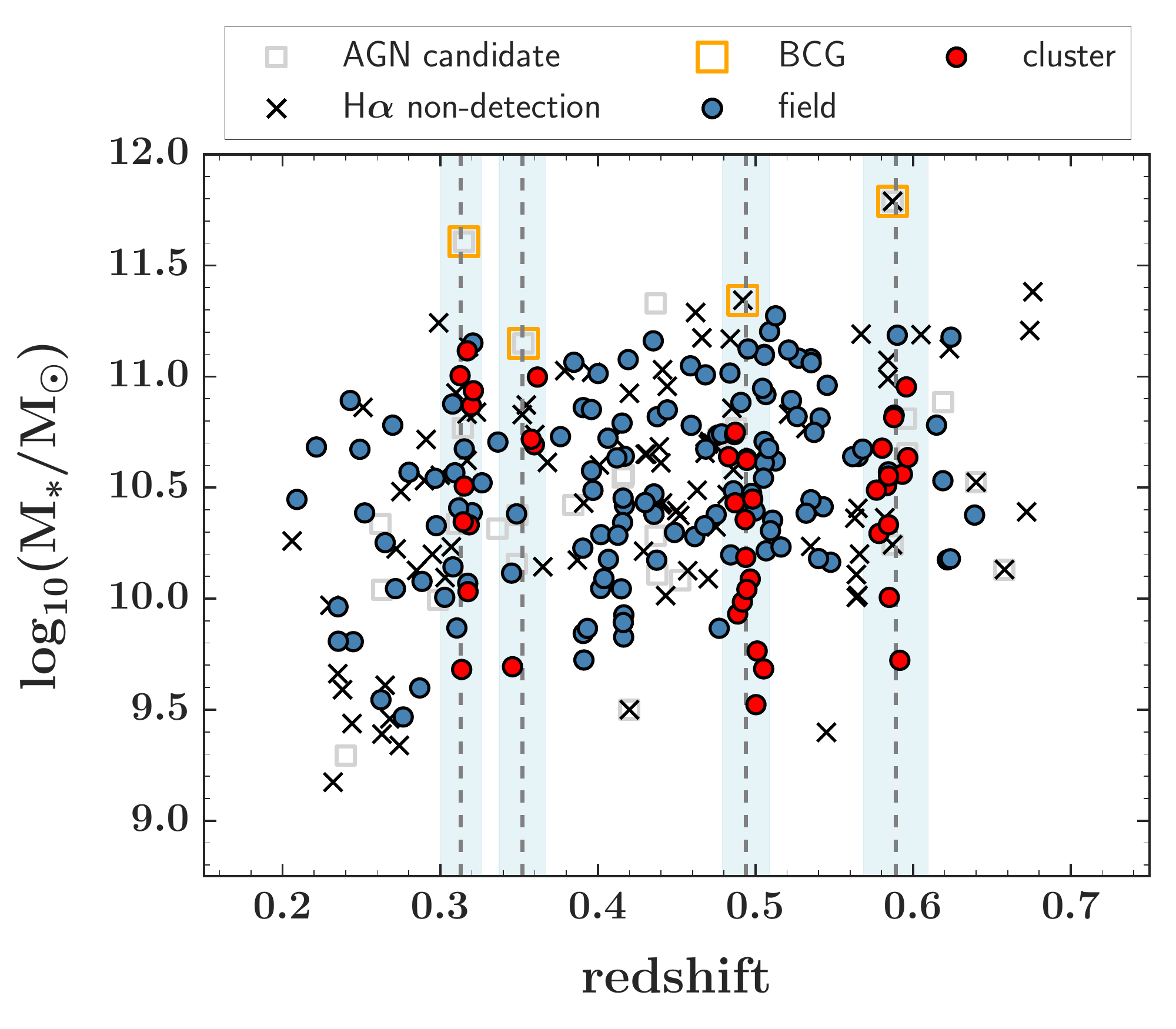}
\end{minipage}
\caption{The position of the K-CLASH galaxies in the stellar mass-redshift plane. For the non-H$\alpha$-detected MACS1311 and MACS2129 BCGs, here we adopt the spectroscopic redshifts tabulated in the \citet{Molino:2017} catalogues available  for each cluster at \url{https://archive.stsci.edu/missions/hlsp/clash/}. Symbols are as described in Figure~\ref{fig:colourmagnitude}. For each cluster we indicate its redshift and the redshift range that corresponds to $\pm3\sigma_{\rm{cluster}}$ with respectively a grey dashed line and a pale-blue vertical band. The stellar masses of those galaxies detected in H$\alpha$ at higher redshifts are slightly larger than those detected at lower redshifts, as expected for an apparent magnitude (i.e.\ flux)-limited sample (given that H$\alpha$ luminosity should correlate with stellar mass for star-forming galaxies). } 
\label{fig:mass_selection}
\end{figure*}

We derive stellar masses for the K-CLASH galaxies by exploiting the wide, multi-wavelength optical ($B$-, $V$-, $R_{\rm{C}}$-, $I_{\rm{C}}$-, $Z$-, and $z'$-band) and infrared ($3.6$ and $4.5$~$\mu$m {\it Spitzer}) imaging available for the four K-CLASH clusters (see \S~\ref{subsec:targetselection}). The bandpasses are consistent across the clusters, with the exception of MACS1311 for which there is no wide $I_{\rm{C}}$-band imaging available, $B$- and $V$-band imaging is from the ESO Wide-Field Imager rather than Subaru, and $z'$-band IMACS imaging is used in place of the Subaru $Z$-band imaging. For the purposes of SED fitting, we measure the flux in each band for each galaxy in a $4''$ radius aperture to ensure we consider the total stellar light in each filter. Stellar mass estimates are then derived by fitting the SED of each galaxy using three independent fitting routines: {\sc Multi-wavelength Analysis of Galaxy Physical Properties} \citep[{\sc magphys};][]{Cunha:2008}, {\sc LePhare} \citep[][]{Arnouts:1999,Ilbert:2006}, and {\sc ProSpect} (Robotham et al., in press; but see \citealt{Lagos:2019} for a detailed description of the generative mode). 

\subsubsection{SED-fitting routines}

Each of the {\sc LePhare}, {\sc magphys}, and {\sc ProSpect} routines allows comparison of a galaxy's observed SED to a suite of template SEDs from \citet[][]{Bruzual:2003aa}, hereafter referred to as BC03. For {\sc magphys} we also fit each galaxy's SED using a second suite of templates from Charlot \& Bruzual (in preparation; commonly referred to in the literature, and hereafter, as CB07). For each routine, a fixed \citet{Chabrier:2003} IMF is assumed. We supply the routines with the spectroscopic redshifts of the galaxies detected in H$\alpha$, and the {\sc bpz} photometric redshifts otherwise. 

The {\sc LePhare} and {\sc magphys} SED fitting codes are well known and both have been widely adopted amongst the extragalactic astronomy community for considerable periods. We therefore only provide a brief summary of the key aspects of each of these two codes. The {\sc LePhare} fitting routine finds the model SED that best fits the observed data and reports the corresponding extinction, metallicity, age, star-formation rate and stellar mass of the model. The suite of model SEDs generated by {\sc LePhare} incorporates a variety of star-formation histories (SFHs) - namely a single burst, an exponential decline, or  a constant star formation. The {\sc magphys} routine, like {\sc LePhare}, compares suites of model SEDs to the observed data. Unlike {\sc LePhare}, {\sc magphys} consistently models the ultraviolet (UV), optical and infrared SEDs of galaxies, allowing for emission in the UV to be absorbed by dust and re-emitted in the infrared according to the \citet{Charlot:2000} model for dust attenuation of starlight. The {\sc magphys} routine also returns variables describing the best-fitting age, SFH, metallicity, and the magnitude of dust attenuation for each observed SED. The SFHs allowed in the {\sc magphys} routine are less varied than {\sc LePhare}, allowing only for continuous star formation with additional ``bursty" episodes. 

{\sc ProSpect} is the newest of the three SED fitting routines. The code is publicly available from GitHub\footnote{\url{https://github.com/asgr/ProSpect}}, including full documentation. For a detailed description of the workings of the code, see \citet{Lagos:2019} (and Robotham et al., in press). Here we provide a brief overview. 

{\sc ProSpect} is similar in concept to {\sc magphys} in so far as its basic approach to fitting a galaxy's observed SED is to take a model galaxy spectrum, attenuate its light and re-emit it at redder wavelengths, place this attenuated spectrum at a given redshift, and then pass it through a chosen set of filters to produce a SED. This model SED is then compared to the observed galaxy SED. This process is iterated over a library of model spectra, with each assigned appropriate age weightings to simulate a desired SFH. 

\begin{figure*}
\centering
\begin{minipage}[]{1.\textwidth}
\centering
\includegraphics[width=0.67\textwidth, trim = 10 0 0 0]{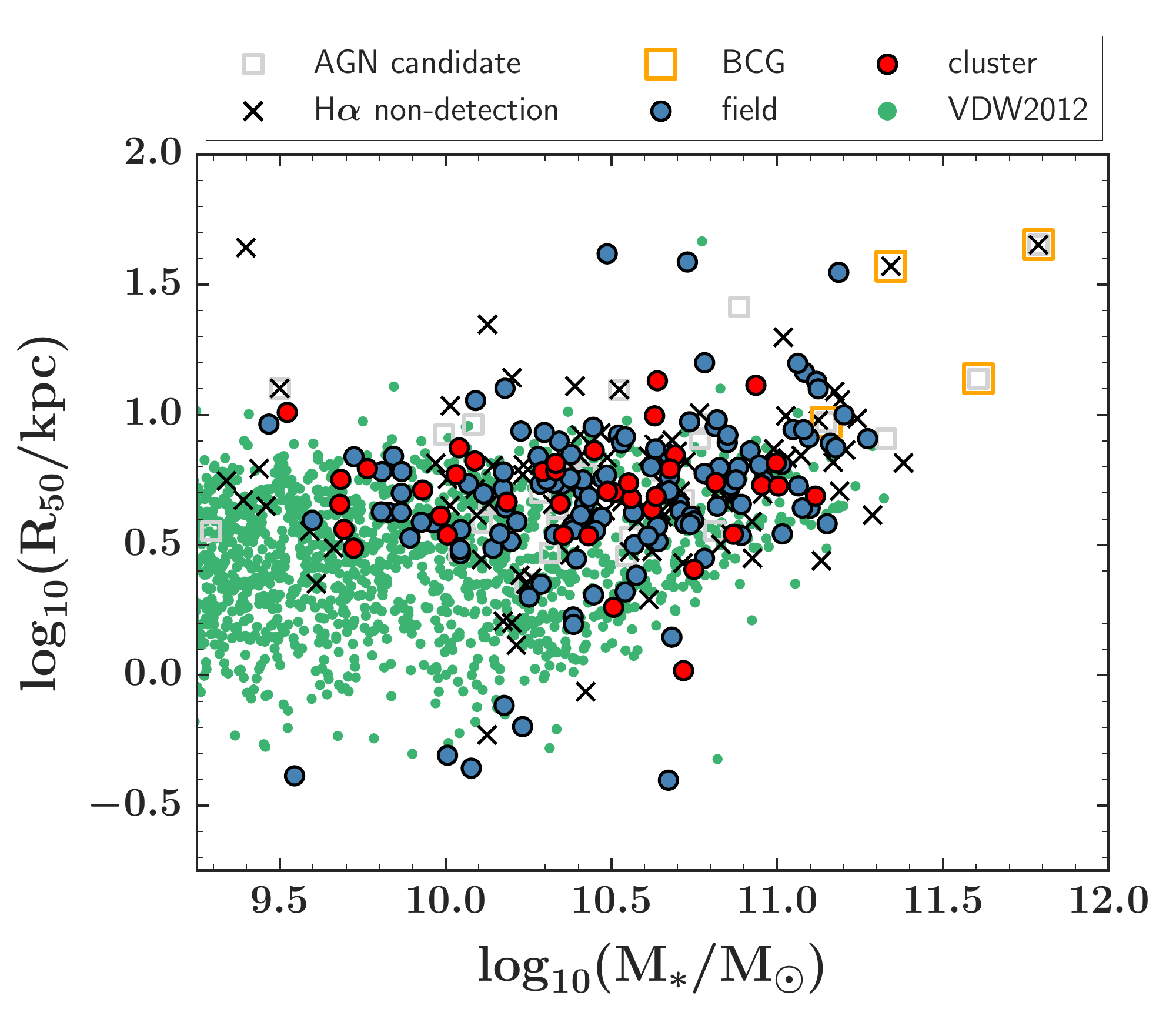}
\end{minipage}
\caption{The positions of the K-CLASH galaxies in the stellar size-mass plane. The K-CLASH galaxies are in close agreement with those of a larger sample of galaxies in the CANDELS fields at the same redshift ($0.2 < z < 0.65$), spatially-resolved in {\it HST} imaging, with sizes measured by \citet{vanderWel:2012}, and with stellar masses from the \citet{Santini:2015} and \citet{Nayyeri:2017} catalogues for respectively the GOODS-S and UDS fields, and the COSMOS field (green circles; the remaining symbols are as described in Figure~\ref{fig:colourmagnitude}). The K-CLASH galaxies thus have ``normal'' stellar sizes for their stellar masses and redshifts.} 
\label{fig:sizeversusmass}
\end{figure*}

The {\sc ProSpect} routine does, however, differ from {\sc magphys} (and {\sc LePhare}) in several key aspects. Firstly, it employs a free form modification of the \citet{Charlot:2000} model for dust attenuation, that allows for separate treatments of light emitted from stellar birth clouds and that emitted from the interstellar medium. The re-emission of this attenuated light is described by the \citet{Dale:2014} library of far-infrared SEDs. Secondly, {\sc ProSpect} allows for broad, user-controlled flexibility in how it processes SFHs, with arbitrary complexity in their functional forms. In this work, we use a skewed Gaussian distribution to describe the model SFHs, with the peak, position, width, and skewness of the distribution left as free fitting parameters (i.e.\ the {\sc massfunc\_norm} function in {\sc ProSpect}). A final key difference between {\sc ProSpect} and previously developed SED-fitting routines is that it allows for a non-constant metallicity history. As for the SFHs, {\sc ProSpect} allows for user-controlled flexibility in the functional form of the metallicity evolution. In this work, for simplicity, we employ a closed box, fixed-yield metallicity mapping (commonly used in the literature), with the starting and finishing metallicity, the fixed yield, and the maximum age of metal evolution describing the metallicity as a function of time (i.e.\ the {\sc Zfunc\_massmap\_box} function in {\sc ProSpect}). Only the finishing metallicity is left as a free parameter in the fitting. 

\subsubsection{SED-fitting results}
\label{subsubsec:sedresults}

In Appendix~\ref{sec:SEDcodecheck}, we provide a detailed comparison between the stellar masses and SFRs derived for the K-CLASH sample galaxies from the different routines. In summary, we find the {\sc ProSpect}- and {\sc magphys}-derived stellar mass and SFR estimates to be in good agreement. We find the {\sc LePhare} estimates of stellar mass to be in reasonable agreement with those derived from {\sc magphys} and {\sc ProSpect}, but with significant trends in the residuals that depend on the mass estimates from the latter. The {\sc LePhare} SFR estimates strongly disagree with those of {\sc magphys} and {\sc ProSpect}, being systematically lower compared to either code. 

That the SED-fitting routines produce similar results (with the exception of the {\sc LePhare} SFRs), is reassuring. Nevertheless, we must select a single set of results with which to proceed with our analysis. In Appendix~\ref{sec:SEDcodecheck}, we therefore also examine the positions of the galaxies in the SFR-stellar mass plane, using the results from each of the SED-fitting codes to decide which is the most self consistent, and thus which likely provides the most robust estimates of the stellar masses and SFRs. We find the {\sc ProSpect} measurements produce the tightest correlation (i.e. the correlation with least scatter) between SFR and stellar mass, suggesting it is the most self-consistent code. We thus adopt the {\sc ProSpect} estimates for the remainder of our analysis, and hereafter ignore the results of the other SED-fitting routines. 

We also note that the {\sc ProSpect} results are in closest agreement with the trend expected for ``main sequence'' star-forming galaxies at the same (average) redshift as the K-CLASH galaxies, as measured by \citet{Schreiber:2015}. Since the K-CLASH galaxies were predominantly selected to be blue (and bright), and the number of {\it cluster} sub-sample galaxies is only a small fraction of the total number of K-CLASH galaxies, we expect the K-CLASH sample to be, on average, comprised of galaxies with SFRs typical of ``normal'' star-forming systems at their epochs (and thus in agreement with the \citealt{Schreiber:2015} measurements). The {\sc ProSpect} results support this hypothesis. However, rather than rely on these results alone, in \S~\ref{subsec:fluxandsfr} we make additional estimates of the SFRs for H$\alpha$-detected K-CLASH galaxies using their H$\alpha$ emission, finding them to be in good agreement with those derived from {\sc ProSpect}. There we use both measurements of SFR to inform us on the likelihood that the K-CLASH galaxies are typically main sequence galaxies for their epoch.

The stellar masses of K-CLASH galaxies derived from {\sc ProSpect} are shown as a function of redshift in Figure~\ref{fig:mass_selection}. Both H$\alpha$-detections and non-detections are biased toward increasingly high stellar masses with increasing redshift, as expected for an apparent magnitude-limited sample. The {\it cluster} galaxies, by definition, have redshifts that cluster around the redshifts of the four K-CLASH clusters, whereas the {\it field} galaxies are spread across a redshift range $0.21 \lesssim z \lesssim 0.64$. As expected, the BCGs reside at the maximum of the stellar mass range for K-CLASH galaxies at the same redshift. Interestingly, for the two highest-redshift clusters, {\it cluster} sub-sample galaxies have, on average, significantly lower stellar masses than those in the {\it field} sub-sample at similar redshifts ($\overline{\log_{10}(\rm{M}_{*}/\rm{M}_{\odot})} = 10.32 \pm 0.08$ versus $10.66 \pm 0.04$ for respectively {\it cluster} and {\it field} galaxies with $z \geq 0.423$). However, the same cannot be said when considering instead the two lowest redshift clusters (i.e.\ there is no significant difference between the mean ($\log_{10}$) stellar masses of {\it cluster} and {\it field} galaxies with $z < 0.423$), although the number of {\it cluster} galaxies in our sample is small for these two clusters.

Finally we note that the H$\alpha$ non-detections appear to occupy a mass range similar to that of H$\alpha$-detected galaxies at the same redshift. However, their redshift distribution is visibly more clustered than the detected galaxies, in line with the explanations offered in \S~\ref{subsec:detectionstats} as to why these systems are not detected, i.e. their redshifted line emission falls within strong telluric absorption or sky line emission features in the KMOS spectra.   

\subsection{Galaxy sizes}
\label{subsec:sizes}

For a full description of the modelling of the stellar continuum images of the K-CLASH sample galaxies see \citet{vaughan:2020}. Here we provide a brief summary. 

Whilst each of the galaxy clusters targeted by K-CLASH benefits from high spatial resolution, multi-wavelength {\it HST} imaging, this only spans the most central regions of each cluster and excludes most of the K-CLASH sample galaxies. We therefore measure the stellar continuum size of each galaxy in our sample from wider $R_{\rm{C}}$-band Subaru imaging, publicly available for all K-CLASH targets from the CLASH archive\footnote{\url{https://archive.stsci.edu/prepds/clash/}}.  We use these $R_{\rm{C}}$-band images, despite the fact that other images are available in redder bands, as this is the only band for which the images have not been convolved to a common, limiting PSF before stacking (known as ``PSF-matching'', a method designed to improve the accuracy of {\it photometric} measurements rather than size measurements). At the typical redshift of the K-CLASH sample, the $R_{\rm{C}}$ band corresponds to the rest-frame $B$ band.  

Our galaxy sizes are measured by fitting a two-dimensional S\'ersic profile to a $6\farcs4 \times 6\farcs4$ $R_{\rm{C}}$-band cutout of each K-CLASH galaxy, using {\sc imfit}\footnote{\url{https://www.mpe.mpg.de/~erwin/code/imfit/}} in {\sc Python} \citep{Erwin:2015}, allowing the S\'ersic index to vary in the range $1 \leq n_{\rm{S}} \leq 10$. For each galaxy, iterations of the model are convolved with the PSF\footnote{The seeing varied in the range $\rm{FWHM}_{\rm{PSF}} \approx 0\farcs6$ -- $0\farcs9$ in the K-CLASH fields.}, constructed via a median stack of $\approx100$ stars taken from the larger $R_{\rm{C}}$-band mosaic image, before being compared to the observed cutout. The size of the cutout is chosen to maximise the accuracy of the fit (by providing a robust measure of the local background level of the image), whilst minimising the number of interloping foreground and background objects. If the latter are present in a cutout, we also simultaneously fit these with additional S\'ersic profiles. We determine the uncertainties via a 1000-step bootstrap resampling of the input data. 

In this work, we measure the intrinsic stellar half-light radius ($\rm{R}_{50}$) from a curve-of-growth analysis for each galaxy, constructed by summing the flux of the {\it intrinsic} (i.e.\ before convolution with the PSF) best-fitting S\'ersic model within elliptical apertures of increasing sizes. The elliptical apertures for a given galaxy each share a common position angle and axial ratio, equal to those of the best-fitting profile. The positions of the K-CLASH galaxies in the stellar size-mass plane are shown in Figure~\ref{fig:sizeversusmass}. For reference, we also include galaxies in the Cosmic Assembly Near-infrared Deep Extragalactic Legacy Survey \citep[CANDELS;][]{Grogin:2011} fields at the same redshift ($0.2 < z < 0.65$), resolved in $H_{\rm{F}160\rm{W}}$-band {\it HST} imaging, with robust size measurements from \citet{vanderWel:2012}, and with stellar mass measurements from \citet{Santini:2015} and \citet{Nayyeri:2017}, for galaxies in the Great Observatories Origins Deep Survey \citep[GOODS][]{Dickinson:2003,Giavalisco:2004} South (GOODS-S) and Ultra-Deep Survey \citep[UDS;][]{Lawrence:2007,Cirasuolo:2007,Galametz:2013} fields, and the Cosmic Evolution Survey \citep[COSMOS;][]{Scoville:2007} field, respectively\footnote{Both stellar mass catalogues are publicly available at \url{https://archive.stsci.edu/prepds/candels/}}. The K-CLASH galaxies are coincident with the galaxies from the larger comparison sample, indicating that they have ``normal'' sizes for their stellar masses and redshifts. We find no systematic offset between the positions of {\it cluster} and {\it field} galaxies in this plane.

\subsection{Total H$\alpha$ fluxes and H$\alpha$ star-formation rates}
\label{subsec:fluxandsfr}

In this sub-section, we describe our methods to calculate total (i.e.\ aperture- and extinction-corrected) nebular H$\alpha$ fluxes for the K-CLASH galaxies. We also outline how we derive an additional measure of the SFRs of the H$\alpha$-detected galaxies in our sample, based on their total H$\alpha$ fluxes.

\subsubsection{Aperture-corrected H$\alpha$ fluxes}
\label{subsubsec:totalfluxes}

\begin{figure*}
\centering
\begin{minipage}[]{1.\textwidth}
\centering
\includegraphics[width=.83\textwidth,trim= 0 0 0 0,clip=True]{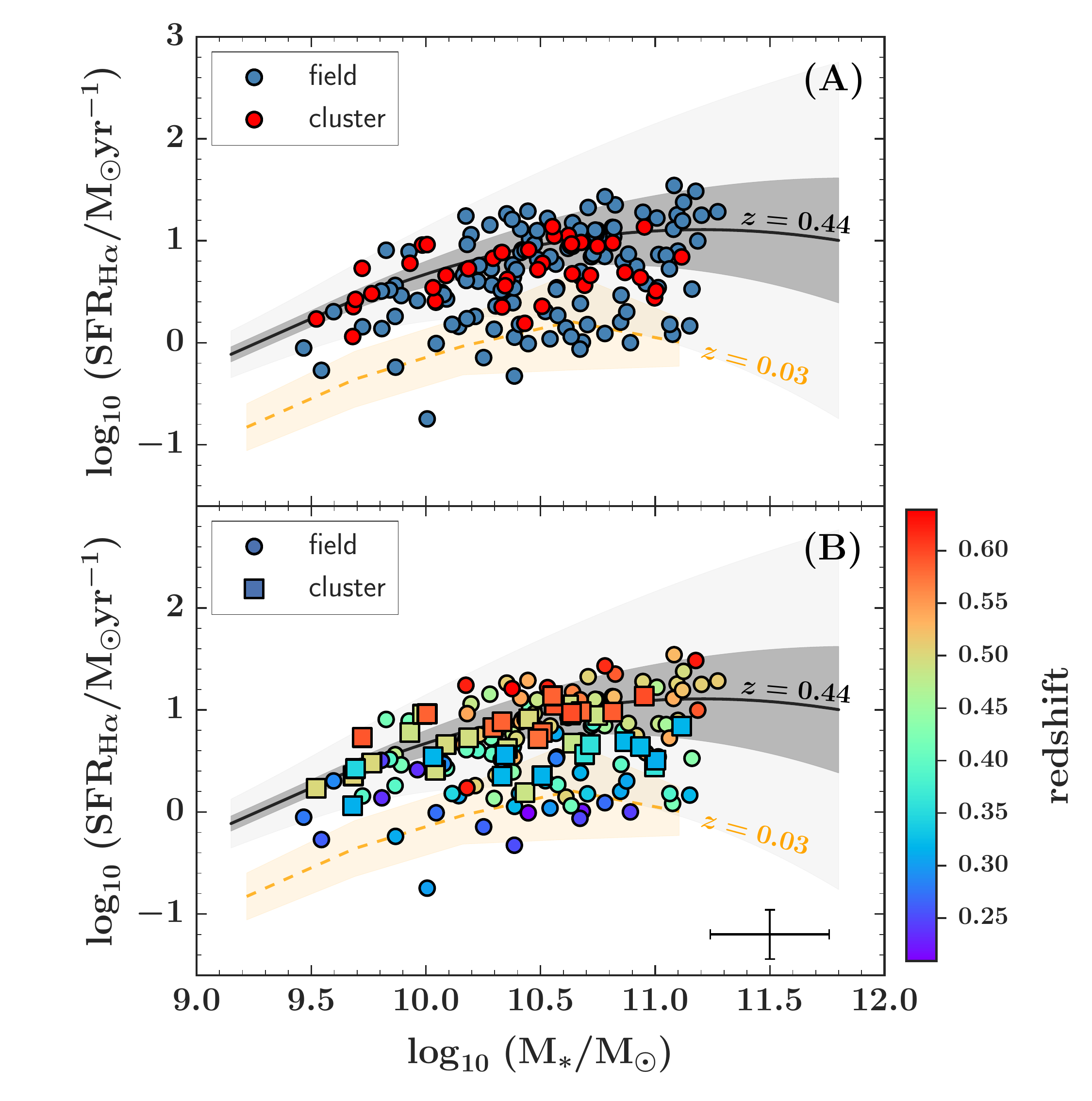}
\end{minipage}
\caption{%
{\bf (A):} The H$\alpha$-derived total SFR (see \S~\ref{subsubsec:halphaSFRs}) of the K-CLASH {\it field} and {\it cluster} sub-sample galaxies (respectively blue and red filled circles; see \S~\ref{subsec:clustermembership}) as a function of their stellar masses (see \S~\ref{subsec:stellarmasses}). The ``main sequence" of star-formation, as defined by \citet{Schreiber:2015} (and converted to a Chabrier IMF), at the median redshift of the K-CLASH galaxies is indicated by a black solid line. The $\pm1\sigma$ and $\pm2.5\sigma$ uncertainties on the \citet{Schreiber:2015} main sequence are indicated by respectively the dark-grey and light-grey filled regions. We also include the main sequence of star-formation (orange dashed line) and associated $\pm1\sigma$ scatter (orange filled region) for a subset of galaxies from the SAMI Galaxy Survey \citep{Bryant:2015} presented in \citet{Tiley:2019a}, with the SAMI stellar masses and star-formation rates from \citet{Bryant:2015} and \citet{Davies:2016}, respectively (see \S~\ref{subsec:redshiftcontext}). The K-CLASH galaxies are consistent with the findings of \citet{Schreiber:2015}, albeit with data points preferentially scattering towards lower SFRs at a given stellar mass. {\bf (B):} As for (A), but the data points are colour-coded according to their H$\alpha$-derived spectroscopic redshift ({\it field} galaxies are shown as filled circles and {\it cluster} galaxies as filled squares). The median error bar for the K-CLASH galaxies is shown in the bottom-right corner. The scatter towards lower SFRs at a given stellar mass is strongly correlated with redshift, indicating the K-CLASH galaxies are typical star-forming systems for their masses and epochs.  %
     }%
\label{fig:mainsequence}
\end{figure*}

In \S~\ref{subsec:measuringlinefluxes} we described how we measured the integrated H$\alpha$ (and [N {\sc ii}]) fluxes of each K-CLASH galaxy within three circular apertures of increasing diameter ($D = 0\farcs6$, $1\farcs2$, and $2\farcs4$, respectively, where the largest aperture is limited by the size of the IFU FOV). However, given the redshift range of the K-CLASH sample galaxies, their {\it physical} sizes, and the finite size of the KMOS IFUs' fields-of-view, in most cases these apertures do not encompass the total {\it angular} extent of a galaxy i.e.\ the integrated H$\alpha$ flux we measure is not the {\it total} H$\alpha$ flux of the galaxy, because even our largest aperture is too small to catch all of the incident flux (each IFU FOV corresponds to $9$, $16$, and $20$ kpc on each side at respectively $z=0.2$, $0.43$, and $0.65$). To estimate the total H$\alpha$ flux for each H$\alpha$-detected galaxy, we therefore apply correction factors to our integrated measurements to account for the mismatch between the aperture size and the galaxy size. 

In Appendix~\ref{subsec:aperturecorrection}, we describe in detail how we calculate these corrections (and verify that they are effective at removing the effects of the finite aperture size). Briefly, for each galaxy we measure the curve-of-growth from the (intrinsic) stellar light model image that best fits the observed $R_{\rm{C}}$-band image (see \S~\ref{subsec:sizes}), where the model image is here also convolved with a two-dimensional Gaussian with width equal to that of the Gaussian that best fits the observed KMOS PSF. Making the explicit assumption that the H$\alpha$ light distribution follows exactly that of the stars, for each galaxy we then simply measure from the curve-of-growth the fractions of the total (model) stellar light contained within our three apertures, and calculate the required aperture corrections as the inverse of these fractions. To minimise extrapolation uncertainty, we calculate the total H$\alpha$ flux of each galaxy by correcting the flux measured from the largest aperture in which we detect H$\alpha$. The median average and $\sigma_{\rm{MAD}}$ ($ = 1.483\ \rm{MAD}$, where MAD is the median absolute deviation from the median itself) spread of the corrections made are respectively $1.42 \pm 0.03$ and $0.36 \pm 0.04$. 

We note here the important caveat that our assumption that the H$\alpha$ light distributions follows that of the stars ignores the possibility of truncation of the H$\alpha$ emission profile (or indeed any significant deviation from the $R_{\rm{C}}$-band radial profile) outside of an IFU's FOV, for example as a result of cluster environmental quenching \citep[e.g.][]{Koopmann:2004a,Koopmann:2004b}. For further discussion of this matter, including a full examination of the role of the cluster environment in the quenching of K-CLASH galaxies, see \citet{vaughan:2020}.

\subsubsection{H$\alpha$ star-formation rates}
\label{subsubsec:halphaSFRs}

In \S~\ref{subsubsec:sedresults}, we discussed estimates of the SFRs of K-CLASH galaxies based on model fits to their observed SEDs spanning the optical bandpasses as well as two {\it Spitzer} infrared channels. However these estimates are not necessarily well constrained by the available K-CLASH photometry, since the K-CLASH SEDs, depending on the redshift, many not cover the UV and/or far-infrared regimes in the {\it rest frame} of the galaxies (vital for an accurate estimate of the SFR from a SED). Therefore we also calculate SFR estimates using the total H$\alpha$ fluxes for those K-CLASH galaxies robustly detected in H$\alpha$ emission. This is to help confirm that these H$\alpha$-detected galaxies are indeed on the ``main sequence'' of star formation for their average redshift, as indicated by the {\sc ProSpect} SED-fitting results.

The H$\alpha$ star-formation rate (SFR$_{\rm{H}\alpha}$) of each galaxy is derived from its extinction-corrected H$\alpha$ luminosity ($L_{\rm{H}\alpha}$) according to the method of \citet{Kennicutt:1998} and is corrected to a Chabrier \citep{Chabrier:2003} IMF such that

\begin{equation}
\frac{\rm{SFR}_{\rm{H}\alpha}}{\rm{M}_{\odot}\ \rm{yr}^{-1}} = \rm{C}_{\rm{IMF}} \ \rm{X}_{\rm{H}\alpha} \ \frac{L_{\rm{H}\alpha}}{\rm{ergs\ s}^{-1}}\,\,,
\label{eq:LHaToSFR}
\end{equation}

\noindent where $\rm{X}_{\rm{H}\alpha} = 7.9 \times 10^{-42}\ \rm{M}_{\odot}\ \rm{yr}^{-1}\ \rm{ergs}^{-1}\ \rm{s}$ is the \citet{Kennicutt:1998} conversion factor between H$\alpha$ luminosity and SFR, assuming a \citet{Salpeter:1955} IMF. We convert to a \citet{Chabrier:2003} IMF with a factor $\rm{C}_{\rm{IMF}} = 10^{-0.215}$, based on offsets measured by \citet{Madau:2014}.

We calculate the extinction-corrected H$\alpha$ luminosity itself as

\begin{equation}
L_{\rm{H}\alpha} = 4 \pi D_{L}^{2}\ 10^{0.4A_{\rm{H}\alpha,\rm{gas}}}\ F_{\rm{H}\alpha}\,\,, 
\label{eq:fHaToLHa}
\end{equation}

\noindent where $D_{L}$ is again the luminosity distance (calculated from the galaxy spectroscopic redshift since we only consider H$\alpha$ detected systems), and $F_{\rm{H}\alpha}$ is the aperture-corrected H$\alpha$ flux calculated in \S~\ref{subsubsec:totalfluxes}. We calculate $A_{\rm{H}\alpha,\rm{gas}}$, the rest-frame nebular extinction at the wavelength of H$\alpha$, according to the method of \citet{Wuyts:2013} as 

\begin{equation}
A_{\rm{H}\alpha,\rm{gas}} = A_{\rm{H}\alpha,\rm{stars}}\ (1.9 - 0.15\ A_{\rm{H}\alpha,\rm{stars}})\,\,,
\label{eq:AstarToAgas}
\end{equation}

\noindent where $A_{\rm{H}\alpha,\rm{stars}}$ is the rest-frame stellar extinction at the wavelength of H$\alpha$, which we calculate by converting the familiar $V$-band stellar extinction ($A_{V}$) assuming a \citet{Calzetti:1994} extinction law. Due to the limited wavelength coverage of our KMOS observations, we are unable to calculate the dust extinction directly, for example via the commonly used nebular emission line flux ratio H$\alpha$/H$\beta$ (i.e.\ the ``Balmer decrement''). Additionally, the rest-frame model $V$-band extinction is not well constrained by the {\sc ProSpect} SED-fitting process for the K-CLASH galaxies, due to the limited wavelength coverage of their photometry. We therefore instead adopt the more robust measurements of $A_{V}$ from \citet[][hereafter referred to as D20]{Dudzevi:2019}, who used {\sc magphys} to model the SEDs of $\approx10^{5}$ galaxies in the UDS extragalactic field, with extensive multi-wavelength photometry ranging from the UV to the infrared. 

Since we are primarily interested in K-CLASH galaxies for which we detect H$\alpha$ (and assume to be actively star forming, after removing potential AGN contaminants; see \S~\ref{subsec:AGNselection}), we wish to know the typical $A_{V}$ of star-forming galaxies at the same redshifts as the K-CLASH sample galaxies. We therefore isolate the main sequence of star formation for galaxies in the D20 sample with $0.2 < z < 0.65$ (to match the redshift range of the K-CLASH galaxies) and stellar masses $\log (\rm{M}_{*}/\rm{M}_{\odot}) < 11.5$ (to match the mass range for the majority of K-CLASH galaxies), using an iterative $1.5\sigma$ clip of the running median in the star formation--stellar mass plane. For each K-CLASH galaxy, we then take its $A_{V}$ as the median $A_{V}$ of the galaxies in the D20 sample main sequence and within an initial stellar mass window of $\pm 0.2$ dex centred on the stellar mass of the K-CLASH galaxy. We take the uncertainty as the $A_{V}$ $\sigma_{\rm{MAD}}$ spread in the selected comparison sub-sample (rather than the standard deviation, since the number of D20 comparison galaxies is typically small). We also require that the stellar mass comparison window contains at least 50 galaxies, iteratively expanding its width by a factor of 2 until this requirement is met. In practice, an expanded stellar mass window is only required for 25 out of 282 (9 per cent of the) K-CLASH galaxies. Of these, 20 galaxies require a stellar mass window of $\pm 0.4$ dex and only 5 require a larger window still (3 of which are BCGs). The number of galaxies in the comparison sub-samples was chosen as a compromise between a robust measurement of the average $A_{V}$, and selecting galaxies with stellar masses as close as possible to that of each K-CLASH galaxy.    

\begin{figure*}
\centering
\begin{minipage}[]{.9\textwidth}
\centering
\includegraphics[width=.83\textwidth,trim= 0 0 0 0,clip=True]{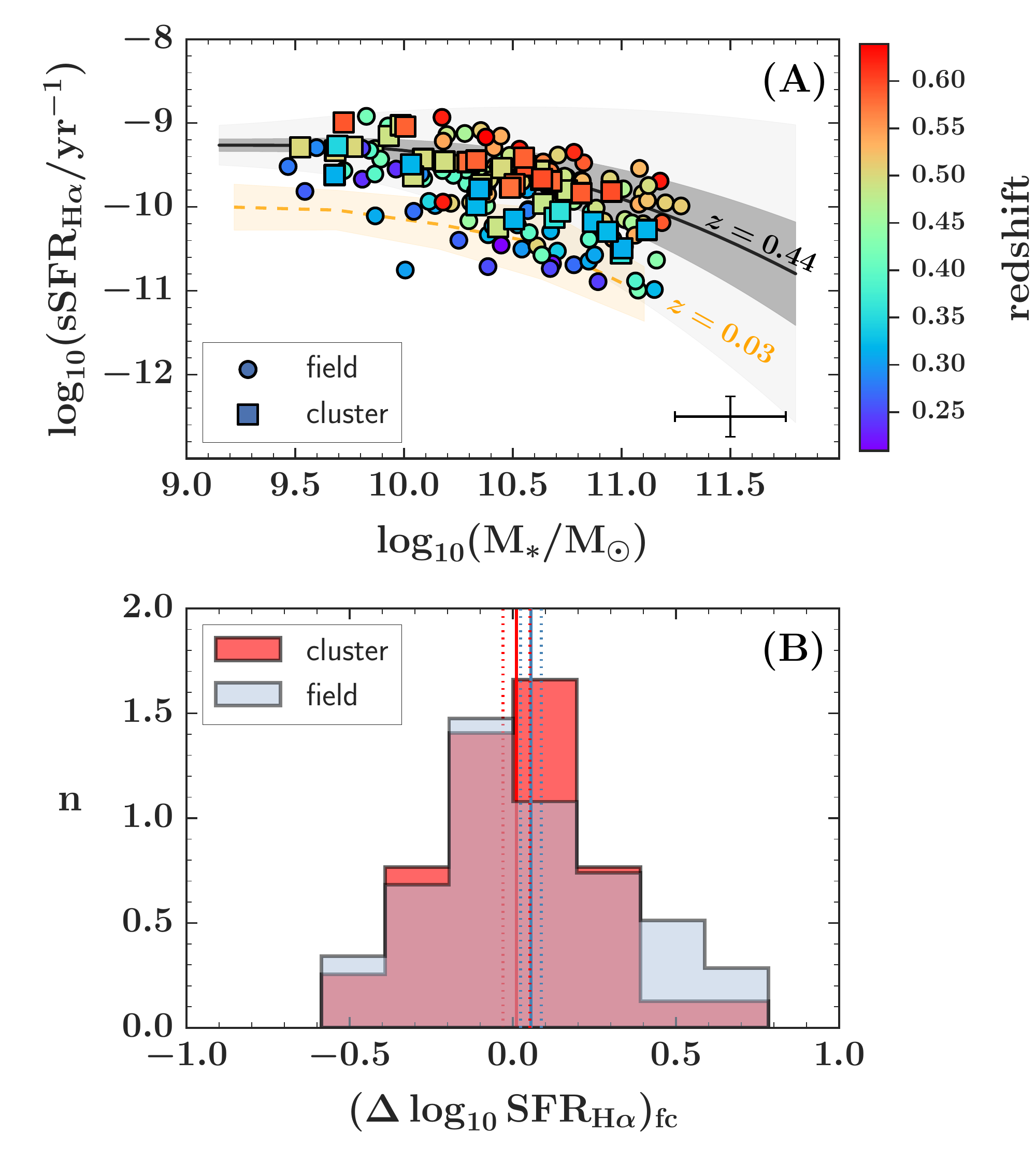}
\end{minipage}
\caption{%
{\bf (A):} The H$\alpha$-derived total sSFRs (see \S~\ref{subsubsec:comparingSFRs}) of the K-CLASH {\it field} and {\it cluster} sub-sample galaxies (respectively filled circles and squares; see \S~\ref{subsec:clustermembership}) as a function of their stellar masses (see \S~\ref{subsec:stellarmasses}). The sSFRs of main sequence galaxies, as defined by \citet{Schreiber:2015} (and converted to a Chabrier IMF), at the median redshift of the K-CLASH galaxies are indicated by a black solid line. The $\pm1\sigma$ and $\pm2.5\sigma$ uncertainties on the \citet{Schreiber:2015} sSFRs are indicated by respectively the dark-grey and light-grey filled region. The trend for main sequence SAMI galaxies at an average redshift of $z = 0.03$ (\S~\ref{subsec:redshiftcontext}) and its $\pm1\sigma$ scatter are indicated as respectively the orange dashed line and filled region. The median error bar for the K-CLASH galaxies is shown in the bottom-right corner. Accounting for redshift and stellar mass, there is no significant difference between the sSFRs of {\it cluster} and {\it field} galaxies. {\bf (B):} The (normalised) distribution of the difference between the ($\log_{10}$) SFR of each individual  {\it cluster} (red) and {\it field} galaxy (in the same stellar mass and redshift range as the {\it cluster} sub-sample galaxies; blue) and the (bootstrap) median ($\log_{10}$) SFR of the five {\it field control} galaxies most similar in stellar mass and redshift (see \S~\ref{subsubsec:comparingSFRs}). The mean of the distribution for the {\it cluster} and {\it field} galaxies is shown as respectively the red and the blue vertical solid line. The uncertainty on each mean is represented by vertical dotted lines in the corresponding colour. There is no evidence of a significant difference between the SFRs of star-forming {\it cluster} galaxies and those of {\it field} galaxies of similar stellar masses and redshifts. %
     }%
\label{fig:SFRcomp}
\end{figure*} 

The mean and standard deviation of the derived $A_{V}$ values of K-CLASH galaxies are $\overline{A_{V}} = 1.26 \pm 0.01$ mag and $\sigma_{A_{V}} = 0.19 \pm 0.01$ mag, with a typical uncertainty on the individual $A_{V}$ values themselves of $\overline{\Delta A_{V}} = 0.5$ mag. In Appendix~\ref{subsec:checkextinctioncorr}, we compare our derived $A_{V}$ to those required for our {\it non-extinction corrected} K-CLASH H$\alpha$ SFRs to match those derived from {\sc ProSpect}, finding good agreement between the two. 

In Figure~\ref{fig:mainsequence} (A), we show the H$\alpha$-derived SFRs of the {\it cluster} and {\it field} sub-samples of H$\alpha$-detected K-CLASH galaxies. As for the {\sc ProSpect}-derived measurements, the position of the majority of K-CLASH galaxies in the SFR--stellar mass plane is in line with the expectation for main sequence systems at the same redshifts (once again as judged by comparison to the measurements of \citealt{Schreiber:2015}), albeit with a large scatter that is preferentially toward lower SFRs at fixed stellar mass. However, Figure~\ref{fig:mainsequence} (B) reveals that this vertical scatter is strongly correlated with the spectroscopic redshift of the galaxy, with galaxies at the lowest redshifts more consistent with the main sequence of star-forming galaxies from SAMI, with an average redshift of $\overline{z}=0.03$ (see \S~\ref{subsec:redshiftcontext}). Since we do not expect significant evolution in the normalisation of the star-formation main sequence between the lowest redshifts of the K-CLASH galaxies ($z\approx0.2$) and $z\approx0$, we conclude that the H$\alpha$-detected K-CLASH galaxies have ``normal'' SFRs for their stellar masses and redshifts.

A strong caveat to the results presented in Figure~\ref{fig:mainsequence} is our assumption that the H$\alpha$ emission from the K-CLASH galaxies (after removing potential AGN contamination) is driven by ongoing star formation, and thus that it is appropriate to adopt $A_{V}$ inferred from mass-matched sub-samples of a much larger main-sequence star-forming parent sample. Nevertheless, it is  revealing that adopting these $A_{V}$ yields H$\alpha$-derived SFRs that place the K-CLASH sample galaxies on the main sequence for their (average) redshift. This need not be the case and, combined with the fact that the {\sc ProSpect} SED-fitting results also place the K-CLASH galaxies on the main sequence, strongly suggests that the K-CLASH sample as a whole is indeed comprised of galaxies typical of star-forming systems at their epoch(s). We also highlight here that the {\sc ProSpect}- and H$\alpha$-derived SFRs are in good agreement for galaxies in the {\it cluster} and {\it field} sub-samples; the median ($\log_{10}$) difference between the two measurements is $-0.11 \pm 0.03$ dex, and the spread of this distribution is $\sigma_{\rm{MAD}} = 0.3 \pm 0.04$ dex. Similarly, we perform a (bisector) straight line fit to the logarithms of the two SFRs measures, of the form $\log_{10}\rm{SFR}_{\sc ProSpect} = M(\log_{10}\rm{SFR}_{\rm{H}\alpha} - \log_{10}\rm{SFR}_{\rm{H}\alpha,0}) + C$, where $ \log_{10}\rm{SFR}_{\rm{H}\alpha,0}$ is the log median of the H$\alpha$ SFRs. We find the best-fitting slope and zero-point to be $M = 0.8 \pm 0.2$ and $C = 0.70 \pm 0.05$ dex, consistent within uncertainties with a 1:1 ratio between the two measures of SFR.

\subsubsection{Comparing cluster and field star-formation rates}
\label{subsubsec:comparingSFRs}

Figure~\ref{fig:mainsequence} suggests there is no clear difference between the SFRs of our {\it cluster} and {\it field} galaxies, accounting for stellar mass and redshift. However, to investigate further, in Figure~\ref{fig:SFRcomp} (A), we present the H$\alpha$-derived {\it specific} SFRs (sSFR$_{\rm{H}\alpha} \equiv\ $SFR$_{\rm{H}\alpha}/\rm{M}_{*}$) as a function of stellar mass for the {\it cluster} and {\it field} galaxies. Again, we find no clear difference between the two sub-samples. 

To formally quantify whether any difference exists between the SFRs of the star-forming K-CLASH galaxies in the two environments, we also make a direct comparison between the SFRs of {\it cluster} galaxies and those of {\it field} galaxies drawn from redshift and stellar mass ranges that are close to, but encompass, the ranges spanned by the {\it cluster} galaxies ($0.3 \leq z \leq 0.64$ and $9.3 \leq \log_{10}(\rm{M}_{*}/\rm{M}_{\odot}) \leq 11.3$). We hereafter refer to this comparison sample as the {\it field control} sub-sample (or {\it field control} galaxies), which comprises 110 {\it field} galaxies. 

For each {\it cluster} galaxy, we compute the difference, $(\Delta \log_{10}\rm{SFR}_{\rm{H}\alpha})_{\rm{fc}}$, between its $\log_{10}$SFR and the average $\log_{10}$SFR of the 5 {\it field control} galaxies that are closest in stellar mass and redshift. We calculate the latter as the median of the distribution of median ($\log_{10}$) SFRs of 1000 bootstrap samples of the 5 paired {\it field control} galaxies in each case. We select the 5 closest {\it field control} galaxies as those that minimise the ``distance'' to the {\it cluster} galaxy in the stellar mass-redshift plane, defined as

\begin{equation}
d_{i} = \sqrt{(0.3\Delta z_{i} / \Delta z)^{2} + (\Delta \log_{10}\rm{M}_{*,i} / \Delta \log_{10}\rm{M}_{*})^{2}}\,\,,
\label{eq:pointdist}
\end{equation}

\noindent where $\Delta z_{i}$ and $\Delta z$ are respectively the difference in redshift between the {\it cluster} galaxy and {\it field control} galaxy, and the range in redshift that the {\it field control} galaxies span. Similarly, $\Delta \log_{10}\rm{M}_{*,i}$ and $\Delta \log_{10}\rm{M}_{*}$ are respectively the difference in ($\log_{10}$) stellar mass between the {\it cluster} and {\it field control} galaxy in each case, and the range in ($\log_{10}$) stellar mass spanned by the {\it field control} sub-sample. A scaling factor of $0.3$ is applied to the redshift term in Equation~\ref{eq:pointdist} to account for the fact that galaxy SFRs are more strongly correlated with stellar mass than redshift within the ranges spanned by the K-CLASH galaxies. It is equal to the inverse of the ratio of the maximum change in SFR across $\Delta \log_{10}\rm{M}_{*}$ at fixed redshift, to the maximum expected across $\Delta z$ at the median stellar mass of the {\it field control} galaxies \citep[according to][]{Schreiber:2015}. For a fair comparison, we repeat the same exercise for the 90 {\it field} galaxies in the same stellar mass and redshift range as the {\it cluster} galaxies, comparing to the five closest {\it field control} galaxies in each case.  

The normalised distributions of $(\Delta \log_{10}\rm{SFR}_{\rm{H}\alpha})_{\rm{fc}}$ for the {\it cluster} and {\it field} galaxies are presented in Figure~\ref{fig:SFRcomp} (B). As expected, the distribution for the {\it field} galaxies has an approximately Gaussian shape and is centred on zero (~$\overline{(\Delta \log_{10}\rm{SFR}_{\rm{H}\alpha})_{\rm{fc}}} = 0.06 \pm 0.03$). Similarly, the distribution for the {\it cluster} galaxies appears approximately Gaussian, and also has a mean consistent with zero (~$\overline{(\Delta \log_{10}\rm{SFR}_{\rm{H}\alpha})_{\rm{fc}}} = 0.01 \pm 0.04$). A two-sample K-S test between the {\it cluster} and {\it field} $(\Delta \log_{10}\rm{SFR}_{\rm{H}\alpha})_{\rm{fc}}$ distributions, returns $p=0.22$. Adopting a critical value of $p = 0.05$, we cannot reject the null hypothesis that the two are subsets drawn from the same underlying distribution. Thus, we find no evidence that the SFRs of the K-CLASH {\it cluster} galaxies differ from those of K-CLASH {\it field} galaxies, after accounting for stellar mass and redshift. 

Finally, we again highlight the important caveat that the total H$\alpha$ SFRs are aperture-corrected under the assumption that the H$\alpha$ light follows the $R_{\rm{C}}$-band light outside of the KMOS IFUs' FOV (see \S~\ref{subsubsec:totalfluxes}). Reassuringly, however, we obtain a similar result (i.e.\ that the SFRs of {\it cluster} and {\it field} K-CLASH galaxies do not significantly differ after accounting for stellar mass and redshift) if we instead consider the {\sc ProSpect}-derived SFRs.

\subsection{Spatially-resolved properties}
\label{subsubsec:maps}

\begin{figure*}
\centering
\begin{minipage}[]{1.\textwidth}
\includegraphics[width=.935\textwidth,trim= -40 0 20 20,clip=True]{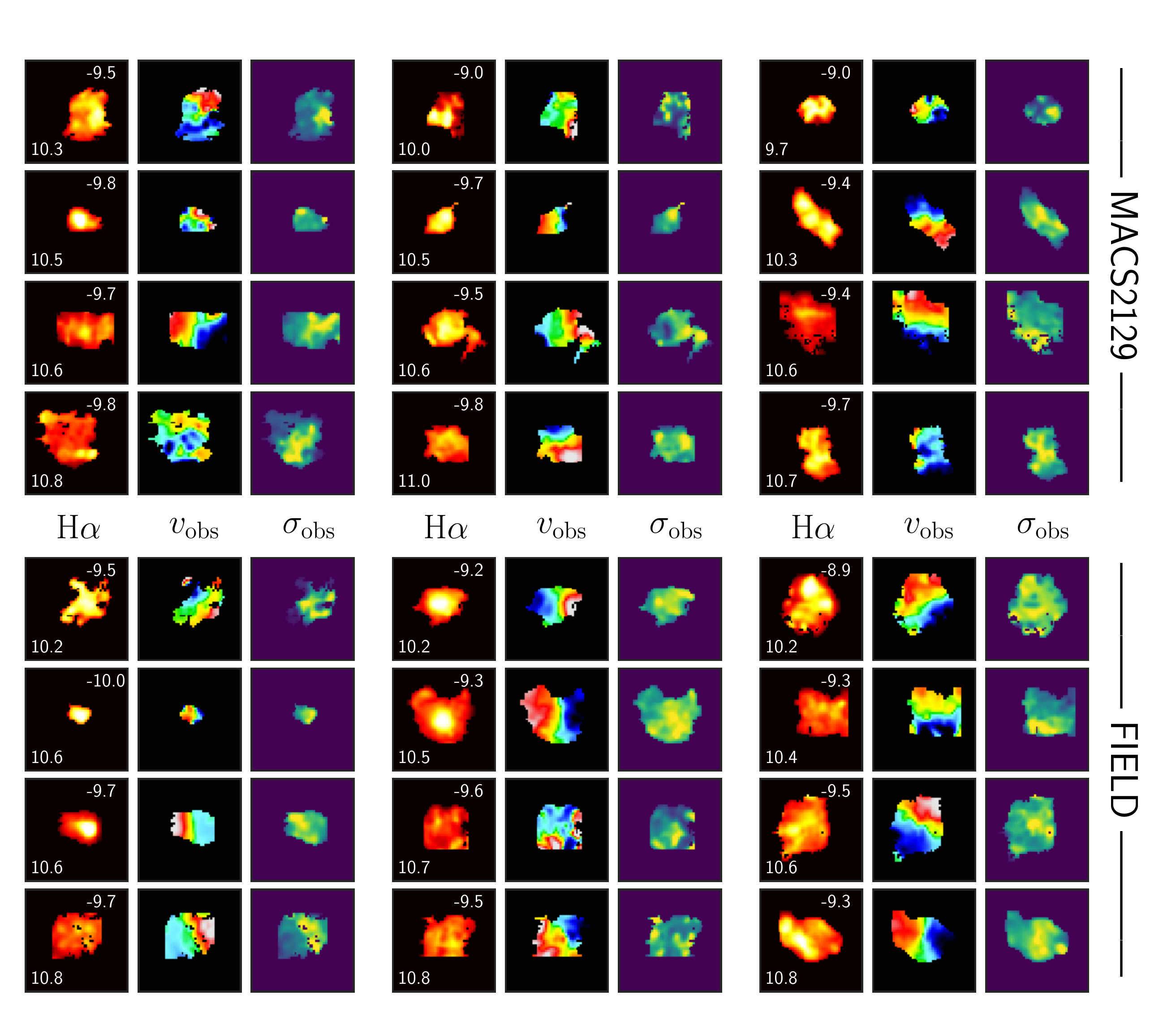}
\end{minipage}
\caption{%
Examples of spatially-resolved gas property and kinematic maps of K-CLASH galaxies measured with KMOS. We present the H$\alpha$ intensity (lower to higher: orange to yellow), mean line-of-sight velocity ($v_{\rm{obs}}$; negative to positive: blue to red), and line-of-sight velocity dispersion ($\sigma_{\rm{obs}}$; lower to higher: green to yellow) maps for 12 {\it cluster} galaxies that belong to the MACS2129 cluster at $z=0.589$, as well as maps of 12 {\it field} galaxies at the same redshift ($|\Delta z| \leq 0.045$, with respect to the cluster redshift) and within the same stellar mass range ($9.7 \leq \log(\rm{M}_{*}/\rm{M}_{\odot} \leq 10.95$). Each panel is $5'' \times 5''$, i.e.\ larger than the $2\farcs8 \times 2\farcs8$ KMOS IFU FOV, to account for dithering and centering offsets between frames (so that the spatial extent of H$\alpha$ can extend beyond the nominal IFU FOV in the final cube). For each galaxy the ($\log_{10}$) stellar mass and ($\log_{10}$) sSFR are indicated respectively in the bottom-left and top-right corner of its H$\alpha$ intensity map. The maps are constructed via an adaptive smoothing process (see \S~\ref{subsubsec:maps}) and do not show individual spaxel values. A two-dimensional Gaussian filter has also been applied to the maps, with a sigma-width of 1 pixel, to mitigate the effects of any remaining ``hot" pixel in the maps. Both the {\it cluster} and {\it field} galaxies display a range of H$\alpha$ morphologies and kinematics, from disc-like systems to those with highly irregular kinematics and complex morphologies. %
     }%
\label{fig:mapsposter_MACS2129}
\end{figure*}

In \S~\ref{subsec:constructingmaps}, we described how we construct two-dimensional maps of the emission line properties and kinematics of K-CLASH galaxies from their KMOS data cubes. In this sub-section, we present a limited number of these maps for a sub-set of K-CLASH galaxies spatially-resolved in H$\alpha$ (see \S~\ref{subsec:resolvednumbers}). We do this to provide a qualitative insight into the gas properties of star-forming systems in a high-redshift cluster as compared to those of star-forming galaxies in the field, and to give examples of the data quality and potential diagnostic power of our KMOS observations. We refrain from a quantitative analysis of the total set of K-CLASH maps in this work. 

In Figure~\ref{fig:mapsposter_MACS2129} we present the H$\alpha$ intensity, mean line-of-sight velocity ($v_{\rm{obs}}$), and line-of-sight velocity dispersion ($\sigma_{\rm{obs}}$) KMOS maps for the 12 K-CLASH galaxies that are spatially-resolved in H$\alpha$ and belong to the highest redshift cluster in our sample, MACS2129 ($z=0.589$). For comparison, we also show maps of the same quantities for 12 {\it field} galaxies with spatially-resolved H$\alpha$ at the same redshift (an absolute redshift difference $\leq 0.045$, with respect to the cluster redshift) and within the same stellar mass range ($9.7 \leq \log(\rm{M}_{*}/\rm{M}_{\odot} \leq 10.95$) as the 12 {\it cluster} galaxies. The {\it cluster} galaxies show a range of kinematic morphologies, from disc-like systems to apparently much more irregular or, perhaps, disturbed systems. Similarly, they also exhibit a variety of H$\alpha$ morphologies, ranging from a relatively smooth and centrally-peaked spatial distribution through to more clumpy morphologies and apparently compact systems. This variety in kinematics and H$\alpha$ morphology is mirrored in the {\it field} comparison galaxies, which also include disc-like systems as well as highly kinematically irregular systems with complex H$\alpha$ morphologies.

It is perhaps surprising to see a similar range of kinematics and H$\alpha$ morphologies amongst galaxies in cluster and field environments at this redshift. Naively, one would expect to see more disturbed star-forming systems within the cluster environment, as the galaxies are expected to interact with the intracluster medium and other cluster members. Conversely, one would expect a higher propensity for ordered, disc-like systems in the {\it field} sub-sample, as the galaxies are more likely to be isolated. However, here we have only considered a small number of galaxies over a wide mass range, so we avoid drawing any firm conclusion from such a limited comparison. Furthermore, we are of course selecting for star-forming systems in both environments, since we require an H$\alpha$ detection to include a galaxy in our analysis. As such, any galaxy that is significantly affected by the cluster environment, i.e.\ with significantly-quenched star formation, will be absent from our sample. For a full and detailed examination of the evidence for quenching in the K-CLASH sample, see \citet{vaughan:2020}.

\section{Discussion}
\label{sec:discussion}

The main aims of the K-CLASH survey are 1) to explore the role of the cluster environment in galaxy quenching and the build-up of today's red sequence of galaxies, and 2) to examine the transition of the star-forming population from prolifically star-forming, turbulent, disc-like systems with high levels of chaotic motions in the distant Universe, into comparatively quiescent, strongly rotation-dominated late-type galaxies in the local Universe. The K-CLASH survey is designed to utilise the substantial diagnostic power of IFS observations to achieve these aims, and the targets are selected with the express intention of bridging the redshift gap between existing large IFS surveys of star-forming galaxies at higher ($z\approx1$--$2$) and lower ($z\approx0$) redshifts. K-CLASH thus provides a galaxy sample at intervening epochs that is similar in its size and selection criteria, whilst simultaneously including star-forming galaxies in cluster environments in the same redshift range. 

In this work, we have shown that K-CLASH has achieved high quality integrated and spatially-resolved measurements of the ionised gas properties of a modestly-sized sample of (40) star-forming cluster members spanning the redshift range $z\approx0.3$--$0.6$ (i.e.\ the {\it cluster} sub-sample), as well as a larger sample of (128) star-forming galaxies not in cluster environments (i.e.\ the {\it field} sub-sample). Having presented the key properties of the sub-sample galaxies in earlier sections, here we discuss their suitability for use in achieving the K-CLASH survey goals.

\subsection{Star-forming galaxies in cluster environments}
\label{subsec:clusterdiscussion}

Given that a detailed exploration of the evidence for cluster quenching in the K-CLASH data is presented in \citet{vaughan:2020}, in this work we have refrained from a comprehensive comparison of the properties of {\it cluster} and {\it field} galaxies in K-CLASH. Nevertheless, we have still highlighted some key differences and similarities between star-forming galaxies in the two environments. These provide useful insight into the nature of our star-forming {\it cluster} galaxies, which we discuss here. 

In \S~\ref{subsubsec:sedresults}, we showed that for the two highest redshift K-CLASH clusters, our star-forming {\it cluster} galaxies have stellar masses systematically lower than those of star-forming galaxies at the same redshift. It is tempting to interpret this as evidence for ``downsizing'' scenarios \citep[e.g][]{Cowie:1996,Guzman:1997,Brinchmann:2000,Kodama:2004,Bell:2005,Juneau:2005,Noeske:2007} in cluster environments \citep[e.g.][]{Stott:2007}, whereby the most massive galaxies in clusters formed and quenched first. In these scenarios, we expect cluster galaxies with ongoing star-formation (i.e.\ those we detect in H$\alpha$ emission) to have stellar masses systematically lower than those of cluster galaxies that are quenched (i.e.\ those we do not detect in H$\alpha$ emission), in line with our findings. 

However, we caution that we only see a disparity between the average stellar masses of star-forming {\it cluster} and {\it field} galaxies in the two highest-redshift clusters; for the two lowest-redshift clusters, the {\it field} and {\it cluster} galaxies have similar average stellar masses (although the number of {\it cluster} galaxies here is small, making a direct comparison less robust). Furthermore, since we cannot confirm cluster membership for those galaxies we do not detect in H$\alpha$ emission, any evidence for downsizing also implicitly assumes that the stellar mass functions of galaxies in field and cluster environments are the same at a given redshift, and that we have sampled this function equally in both environments. 

Finally, we note that we do not expect a strong environmental dependence of the relation between (central) dust obscuration and SFR. So while we may miss some highly dust-obscured star-forming systems from our sample, this is an unlikely explanation for the dearth of high stellar mass, star-forming cluster galaxies in our higher redshift clusters.  

In \S~\ref{subsubsec:comparingSFRs}, we found that star-forming {\it cluster} galaxies in K-CLASH have the same SFRs as those of star-forming {\it field} galaxies, after accounting for stellar mass and redshift. This is a somewhat surprising result. Naively we would expect {\it cluster} galaxies to have SFRs lower than those of their field counterparts, since they should be affected by environmental processes \citep[e.g. strangulation, ram pressure stripping, or harrassment; e.g.][]{Gunn:1972,Larson:1980,Moore:1996,Balogh:2000,vanGorkom:2004,Bekki:2009}. However, combined with the fact that in K-CLASH we are explicitly considering star-forming galaxies only, the most likely explanation of this similarity is that the {\it cluster} galaxies have only recently entered the cluster environment, and insufficient time has passed for their star-formation to have been significantly quenched (and for them to have dropped out of our sample). The spatially-resolved H$\alpha$ maps of K-CLASH galaxies (see \S~\ref{subsubsec:maps}) are also qualitatively consistent with this hypothesis, as we see similar ranges of H$\alpha$ morphologies and kinematics in the {\it field} and {\it cluster} galaxies. 

Cosmological N-body simulations of galaxy clusters suggest that quenching of star-forming (disc) galaxies should occur within a timescale of $\approx 3$ Gyr \citep[e.g.][]{Taranu:2014}. Since we see no difference at all between {\it cluster} and {\it field} galaxy SFRs, this suggests either that the quenching of star-forming galaxies in clusters is actually very sudden (such that quenching galaxies immediately drop out of our sample), or that the K-CLASH {\it cluster} galaxies have entered their respective cluster environments $\ll 3$ Gyr ago. In the latter case, it follows that the we may even observe the {\it cluster} galaxies as they are undergoing their first infall into their cluster. We note that our results are also consistent with previous findings of ``delayed-then-rapid'' quenching in clusters, whereby the SFRs of cluster galaxies are unaffected for up to $2$--$4$ Gyr after infall, before they are rapidly quenched \citep[e.g.][]{Wetzel:2013}. 

Thus, despite its relatively small size, the K-CLASH star-forming {\it cluster} sub-sample, in combination with the large number of star-forming {\it field} galaxies (providing a high-quality control sample at similar redshifts), is well suited to study the effects of the cluster environment on the properties of star-forming galaxies across a key $\approx3$ Gyr period of cosmic history, during which a large fraction of today's red sequence of galaxies was assembled, perhaps even capturing snapshots of star-forming galaxies at the very start of their quenching in clusters.

\begin{figure*}
\centering
\begin{minipage}[]{1.\textwidth}
\centering
\includegraphics[width=.715\textwidth,trim= 10 15 0 0,clip=True]{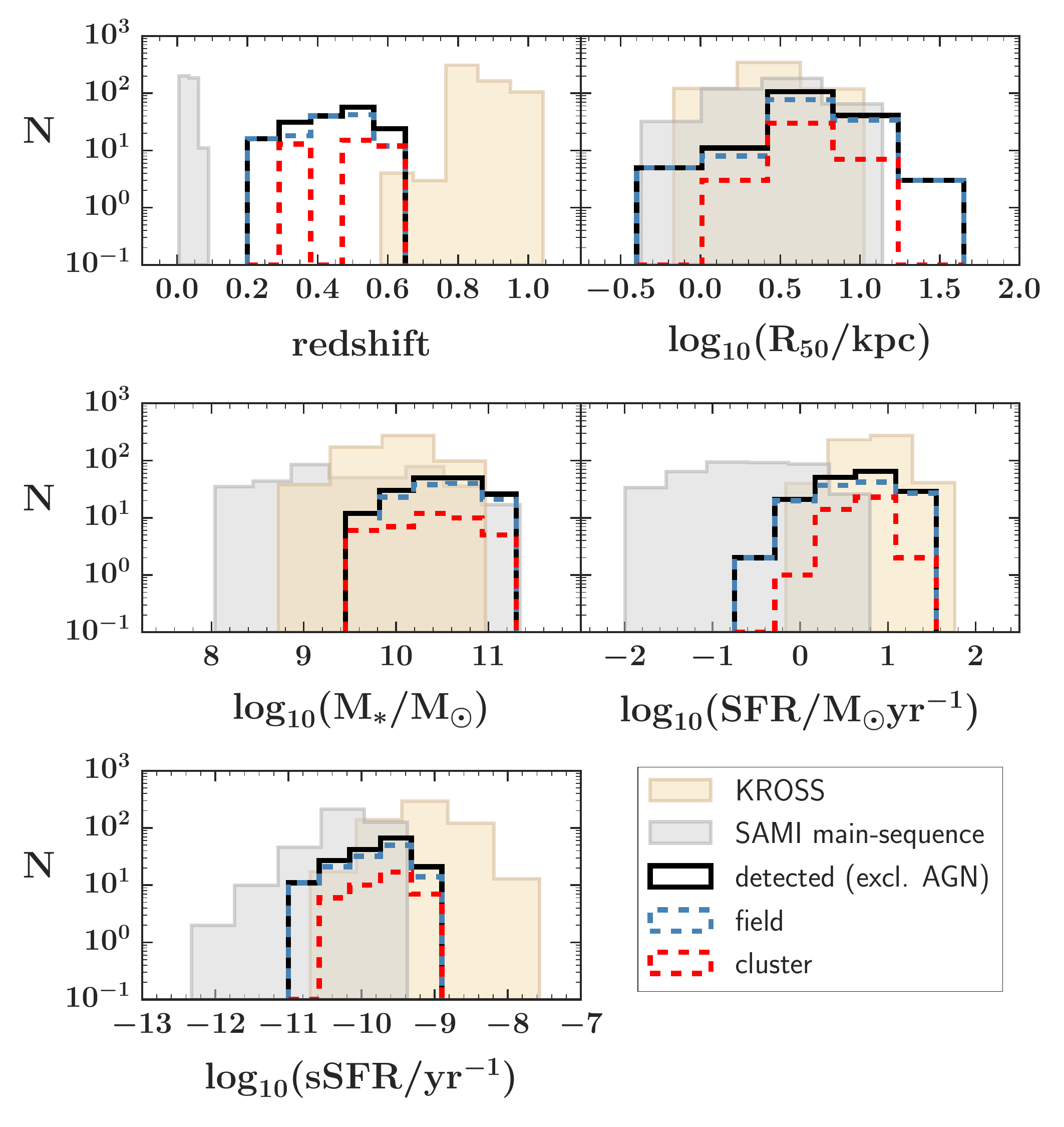}
\end{minipage}
\caption{%
The distributions of key properties of the H$\alpha$-detected K-CLASH galaxies (black solid, and blue and red dashed histograms), compared to the same properties for galaxies observed as part of other exisiting IFS surveys at higher and lower redshifts (orange and grey shaded histograms). K-CLASH galaxies detected in H$\alpha$ are typical star-forming galaxies for their epochs. %
     }%
\label{fig:contexthists}
\end{figure*}

\subsection{K-CLASH in context}
\label{subsec:redshiftcontext}

In this work, we have shown that the sample of 191 H$\alpha$-detected K-CLASH galaxies is predominantly comprised of bright and blue systems (see \S~\ref{sec:sampleoverview}) that reside in the field (the {\it field} sub-sample), with a smaller number of galaxies that are members of one of the four CLASH clusters targeted (the {\it cluster} sub-sample, discussed in \S~\ref{subsec:clusterdiscussion}). The positions of the K-CLASH galaxies within the SFR--stellar mass plane are, on average, coincident with the main sequence of star-formation of galaxies at the median redshift of our sample (according to \citealt{Schreiber:2015}; see \S~\ref{subsec:stellarmasses} and \S~\ref{subsec:fluxandsfr}). This is true whether we consider total SFRs derived from {\sc ProSpect} SED fitting (see Figure~\ref{fig:MSpanelplot}) or SFRs derived from the galaxies' H$\alpha$ emission, accounting for aperture and dust extinction effects (see Figure~\ref{fig:mainsequence}). We also demonstrated that the K-CLASH galaxies have stellar light sizes that are typical for their stellar masses and redshifts (see Figure~\ref{fig:sizeversusmass}). And in \S~\ref{subsubsec:maps}, we showed examples of the spatially-resolved properties of K-CLASH galaxies, discussing in \S~\ref{subsec:resolvednumbers} the fact that we spatially-resolve the H$\alpha$ emission from 146 galaxies, including the majority of the {\it cluster} and {\it field} sub-sample galaxies (85 per cent and 73 per cent, respectively).

Given these key properties of the K-CLASH galaxies, we therefore conclude that the K-CLASH survey provides an appropriate intermediate-epoch IFS sample of ``normal'' star-forming galaxies. K-CLASH bridges the gap in redshift between exisiting similarly-constructed samples and provides the means to further our understanding of the evolution of the properties of star-forming galaxies over a large fraction of the history of our Universe. In this sub-section, we thus quantitatively place the K-CLASH sample in its cosmological context, comparing the key properties of K-CLASH galaxies to those of star-forming galaxies at higher and lower redshifts observed as part of previous large IFS surveys.

For a comparison sample in the local Universe, we use the sub-sample of star-forming (i.e.\ H$\alpha$-emitting) galaxies from the SAMI Galaxy Survey \citep{Bryant:2015} presented in \citet{Tiley:2019a}. The SAMI Galaxy Survey used the Sydney-AAO Multi-object Integral-field spectrograph \citep[SAMI;][]{Croom:2012} to measure the spatially-resolved kinematics of $\approx3000$ galaxies at $0.004 < z < 0.095$, encompassing a large range of environments. Each SAMI IFU observation has a $15''$ diameter effective FOV. SAMI observations have a spectral resolution of $R\approx4500$ (corresponding to a velocity resolution $\sigma=29$ km s$^{-1}$) in the red band ($6250$--$7350$~\AA). The SAMI observations were carried out in natural seeing conditions, with a typical range of $0\farcs9$ -- $3\farcs0$.  To compare with the K-CLASH sample, we extract a subset of 397 main sequence galaxies (spatially-resolved in H$\alpha$ by SAMI) from the sample presented in \citet{Tiley:2019a} by applying an iterative $1.5\sigma$ clip to their running median in the SFR--stellar mass plane. In this work, we use the SAMI galaxy stellar masses described in \citet{Bryant:2015}, derived from $g - i$ colours and $i$-band magnitudes from the Galaxy And Mass Assembly survey \citep[GAMA;][]{Driver:2011} according to the method of \citet{Taylor:2011} and assuming a \citet{Chabrier:2003} IMF. The SAMI galaxy SFRs used in this work were derived by \citet{Davies:2016} via the application of {\sc magphys} to extensive multi-wavelength GAMA photometry.

To compare to star-forming galaxies at higher redshift, we consider the 586 star-forming galaxies at $z\approx0.9$ spatially-resolved in H$\alpha$ emission as part of KROSS \citep[][]{Stott:2016,Harrison:2017}. KROSS comprises IFS observations of the H$\alpha$ and [N {\sc ii}] emission from galaxies at $0.6\lesssim z \lesssim 1$ with KMOS in the $YJ$-band ($\approx 1.02$--$1.36$~$\mu$m). The spectral resolving power of KMOS in this band ranges from $R \approx 3000$ to $\approx4000$ ($\sigma \approx 30$ -- $40$ km s$^{-1}$ depending on the wavelength). The KROSS targets were selected to have $K < 22.5$, with preference given to blue ($r - z <1.5$) systems. The KROSS galaxies reside on the main-sequence of star-formation at their epoch. The median seeing in the $YJ$-band for KROSS observations was $0\farcs7$.

In Figure~\ref{fig:contexthists}, we compare key properties of the K-CLASH galaxies detected in H$\alpha$ emission (excluding candidate AGN hosts) with those of the star-forming galaxies selected from SAMI and KROSS. Figure~\ref{fig:contexthists} shows that the K-CLASH galaxies conveniently bridge the gap in redshift coverage between SAMI and KROSS. It also shows that the K-CLASH galaxies have stellar masses higher, on average, than those of the other IFS samples (as a result of the limited depth of the K-CLASH observations, we preferentially selected high-mass systems that are bright in H$\alpha$). Correspondingly \citep[since stellar size correlates with stellar mass; e.g.][]{Shen:2003,Bernardi:2011,Lange:2015}, the K-CLASH galaxies have a larger average stellar size than that of galaxies in the other IFS samples. The star-formation rates of the K-CLASH galaxies are also much larger than those of the SAMI galaxies, and moderately smaller than those of the KROSS galaxies. Thus we find that the sSFRs of the K-CLASH galaxies are, on average, much smaller than those of the KROSS galaxies and only moderately larger than those of the SAMI galaxies.

Figure~\ref{fig:contexthists} also shows that the {\it cluster} and {\it field} sub-sample galaxies are well-matched in their key properties; the {\it field} sub-sample comfortably encompasses the {\it cluster} sample in terms of its redshift and stellar mass ranges, allowing for a fair comparison between the properties of star-forming galaxies in the cluster and field environments in the range $z\approx0.3$ -- $0.6$.

In agreement with our previous conclusions, we thus find that K-CLASH galaxies detected in H$\alpha$ emission constitute an ideal sample with which to bridge the gap in redshift between IFS studies at $z\approx1$--$3$ and $z\approx0$.

\section{Conclusions and Future Work}
\label{sec:conclusions}

We have presented the K-CLASH survey, a University of Oxford guaranteed time KMOS programme to study the spatially-resolved ionised gas properties of (predominantly blue) star-forming galaxies in cluster and field environments in the redshift range $0.2 \lesssim z\lesssim 0.6$, in CLASH cluster fields. We detected H$\alpha$ from 191 out of 282 galaxies targeted with KMOS, spatially-resolving H$\alpha$ in 146 of those detected. Excluding candidate AGN hosts, we defined {\it cluster} and {\it field} galaxy sub-samples, comprising respectively 40 galaxies detected in H$\alpha$ (34 spatially-resolved in H$\alpha$) and within $3\sigma_{\rm{cluster}}$ and $2\rm{R}_{200}$ of one of the four CLASH clusters targeted, and 128 galaxies detected in H$\alpha$ (94 spatially-resolved in H$\alpha$) and not in a CLASH cluster. 

We showed that the K-CLASH galaxies, on average, are on the main sequence of star formation and have normal stellar sizes given their stellar masses and redshifts. We showed that, depending on the cluster redshift, star-forming (i.e.\ H$\alpha$-detected) K-CLASH galaxies in cluster environments have stellar masses either similar to or systematically lower than those of star-forming galaxies in the field at the same redshift. Star-forming cluster members also have the same SFRs as star-forming field galaxies after accounting for stellar mass and redshift. We however noted the caveat that our selection for blue galaxies implies that, by construction, we are more likely to detect those galaxies with a higher H$\alpha$ luminosity, and thus a higher SFR (necessary to effectively trace the spatially-resolved gas properties and kinematics of galaxies at $0.3 \lesssim z \lesssim 0.6$ with KMOS). Thus we are also likely to simultaneously preferentially exclude from our sample the most quenched galaxies in the cluster environments, that should have much lower H$\alpha$ luminosities. 

We then demonstrated the diagnostic power of our spatially-resolved KMOS observations, presenting KMOS maps for example {\it cluster} and {\it field} galaxies and showing that star-forming galaxies in either environment exhibit a similar range of H$\alpha$ morphologies and kinematics. We reasoned that our results suggest either that quenching of star-forming galaxies in clusters at $z\approx0.3$--$0.6$ is very rapid, or that star-forming K-CLASH galaxies in cluster environments have only recently arrived there and have not yet undergone significant quenching. A more detailed discussion of the properties of the K-CLASH star-forming {\it cluster} galaxies, and the role of the cluster environment in galaxy quenching between $z\approx0.3$ and $\approx0.6$, is presented in \citet{vaughan:2020}. 

Finally, we placed the K-CLASH survey in the context of other large IFS surveys at higher and lower redshifts, showing that the K-CLASH galaxies occupy a unique gap in redshift and sSFR not covered by the large samples of star-forming galaxies constructed by e.g. the KROSS ($z\approx0.9$) and SAMI ($z\approx0$) surveys. Combined with these exisiting data sets (and future ones), K-CLASH thus provides the ideal sample to help further our understanding of how, why, and when  in the past 10 Gyr the star-forming galaxy population in the Universe underwent such substantial transformations in morphology and kinematics, and what role clusters have played in the build-up of the red sequence over the same period. 

Future K-CLASH works will examine the redshift evolution of the properties of the star-forming population since $z\approx1$--$3$, presenting the kinematics properties of the K-CLASH sample and combining the K-CLASH data with public data from suitable IFS samples at higher and lower redshifts. Additional studies will include an examination of the properties of the candidate AGN hosts and BCGs in the K-CLASH sample, which are not discussed in this work. 

\section*{Acknowledgments}

ALT acknowledges support from a Forrest Research Foundation Fellowship, Science and Technology Facilities Council (STFC) grants ST/L00075X/1 and ST/P000541/1, the ERC Advanced Grant DUSTYGAL (321334), and an STFC Studentship. SPV acknowledges support from a doctoral studentship from the STFC grant ST/N504233/1. MB acknowledges support from STFC rolling grant Astrophysics at Oxford PP/E001114/1. LC is the recipient of an Australian Research Council Future Fellowship (FT180100066) funded by the Australian Government. Parts of this research were supported by the Australian Research Council Centre of Excellence for All Sky Astrophysics in 3 Dimensions (ASTRO 3D), through project number CE170100013. This work is based on observations made with ESO Telescopes at the La Silla Paranal Observatory under the programme IDs 097.A-0397, 098.A-0224, 099.A-0207, and 0100.A-0296. The SAMI Galaxy Survey is based on observations made at the Anglo-Australian Telescope. The Sydney-AAO Multi-object Integral-field spectrograph (SAMI) was developed jointly by the University of Sydney and the Australian Astronomical Observatory. The SAMI input catalogue is based on data taken from the Sloan Digital Sky Survey, the GAMA Survey and the VST ATLAS Survey. The SAMI Galaxy Survey is funded by the Australian Research Council Centre of Excellence for All-sky Astrophysics (CAASTRO), through project number CE110001020, and other participating institutions. The SAMI Galaxy Survey website is \url{http://sami-survey.org/}.




\bibliographystyle{mnras}
\bibliography{kclash_survey_paper.bib} 

\begin{thebibliography}{}
\makeatletter
\relax
\def\mn@urlcharsother{\let\do\@makeother \do\$\do\&\do\#\do\^\do\_\do\%\do\~}
\def\mn@doi{\begingroup\mn@urlcharsother \@ifnextchar [ {\mn@doi@}
  {\mn@doi@[]}}
\def\mn@doi@[#1]#2{\def\@tempa{#1}\ifx\@tempa\@empty \href
  {http://dx.doi.org/#2} {doi:#2}\else \href {http://dx.doi.org/#2} {#1}\fi
  \endgroup}
\def\mn@eprint#1#2{\mn@eprint@#1:#2::\@nil}
\def\mn@eprint@arXiv#1{\href {http://arxiv.org/abs/#1} {{\tt arXiv:#1}}}
\def\mn@eprint@dblp#1{\href {http://dblp.uni-trier.de/rec/bibtex/#1.xml}
  {dblp:#1}}
\def\mn@eprint@#1:#2:#3:#4\@nil{\def\@tempa {#1}\def\@tempb {#2}\def\@tempc
  {#3}\ifx \@tempc \@empty \let \@tempc \@tempb \let \@tempb \@tempa \fi \ifx
  \@tempb \@empty \def\@tempb {arXiv}\fi \@ifundefined
  {mn@eprint@\@tempb}{\@tempb:\@tempc}{\expandafter \expandafter \csname
  mn@eprint@\@tempb\endcsname \expandafter{\@tempc}}}

\bibitem[\protect\citeauthoryear{{Acker}, {K{\"o}ppen}, {Samland}  \&
  {Stenholm}}{{Acker} et~al.}{1989}]{Acker:1989}
{Acker} A.,  {K{\"o}ppen} J.,  {Samland} M.,   {Stenholm} B.,  1989, The
  Messenger, \href {https://ui.adsabs.harvard.edu/abs/1989Msngr..58...44A} {58,
  44}

\bibitem[\protect\citeauthoryear{{Arnouts}, {Cristiani}, {Moscardini},
  {Matarrese}, {Lucchin}, {Fontana}  \& {Giallongo}}{{Arnouts}
  et~al.}{1999}]{Arnouts:1999}
{Arnouts} S.,  {Cristiani} S.,  {Moscardini} L.,  {Matarrese} S.,  {Lucchin}
  F.,  {Fontana} A.,   {Giallongo} E.,  1999, \mn@doi [\mnras]
  {10.1046/j.1365-8711.1999.02978.x}, \href
  {http://adsabs.harvard.edu/abs/1999MNRAS.310..540A} {310, 540}

\bibitem[\protect\citeauthoryear{{Baade} et~al.,}{{Baade}
  et~al.}{1999}]{Baade:1999}
{Baade} D.,  et~al., 1999, The Messenger, \href
  {http://adsabs.harvard.edu/abs/1999Msngr..95...15B} {95, 15}

\bibitem[\protect\citeauthoryear{{Bacon} et~al.,}{{Bacon}
  et~al.}{2010}]{Bacon:2010}
{Bacon} R.,  et~al., 2010, in Ground-based and Airborne Instrumentation for
  Astronomy III. p. 773508, \mn@doi{10.1117/12.856027}

\bibitem[\protect\citeauthoryear{{Baldwin}, {Phillips}  \&
  {Terlevich}}{{Baldwin} et~al.}{1981}]{BPT}
{Baldwin} J.~A.,  {Phillips} M.~M.,   {Terlevich} R.,  1981, \mn@doi [\pasp]
  {10.1086/130766}, \href {http://adsabs.harvard.edu/abs/1981PASP...93....5B}
  {93, 5}

\bibitem[\protect\citeauthoryear{{Balogh}, {Navarro}  \& {Morris}}{{Balogh}
  et~al.}{2000}]{Balogh:2000}
{Balogh} M.~L.,  {Navarro} J.~F.,   {Morris} S.~L.,  2000, \mn@doi [\apj]
  {10.1086/309323}, \href
  {https://ui.adsabs.harvard.edu/abs/2000ApJ...540..113B} {540, 113}

\bibitem[\protect\citeauthoryear{{Beifiori} et~al.,}{{Beifiori}
  et~al.}{2017}]{Beifiori:2017}
{Beifiori} A.,  et~al., 2017, \mn@doi [\apj] {10.3847/1538-4357/aa8368}, \href
  {http://adsabs.harvard.edu/abs/2017ApJ...846..120B} {846, 120}

\bibitem[\protect\citeauthoryear{{Bekki}}{{Bekki}}{2009}]{Bekki:2009}
{Bekki} K.,  2009, \mn@doi [\mnras] {10.1111/j.1365-2966.2009.15431.x}, \href
  {https://ui.adsabs.harvard.edu/abs/2009MNRAS.399.2221B} {399, 2221}

\bibitem[\protect\citeauthoryear{{Bell} et~al.,}{{Bell}
  et~al.}{2004}]{Bell:2004}
{Bell} E.~F.,  et~al., 2004, \mn@doi [\apj] {10.1086/420778}, \href
  {http://adsabs.harvard.edu/abs/2004ApJ...608..752B} {608, 752}

\bibitem[\protect\citeauthoryear{{Bell} et~al.,}{{Bell}
  et~al.}{2005}]{Bell:2005}
{Bell} E.~F.,  et~al., 2005, \mn@doi [\apj] {10.1086/429552}, \href
  {https://ui.adsabs.harvard.edu/abs/2005ApJ...625...23B} {625, 23}

\bibitem[\protect\citeauthoryear{{Ben{\'{\i}}tez}}{{Ben{\'{\i}}tez}}{2000}]{Benitez:2000}
{Ben{\'{\i}}tez} N.,  2000, \mn@doi [\apj] {10.1086/308947}, \href
  {http://adsabs.harvard.edu/abs/2000ApJ...536..571B} {536, 571}

\bibitem[\protect\citeauthoryear{{Bernardi}, {Roche}, {Shankar}  \&
  {Sheth}}{{Bernardi} et~al.}{2011}]{Bernardi:2011}
{Bernardi} M.,  {Roche} N.,  {Shankar} F.,   {Sheth} R.~K.,  2011, \mn@doi
  [\mnras] {10.1111/j.1745-3933.2010.00982.x}, \href
  {http://adsabs.harvard.edu/abs/2011MNRAS.412L...6B} {412, L6}

\bibitem[\protect\citeauthoryear{Bollen}{Bollen}{1989}]{Bollen:1989}
Bollen K.~A.,  1989, Structural equations with latent variables.
Wiley, New York

\bibitem[\protect\citeauthoryear{{Brinchmann} \& {Ellis}}{{Brinchmann} \&
  {Ellis}}{2000}]{Brinchmann:2000}
{Brinchmann} J.,  {Ellis} R.~S.,  2000, \mn@doi [\apjl] {10.1086/312738}, \href
  {https://ui.adsabs.harvard.edu/abs/2000ApJ...536L..77B} {536, L77}

\bibitem[\protect\citeauthoryear{{Bruzual} \& {Charlot}}{{Bruzual} \&
  {Charlot}}{2003}]{Bruzual:2003aa}
{Bruzual} G.,  {Charlot} S.,  2003, \mn@doi [\mnras]
  {10.1046/j.1365-8711.2003.06897.x}, \href
  {http://adsabs.harvard.edu/abs/2003MNRAS.344.1000B} {344, 1000}

\bibitem[\protect\citeauthoryear{{Bryant} et~al.,}{{Bryant}
  et~al.}{2015}]{Bryant:2015}
{Bryant} J.~J.,  et~al., 2015, \mn@doi [\mnras] {10.1093/mnras/stu2635}, \href
  {http://adsabs.harvard.edu/abs/2015MNRAS.447.2857B} {447, 2857}

\bibitem[\protect\citeauthoryear{{Bundy} et~al.,}{{Bundy}
  et~al.}{2015}]{Bundy:2015}
{Bundy} K.,  et~al., 2015, \mn@doi [\apj] {10.1088/0004-637X/798/1/7}, \href
  {http://adsabs.harvard.edu/abs/2015ApJ...798....7B} {798, 7}

\bibitem[\protect\citeauthoryear{{Butcher} \& {Oemler}}{{Butcher} \&
  {Oemler}}{1978}]{Butcher:1978}
{Butcher} H.,  {Oemler} Jr. A.,  1978, \mn@doi [\apj] {10.1086/156640}, \href
  {http://adsabs.harvard.edu/abs/1978ApJ...226..559B} {226, 559}

\bibitem[\protect\citeauthoryear{{Calzetti}, {Kinney}  \&
  {Storchi-Bergmann}}{{Calzetti} et~al.}{1994}]{Calzetti:1994}
{Calzetti} D.,  {Kinney} A.~L.,   {Storchi-Bergmann} T.,  1994, \mn@doi [\apj]
  {10.1086/174346}, \href
  {https://ui.adsabs.harvard.edu/abs/1994ApJ...429..582C} {429, 582}

\bibitem[\protect\citeauthoryear{{Cappellari}}{{Cappellari}}{2016}]{Cappellari:2016}
{Cappellari} M.,  2016, \mn@doi [\araa] {10.1146/annurev-astro-082214-122432},
  \href {https://ui.adsabs.harvard.edu/abs/2016ARA&A..54..597C} {54, 597}

\bibitem[\protect\citeauthoryear{{Cappellari} et~al.,}{{Cappellari}
  et~al.}{2011}]{Cappellari:2011}
{Cappellari} M.,  et~al., 2011, \mn@doi [\mnras]
  {10.1111/j.1365-2966.2010.18174.x}, \href
  {http://adsabs.harvard.edu/abs/2011MNRAS.413..813C} {413, 813}

\bibitem[\protect\citeauthoryear{{Chabrier}}{{Chabrier}}{2003}]{Chabrier:2003}
{Chabrier} G.,  2003, \mn@doi [\pasp] {10.1086/376392}, \href
  {http://adsabs.harvard.edu/abs/2003PASP..115..763C} {115, 763}

\bibitem[\protect\citeauthoryear{{Charlot} \& {Fall}}{{Charlot} \&
  {Fall}}{2000}]{Charlot:2000}
{Charlot} S.,  {Fall} S.~M.,  2000, \mn@doi [\apj] {10.1086/309250}, \href
  {https://ui.adsabs.harvard.edu/abs/2000ApJ...539..718C} {539, 718}

\bibitem[\protect\citeauthoryear{{Cirasuolo} et~al.,}{{Cirasuolo}
  et~al.}{2007}]{Cirasuolo:2007}
{Cirasuolo} M.,  et~al., 2007, \mn@doi [\mnras]
  {10.1111/j.1365-2966.2007.12038.x}, \href
  {http://adsabs.harvard.edu/abs/2007MNRAS.380..585C} {380, 585}

\bibitem[\protect\citeauthoryear{{Coe}, {Ben{\'{\i}}tez}, {S{\'a}nchez}, {Jee},
  {Bouwens}  \& {Ford}}{{Coe} et~al.}{2006}]{Coe:2006}
{Coe} D.,  {Ben{\'{\i}}tez} N.,  {S{\'a}nchez} S.~F.,  {Jee} M.,  {Bouwens} R.,
    {Ford} H.,  2006, \mn@doi [\aj] {10.1086/505530}, \href
  {http://adsabs.harvard.edu/abs/2006AJ....132..926C} {132, 926}

\bibitem[\protect\citeauthoryear{{Cowie}, {Songaila}, {Hu}  \& {Cohen}}{{Cowie}
  et~al.}{1996}]{Cowie:1996}
{Cowie} L.~L.,  {Songaila} A.,  {Hu} E.~M.,   {Cohen} J.~G.,  1996, \mn@doi
  [\aj] {10.1086/118058}, \href
  {https://ui.adsabs.harvard.edu/abs/1996AJ....112..839C} {112, 839}

\bibitem[\protect\citeauthoryear{{Croom} et~al.,}{{Croom}
  et~al.}{2012}]{Croom:2012}
{Croom} S.~M.,  et~al., 2012, \mn@doi [\mnras]
  {10.1111/j.1365-2966.2011.20365.x}, \href
  {http://adsabs.harvard.edu/abs/2012MNRAS.421..872C} {421, 872}

\bibitem[\protect\citeauthoryear{{Dale}, {Helou}, {Magdis}, {Armus},
  {D{\'\i}az-Santos}  \& {Shi}}{{Dale} et~al.}{2014}]{Dale:2014}
{Dale} D.~A.,  {Helou} G.,  {Magdis} G.~E.,  {Armus} L.,  {D{\'\i}az-Santos}
  T.,   {Shi} Y.,  2014, \mn@doi [\apj] {10.1088/0004-637X/784/1/83}, \href
  {https://ui.adsabs.harvard.edu/abs/2014ApJ...784...83D} {784, 83}

\bibitem[\protect\citeauthoryear{{Davies} et~al.,}{{Davies}
  et~al.}{2016}]{Davies:2016}
{Davies} L.~J.~M.,  et~al., 2016, \mn@doi [\mnras] {10.1093/mnras/stw1342},
  \href {https://ui.adsabs.harvard.edu/abs/2016MNRAS.461..458D} {461, 458}

\bibitem[\protect\citeauthoryear{{Dickinson}, {Giavalisco}  \& {GOODS
  Team}}{{Dickinson} et~al.}{2003}]{Dickinson:2003}
{Dickinson} M.,  {Giavalisco} M.,   {GOODS Team} 2003, in {Bender} R.,
  {Renzini} A.,  eds, The Mass of Galaxies at Low and High Redshift. p.~324
  (\mn@eprint {} {astro-ph/0204213}), \mn@doi{10.1007/10899892_78}

\bibitem[\protect\citeauthoryear{{Donley} et~al.,}{{Donley}
  et~al.}{2012}]{Donley:2012}
{Donley} J.~L.,  et~al., 2012, \mn@doi [\apj] {10.1088/0004-637X/748/2/142},
  \href {https://ui.adsabs.harvard.edu/\#abs/2012ApJ...748..142D} {748, 142}

\bibitem[\protect\citeauthoryear{{Donnarumma}, {Ettori}, {Meneghetti}  \&
  {Moscardini}}{{Donnarumma} et~al.}{2009}]{Donnarumma:2009}
{Donnarumma} A.,  {Ettori} S.,  {Meneghetti} M.,   {Moscardini} L.,  2009,
  \mn@doi [\mnras] {10.1111/j.1365-2966.2009.15165.x}, \href
  {https://ui.adsabs.harvard.edu/abs/2009MNRAS.398..438D} {398, 438}

\bibitem[\protect\citeauthoryear{{Dressler} et~al.,}{{Dressler}
  et~al.}{2011}]{Dressler:2011}
{Dressler} A.,  et~al., 2011, \mn@doi [\pasp] {10.1086/658908}, \href
  {https://ui.adsabs.harvard.edu/abs/2011PASP..123..288D} {123, 288}

\bibitem[\protect\citeauthoryear{{Driver} et~al.,}{{Driver}
  et~al.}{2011}]{Driver:2011}
{Driver} S.~P.,  et~al., 2011, \mn@doi [\mnras]
  {10.1111/j.1365-2966.2010.18188.x}, \href
  {http://adsabs.harvard.edu/abs/2011MNRAS.413..971D} {413, 971}

\bibitem[\protect\citeauthoryear{{Dudzevi{\v{c}}i{\={u}}t{\.{e}}}
  et~al.,}{{Dudzevi{\v{c}}i{\={u}}t{\.{e}}} et~al.}{2019}]{Dudzevi:2019}
{Dudzevi{\v{c}}i{\={u}}t{\.{e}}} U.,  et~al., 2019, arXiv e-prints, \href
  {https://ui.adsabs.harvard.edu/abs/2019arXiv191007524D} {p. arXiv:1910.07524}

\bibitem[\protect\citeauthoryear{{Ebeling}, {Barrett}, {Donovan}, {Ma}, {Edge}
  \& {van Speybroeck}}{{Ebeling} et~al.}{2007}]{Ebeling:2007}
{Ebeling} H.,  {Barrett} E.,  {Donovan} D.,  {Ma} C.~J.,  {Edge} A.~C.,   {van
  Speybroeck} L.,  2007, \mn@doi [\apjl] {10.1086/518603}, \href
  {https://ui.adsabs.harvard.edu/abs/2007ApJ...661L..33E} {661, L33}

\bibitem[\protect\citeauthoryear{{Ebeling}, {Edge}, {Mantz}, {Barrett},
  {Henry}, {Ma}  \& {van Speybroeck}}{{Ebeling} et~al.}{2010}]{Ebeling:2010}
{Ebeling} H.,  {Edge} A.~C.,  {Mantz} A.,  {Barrett} E.,  {Henry} J.~P.,  {Ma}
  C.~J.,   {van Speybroeck} L.,  2010, \mn@doi [\mnras]
  {10.1111/j.1365-2966.2010.16920.x}, \href
  {https://ui.adsabs.harvard.edu/abs/2010MNRAS.407...83E} {407, 83}

\bibitem[\protect\citeauthoryear{{Ehlert} et~al.,}{{Ehlert}
  et~al.}{2011}]{Ehlert:2011}
{Ehlert} S.,  et~al., 2011, \mn@doi [\mnras]
  {10.1111/j.1365-2966.2010.17801.x}, \href
  {http://adsabs.harvard.edu/abs/2011MNRAS.411.1641E} {411, 1641}

\bibitem[\protect\citeauthoryear{{Erwin}}{{Erwin}}{2015}]{Erwin:2015}
{Erwin} P.,  2015, \mn@doi [\apj] {10.1088/0004-637X/799/2/226}, \href
  {https://ui.adsabs.harvard.edu/abs/2015ApJ...799..226E} {799, 226}

\bibitem[\protect\citeauthoryear{{Fanidakis} et~al.,}{{Fanidakis}
  et~al.}{2012}]{Fanidakis:2012}
{Fanidakis} N.,  et~al., 2012, \mn@doi [\mnras]
  {10.1111/j.1365-2966.2011.19931.x}, \href
  {https://ui.adsabs.harvard.edu/abs/2012MNRAS.419.2797F} {419, 2797}

\bibitem[\protect\citeauthoryear{{Fazio} et~al.,}{{Fazio}
  et~al.}{2004}]{Fazio:2004}
{Fazio} G.~G.,  et~al., 2004, \mn@doi [\apjs] {10.1086/422843}, \href
  {http://adsabs.harvard.edu/abs/2004ApJS..154...10F} {154, 10}

\bibitem[\protect\citeauthoryear{{Folkes} et~al.,}{{Folkes}
  et~al.}{1999}]{Folkes:1999}
{Folkes} S.,  et~al., 1999, \mn@doi [\mnras]
  {10.1046/j.1365-8711.1999.02721.x}, \href
  {https://ui.adsabs.harvard.edu/abs/1999MNRAS.308..459F} {308, 459}

\bibitem[\protect\citeauthoryear{{F{\"o}rster Schreiber} et~al.,}{{F{\"o}rster
  Schreiber} et~al.}{2009}]{ForsterSchreiber:2009}
{F{\"o}rster Schreiber} N.~M.,  et~al., 2009, \mn@doi [\apj]
  {10.1088/0004-637X/706/2/1364}, \href
  {http://adsabs.harvard.edu/abs/2009ApJ...706.1364F} {706, 1364}

\bibitem[\protect\citeauthoryear{{F{\"o}rster Schreiber} et~al.,}{{F{\"o}rster
  Schreiber} et~al.}{2018}]{ForsterSchreiber:2018}
{F{\"o}rster Schreiber} N.~M.,  et~al., 2018, arXiv e-prints, \href
  {https://ui.adsabs.harvard.edu/\#abs/2018arXiv180704738F} {p.
  arXiv:1807.04738}

\bibitem[\protect\citeauthoryear{{Galametz} et~al.,}{{Galametz}
  et~al.}{2013}]{Galametz:2013}
{Galametz} A.,  et~al., 2013, \mn@doi [\apjs] {10.1088/0067-0049/206/2/10},
  \href {https://ui.adsabs.harvard.edu/abs/2013ApJS..206...10G} {206, 10}

\bibitem[\protect\citeauthoryear{{Giavalisco} et~al.,}{{Giavalisco}
  et~al.}{2004}]{Giavalisco:2004}
{Giavalisco} M.,  et~al., 2004, \mn@doi [\apjl] {10.1086/379232}, \href
  {https://ui.adsabs.harvard.edu/abs/2004ApJ...600L..93G} {600, L93}

\bibitem[\protect\citeauthoryear{{Girardi}, {Fadda}, {Giuricin},
  {Mardirossian}, {Mezzetti}  \& {Biviano}}{{Girardi}
  et~al.}{1996}]{Girardi:1996}
{Girardi} M.,  {Fadda} D.,  {Giuricin} G.,  {Mardirossian} F.,  {Mezzetti} M.,
   {Biviano} A.,  1996, \mn@doi [\apj] {10.1086/176711}, \href
  {http://adsabs.harvard.edu/abs/1996ApJ...457...61G} {457, 61}

\bibitem[\protect\citeauthoryear{{Glenn} et~al.,}{{Glenn}
  et~al.}{1998}]{Glenn:1998}
{Glenn} J.,  et~al., 1998, {Bolocam: a millimeter-wave bolometric camera}.
pp 326--334, \mn@doi{10.1117/12.317418}

\bibitem[\protect\citeauthoryear{{Gnerucci} et~al.,}{{Gnerucci}
  et~al.}{2011}]{Gnerucci:2011}
{Gnerucci} A.,  et~al., 2011, \mn@doi [\aap] {10.1051/0004-6361/201015465},
  \href {http://adsabs.harvard.edu/abs/2011A%26A...528A..88G} {528, A88}

\bibitem[\protect\citeauthoryear{{Grant}, {Bautz}, {Ford}  \&
  {Plucinsky}}{{Grant} et~al.}{2014}]{Grant:2014}
{Grant} C.~E.,  {Bautz} M.~W.,  {Ford} P.~G.,   {Plucinsky} P.~P.,  2014,
  {Fifteen years of the Advanced CCD Imaging Spectrometer}.
p. 91443Q, \mn@doi{10.1117/12.2055652}

\bibitem[\protect\citeauthoryear{{Grogin} et~al.,}{{Grogin}
  et~al.}{2011}]{Grogin:2011}
{Grogin} N.~A.,  et~al., 2011, \mn@doi [\apjs] {10.1088/0067-0049/197/2/35},
  \href {http://adsabs.harvard.edu/abs/2011ApJS..197...35G} {197, 35}

\bibitem[\protect\citeauthoryear{{Gunn} \& {Gott}}{{Gunn} \&
  {Gott}}{1972}]{Gunn:1972}
{Gunn} J.~E.,  {Gott} J.~Richard I.,  1972, \mn@doi [\apj] {10.1086/151605},
  \href {https://ui.adsabs.harvard.edu/abs/1972ApJ...176....1G} {176, 1}

\bibitem[\protect\citeauthoryear{{Guzm{\'a}n}, {Gallego}, {Koo}, {Phillips},
  {Lowenthal}, {Faber}, {Illingworth}  \& {Vogt}}{{Guzm{\'a}n}
  et~al.}{1997}]{Guzman:1997}
{Guzm{\'a}n} R.,  {Gallego} J.,  {Koo} D.~C.,  {Phillips} A.~C.,  {Lowenthal}
  J.~D.,  {Faber} S.~M.,  {Illingworth} G.~D.,   {Vogt} N.~P.,  1997, \mn@doi
  [\apj] {10.1086/304797}, \href
  {https://ui.adsabs.harvard.edu/abs/1997ApJ...489..559G} {489, 559}

\bibitem[\protect\citeauthoryear{{Harrison} et~al.,}{{Harrison}
  et~al.}{2017}]{Harrison:2017}
{Harrison} C.~M.,  et~al., 2017, \mn@doi [\mnras] {10.1093/mnras/stx217}, \href
  {http://adsabs.harvard.edu/abs/2017MNRAS.467.1965H} {467, 1965}

\bibitem[\protect\citeauthoryear{{Hilton} et~al.,}{{Hilton}
  et~al.}{2010}]{Hilton:2010}
{Hilton} M.,  et~al., 2010, \mn@doi [\apj] {10.1088/0004-637X/718/1/133}, \href
  {http://adsabs.harvard.edu/abs/2010ApJ...718..133H} {718, 133}

\bibitem[\protect\citeauthoryear{{Hinshaw} et~al.,}{{Hinshaw}
  et~al.}{2013}]{Hinshaw:2013}
{Hinshaw} G.,  et~al., 2013, \mn@doi [\apjs] {10.1088/0067-0049/208/2/19},
  \href {http://adsabs.harvard.edu/abs/2013ApJS..208...19H} {208, 19}

\bibitem[\protect\citeauthoryear{{Hubble}}{{Hubble}}{1926}]{Hubble:1926}
{Hubble} E.~P.,  1926, \mn@doi [\apj] {10.1086/143018}, \href
  {http://adsabs.harvard.edu/abs/1926ApJ....64..321H} {64}

\bibitem[\protect\citeauthoryear{{Hubble}}{{Hubble}}{1936}]{Hubble:1936}
{Hubble} E.~P.,  1936, {Realm of the Nebulae}

\bibitem[\protect\citeauthoryear{{Ilbert} et~al.,}{{Ilbert}
  et~al.}{2006}]{Ilbert:2006}
{Ilbert} O.,  et~al., 2006, \mn@doi [\aap] {10.1051/0004-6361:20065138}, \href
  {http://adsabs.harvard.edu/abs/2006A%26A...457..841I} {457, 841}

\bibitem[\protect\citeauthoryear{{Juneau} et~al.,}{{Juneau}
  et~al.}{2005}]{Juneau:2005}
{Juneau} S.,  et~al., 2005, \mn@doi [\apjl] {10.1086/427937}, \href
  {https://ui.adsabs.harvard.edu/abs/2005ApJ...619L.135J} {619, L135}

\bibitem[\protect\citeauthoryear{{Kartaltepe} et~al.,}{{Kartaltepe}
  et~al.}{2010}]{Kartaltepe:2010}
{Kartaltepe} J.~S.,  et~al., 2010, \mn@doi [\apj]
  {10.1088/0004-637X/709/2/572}, \href
  {https://ui.adsabs.harvard.edu/\#abs/2010ApJ...709..572K} {709, 572}

\bibitem[\protect\citeauthoryear{{Kauffmann} et~al.,}{{Kauffmann}
  et~al.}{2003}]{Kauffmann:2003}
{Kauffmann} G.,  et~al., 2003, \mn@doi [\mnras]
  {10.1111/j.1365-2966.2003.07154.x}, \href
  {http://adsabs.harvard.edu/abs/2003MNRAS.346.1055K} {346, 1055}

\bibitem[\protect\citeauthoryear{{Kennicutt}}{{Kennicutt}}{1998}]{Kennicutt:1998}
{Kennicutt} Jr. R.~C.,  1998, \mn@doi [\araa] {10.1146/annurev.astro.36.1.189},
  \href {http://adsabs.harvard.edu/abs/1998ARA%26A..36..189K} {36, 189}

\bibitem[\protect\citeauthoryear{{Kewley}, {Groves}, {Kauffmann}  \&
  {Heckman}}{{Kewley} et~al.}{2006}]{Kewley:2006}
{Kewley} L.~J.,  {Groves} B.,  {Kauffmann} G.,   {Heckman} T.,  2006, \mn@doi
  [\mnras] {10.1111/j.1365-2966.2006.10859.x}, \href
  {http://adsabs.harvard.edu/abs/2006MNRAS.372..961K} {372, 961}

\bibitem[\protect\citeauthoryear{{Kodama} et~al.,}{{Kodama}
  et~al.}{2004}]{Kodama:2004}
{Kodama} T.,  et~al., 2004, \mn@doi [\mnras]
  {10.1111/j.1365-2966.2004.07711.x}, \href
  {https://ui.adsabs.harvard.edu/abs/2004MNRAS.350.1005K} {350, 1005}

\bibitem[\protect\citeauthoryear{{Koopmann} \& {Kenney}}{{Koopmann} \&
  {Kenney}}{2004a}]{Koopmann:2004a}
{Koopmann} R.~A.,  {Kenney} J. D.~P.,  2004a, \mn@doi [\apj] {10.1086/423190},
  \href {https://ui.adsabs.harvard.edu/abs/2004ApJ...613..851K} {613, 851}

\bibitem[\protect\citeauthoryear{{Koopmann} \& {Kenney}}{{Koopmann} \&
  {Kenney}}{2004b}]{Koopmann:2004b}
{Koopmann} R.~A.,  {Kenney} J. D.~P.,  2004b, \mn@doi [\apj] {10.1086/423191},
  \href {https://ui.adsabs.harvard.edu/abs/2004ApJ...613..866K} {613, 866}

\bibitem[\protect\citeauthoryear{{Labatie}, {Starck}  \&
  {Lachi{\`e}ze-Rey}}{{Labatie} et~al.}{2012}]{Labatie:2012}
{Labatie} A.,  {Starck} J.~L.,   {Lachi{\`e}ze-Rey} M.,  2012, \mn@doi [\apj]
  {10.1088/0004-637X/746/2/172}, \href
  {http://adsabs.harvard.edu/abs/2012ApJ...746..172L} {746, 172}

\bibitem[\protect\citeauthoryear{{Lagos} et~al.,}{{Lagos}
  et~al.}{2019}]{Lagos:2019}
{Lagos} C. d.~P.,  et~al., 2019, \mn@doi [\mnras] {10.1093/mnras/stz2427},
  \href {https://ui.adsabs.harvard.edu/abs/2019MNRAS.489.4196L} {489, 4196}

\bibitem[\protect\citeauthoryear{{Lange} et~al.,}{{Lange}
  et~al.}{2015}]{Lange:2015}
{Lange} R.,  et~al., 2015, \mn@doi [\mnras] {10.1093/mnras/stu2467}, \href
  {https://ui.adsabs.harvard.edu/abs/2015MNRAS.447.2603L} {447, 2603}

\bibitem[\protect\citeauthoryear{{Larson}, {Tinsley}  \& {Caldwell}}{{Larson}
  et~al.}{1980}]{Larson:1980}
{Larson} R.~B.,  {Tinsley} B.~M.,   {Caldwell} C.~N.,  1980, \mn@doi [\apj]
  {10.1086/157917}, \href
  {https://ui.adsabs.harvard.edu/abs/1980ApJ...237..692L} {237, 692}

\bibitem[\protect\citeauthoryear{{Lawrence} et~al.,}{{Lawrence}
  et~al.}{2007}]{Lawrence:2007}
{Lawrence} A.,  et~al., 2007, \mn@doi [\mnras]
  {10.1111/j.1365-2966.2007.12040.x}, \href
  {http://adsabs.harvard.edu/abs/2007MNRAS.379.1599L} {379, 1599}

\bibitem[\protect\citeauthoryear{{Le Fevre} et~al.,}{{Le Fevre}
  et~al.}{2000}]{LeFevre:2000}
{Le Fevre} O.,  et~al., 2000, {VIMOS and NIRMOS multi-object spectrographs for
  the ESO VLT}.
pp 546--557, \mn@doi{10.1117/12.395513}

\bibitem[\protect\citeauthoryear{{Le F{\`e}vre} et~al.,}{{Le F{\`e}vre}
  et~al.}{2004}]{LeFevre:2004}
{Le F{\`e}vre} O.,  et~al., 2004, \mn@doi [\aap] {10.1051/0004-6361:20048072},
  \href {https://ui.adsabs.harvard.edu/abs/2004A&A...428.1043L} {428, 1043}

\bibitem[\protect\citeauthoryear{{Madau} \& {Dickinson}}{{Madau} \&
  {Dickinson}}{2014}]{Madau:2014}
{Madau} P.,  {Dickinson} M.,  2014, \mn@doi [\araa]
  {10.1146/annurev-astro-081811-125615}, \href
  {http://adsabs.harvard.edu/abs/2014ARA%26A..52..415M} {52, 415}

\bibitem[\protect\citeauthoryear{{Mainzer} et~al.,}{{Mainzer}
  et~al.}{2011}]{NeoWISE}
{Mainzer} A.,  et~al., 2011, \mn@doi [\apj] {10.1088/0004-637X/731/1/53}, \href
  {http://adsabs.harvard.edu/abs/2011ApJ...731...53M} {731, 53}

\bibitem[\protect\citeauthoryear{{Markwardt}}{{Markwardt}}{2009}]{Markwardt:2009}
{Markwardt} C.~B.,  2009, in {Bohlender} D.~A.,  {Durand} D.,   {Dowler} P.,
  eds,  Astronomical Society of the Pacific Conference Series Vol. 411,
  Astronomical Data Analysis Software and Systems XVIII. p.~251 (\mn@eprint
  {arXiv} {0902.2850})

\bibitem[\protect\citeauthoryear{{Medezinski} et~al.,}{{Medezinski}
  et~al.}{2013}]{Medezinski:2013}
{Medezinski} E.,  et~al., 2013, \mn@doi [\apj] {10.1088/0004-637X/777/1/43},
  \href {http://adsabs.harvard.edu/abs/2013ApJ...777...43M} {777, 43}

\bibitem[\protect\citeauthoryear{{Miyazaki} et~al.,}{{Miyazaki}
  et~al.}{2002}]{Miyazaki:2002}
{Miyazaki} S.,  et~al., 2002, \mn@doi [\pasj] {10.1093/pasj/54.6.833}, \href
  {http://adsabs.harvard.edu/abs/2002PASJ...54..833M} {54, 833}

\bibitem[\protect\citeauthoryear{{Mobasher} et~al.,}{{Mobasher}
  et~al.}{2015}]{Mobasher:2015}
{Mobasher} B.,  et~al., 2015, \mn@doi [\apj] {10.1088/0004-637X/808/1/101},
  \href {http://adsabs.harvard.edu/abs/2015ApJ...808..101M} {808, 101}

\bibitem[\protect\citeauthoryear{{Molino} et~al.,}{{Molino}
  et~al.}{2017}]{Molino:2017}
{Molino} A.,  et~al., 2017, \mn@doi [\mnras] {10.1093/mnras/stx1243}, \href
  {https://ui.adsabs.harvard.edu/\#abs/2017MNRAS.470...95M} {470, 95}

\bibitem[\protect\citeauthoryear{{Monna} et~al.,}{{Monna}
  et~al.}{2017}]{Monna:2017}
{Monna} A.,  et~al., 2017, \mn@doi [\mnras] {10.1093/mnras/stx015}, \href
  {http://adsabs.harvard.edu/abs/2017MNRAS.466.4094M} {466, 4094}

\bibitem[\protect\citeauthoryear{{Moore}, {Katz}, {Lake}, {Dressler}  \&
  {Oemler}}{{Moore} et~al.}{1996}]{Moore:1996}
{Moore} B.,  {Katz} N.,  {Lake} G.,  {Dressler} A.,   {Oemler} A.,  1996,
  \mn@doi [\nat] {10.1038/379613a0}, \href
  {https://ui.adsabs.harvard.edu/abs/1996Natur.379..613M} {379, 613}

\bibitem[\protect\citeauthoryear{{Navarro}, {Frenk}  \& {White}}{{Navarro}
  et~al.}{1996}]{Navarro:1996}
{Navarro} J.~F.,  {Frenk} C.~S.,   {White} S. D.~M.,  1996, \mn@doi [\apj]
  {10.1086/177173}, \href
  {https://ui.adsabs.harvard.edu/abs/1996ApJ...462..563N} {462, 563}

\bibitem[\protect\citeauthoryear{{Navarro}, {Frenk}  \& {White}}{{Navarro}
  et~al.}{1997}]{Navarro:1997}
{Navarro} J.~F.,  {Frenk} C.~S.,   {White} S.~D.~M.,  1997, \apj, \href
  {http://adsabs.harvard.edu/abs/1997ApJ...490..493N} {490, 493}

\bibitem[\protect\citeauthoryear{{Nayyeri} et~al.,}{{Nayyeri}
  et~al.}{2017}]{Nayyeri:2017}
{Nayyeri} H.,  et~al., 2017, \mn@doi [\apjs] {10.3847/1538-4365/228/1/7}, \href
  {https://ui.adsabs.harvard.edu/abs/2017ApJS..228....7N} {228, 7}

\bibitem[\protect\citeauthoryear{{Neyman} \& {Pearson}}{{Neyman} \&
  {Pearson}}{1933}]{Neyman:1933}
{Neyman} J.,  {Pearson} E.~S.,  1933, \mn@doi [Philosophical Transactions of
  the Royal Society of London Series A] {10.1098/rsta.1933.0009}, \href
  {http://adsabs.harvard.edu/abs/1933RSPTA.231..289N} {231, 289}

\bibitem[\protect\citeauthoryear{{Noeske} et~al.,}{{Noeske}
  et~al.}{2007}]{Noeske:2007}
{Noeske} K.~G.,  et~al., 2007, \mn@doi [\apjl] {10.1086/517927}, \href
  {https://ui.adsabs.harvard.edu/abs/2007ApJ...660L..47N} {660, L47}

\bibitem[\protect\citeauthoryear{{Nonino} et~al.,}{{Nonino}
  et~al.}{2009}]{Nonino:2009}
{Nonino} M.,  et~al., 2009, \mn@doi [\apjs] {10.1088/0067-0049/183/2/244},
  \href {http://adsabs.harvard.edu/abs/2009ApJS..183..244N} {183, 244}

\bibitem[\protect\citeauthoryear{{Oke} \& {Gunn}}{{Oke} \&
  {Gunn}}{1983}]{Oke:1983}
{Oke} J.~B.,  {Gunn} J.~E.,  1983, \mn@doi [\apj] {10.1086/160817}, \href
  {https://ui.adsabs.harvard.edu/abs/1983ApJ...266..713O} {266, 713}

\bibitem[\protect\citeauthoryear{{Owers} et~al.,}{{Owers}
  et~al.}{2017}]{Owers:2017}
{Owers} M.~S.,  et~al., 2017, \mn@doi [\mnras] {10.1093/mnras/stx562}, \href
  {http://adsabs.harvard.edu/abs/2017MNRAS.468.1824O} {468, 1824}

\bibitem[\protect\citeauthoryear{{P{\'e}rez-Gonz{\'a}lez}
  et~al.,}{{P{\'e}rez-Gonz{\'a}lez} et~al.}{2008}]{PerezGonzalez:2008}
{P{\'e}rez-Gonz{\'a}lez} P.~G.,  et~al., 2008, \mn@doi [\apj] {10.1086/523690},
  \href {http://adsabs.harvard.edu/abs/2008ApJ...675..234P} {675, 234}

\bibitem[\protect\citeauthoryear{{Pintos-Castro} et~al.,}{{Pintos-Castro}
  et~al.}{2013}]{PintosCastro:2013}
{Pintos-Castro} I.,  et~al., 2013, \mn@doi [\aap]
  {10.1051/0004-6361/201321474}, \href
  {http://adsabs.harvard.edu/abs/2013A%26A...558A.100P} {558, A100}

\bibitem[\protect\citeauthoryear{{Postman} et~al.,}{{Postman}
  et~al.}{2012}]{Postman:2012}
{Postman} M.,  et~al., 2012, \mn@doi [\apjs] {10.1088/0067-0049/199/2/25},
  \href {http://adsabs.harvard.edu/abs/2012ApJS..199...25P} {199, 25}

\bibitem[\protect\citeauthoryear{{Pratt}, {Croston}, {Arnaud}  \&
  {B{\"o}hringer}}{{Pratt} et~al.}{2009}]{Pratt:2009}
{Pratt} G.~W.,  {Croston} J.~H.,  {Arnaud} M.,   {B{\"o}hringer} H.,  2009,
  \mn@doi [\aap] {10.1051/0004-6361/200810994}, \href
  {https://ui.adsabs.harvard.edu/abs/2009A&A...498..361P} {498, 361}

\bibitem[\protect\citeauthoryear{{Rosati} et~al.,}{{Rosati}
  et~al.}{2014}]{Rosati:2014}
{Rosati} P.,  et~al., 2014, The Messenger, \href
  {https://ui.adsabs.harvard.edu/abs/2014Msngr.158...48R} {158, 48}

\bibitem[\protect\citeauthoryear{{Salpeter}}{{Salpeter}}{1955}]{Salpeter:1955}
{Salpeter} E.~E.,  1955, \mn@doi [\apj] {10.1086/145971}, \href
  {http://adsabs.harvard.edu/abs/1955ApJ...121..161S} {121, 161}

\bibitem[\protect\citeauthoryear{{S{\'a}nchez} et~al.,}{{S{\'a}nchez}
  et~al.}{2012}]{Sanchez:2012}
{S{\'a}nchez} S.~F.,  et~al., 2012, \mn@doi [\aap]
  {10.1051/0004-6361/201117353}, \href
  {http://adsabs.harvard.edu/abs/2012A%26A...538A...8S} {538, A8}

\bibitem[\protect\citeauthoryear{{Santini} et~al.,}{{Santini}
  et~al.}{2015}]{Santini:2015}
{Santini} P.,  et~al., 2015, \mn@doi [\apj] {10.1088/0004-637X/801/2/97}, \href
  {https://ui.adsabs.harvard.edu/abs/2015ApJ...801...97S} {801, 97}

\bibitem[\protect\citeauthoryear{{Schreiber} et~al.,}{{Schreiber}
  et~al.}{2015}]{Schreiber:2015}
{Schreiber} C.,  et~al., 2015, \mn@doi [\aap] {10.1051/0004-6361/201425017},
  \href {http://adsabs.harvard.edu/abs/2015A%26A...575A..74S} {575, A74}

\bibitem[\protect\citeauthoryear{{Scoville}}{{Scoville}}{2007}]{Scoville:2007}
{Scoville} N.,  2007, in {Baker} A.~J.,  {Glenn} J.,  {Harris} A.~I.,  {Mangum}
  J.~G.,   {Yun} M.~S.,  eds,  Astronomical Society of the Pacific Conference
  Series Vol. 375, From Z-Machines to ALMA: (Sub)Millimeter Spectroscopy of
  Galaxies. p.~166

\bibitem[\protect\citeauthoryear{{Sharples} et~al.,}{{Sharples}
  et~al.}{2013}]{Sharples:2013}
{Sharples} R.,  et~al., 2013, The Messenger, \href
  {http://adsabs.harvard.edu/abs/2013Msngr.151...21S} {151, 21}

\bibitem[\protect\citeauthoryear{{Shen}, {Mo}, {White}, {Blanton}, {Kauffmann},
  {Voges}, {Brinkmann}  \& {Csabai}}{{Shen} et~al.}{2003}]{Shen:2003}
{Shen} S.,  {Mo} H.~J.,  {White} S.~D.~M.,  {Blanton} M.~R.,  {Kauffmann} G.,
  {Voges} W.,  {Brinkmann} J.,   {Csabai} I.,  2003, \mn@doi [\mnras]
  {10.1046/j.1365-8711.2003.06740.x}, \href
  {http://adsabs.harvard.edu/abs/2003MNRAS.343..978S} {343, 978}

\bibitem[\protect\citeauthoryear{{Stern}, {Jimenez}, {Verde}, {Stanford}  \&
  {Kamionkowski}}{{Stern} et~al.}{2010}]{Stern:2010}
{Stern} D.,  {Jimenez} R.,  {Verde} L.,  {Stanford} S.~A.,   {Kamionkowski} M.,
   2010, \mn@doi [\apjs] {10.1088/0067-0049/188/1/280}, \href
  {https://ui.adsabs.harvard.edu/abs/2010ApJS..188..280S} {188, 280}

\bibitem[\protect\citeauthoryear{{Stern} et~al.,}{{Stern}
  et~al.}{2012}]{Stern:2012aa}
{Stern} D.,  et~al., 2012, \mn@doi [\apj] {10.1088/0004-637X/753/1/30}, \href
  {http://adsabs.harvard.edu/abs/2012ApJ...753...30S} {753, 30}

\bibitem[\protect\citeauthoryear{{Stocke}, {Morris}, {Gioia}, {Maccacaro},
  {Schild}, {Wolter}, {Fleming}  \& {Henry}}{{Stocke}
  et~al.}{1991}]{Stocke:1991}
{Stocke} J.~T.,  {Morris} S.~L.,  {Gioia} I.~M.,  {Maccacaro} T.,  {Schild} R.,
   {Wolter} A.,  {Fleming} T.~A.,   {Henry} J.~P.,  1991, \mn@doi [\apjs]
  {10.1086/191582}, \href
  {https://ui.adsabs.harvard.edu/abs/1991ApJS...76..813S} {76, 813}

\bibitem[\protect\citeauthoryear{{Stott}, {Smail}, {Edge}, {Ebeling}, {Smith},
  {Kneib}  \& {Pimbblet}}{{Stott} et~al.}{2007}]{Stott:2007}
{Stott} J.~P.,  {Smail} I.,  {Edge} A.~C.,  {Ebeling} H.,  {Smith} G.~P.,
  {Kneib} J.-P.,   {Pimbblet} K.~A.,  2007, \mn@doi [\apj] {10.1086/514329},
  \href {http://adsabs.harvard.edu/abs/2007ApJ...661...95S} {661, 95}

\bibitem[\protect\citeauthoryear{{Stott} et~al.,}{{Stott}
  et~al.}{2016}]{Stott:2016}
{Stott} J.~P.,  et~al., 2016, \mn@doi [\mnras] {10.1093/mnras/stw129}, \href
  {http://adsabs.harvard.edu/abs/2016MNRAS.457.1888S} {457, 1888}

\bibitem[\protect\citeauthoryear{{Swinbank} et~al.,}{{Swinbank}
  et~al.}{2017}]{Swinbank:2017}
{Swinbank} A.~M.,  et~al., 2017, \mn@doi [\mnras] {10.1093/mnras/stx201}, \href
  {http://adsabs.harvard.edu/abs/2017MNRAS.467.3140S} {467, 3140}

\bibitem[\protect\citeauthoryear{{Taranu}, {Hudson}, {Balogh}, {Smith},
  {Power}, {Oman}  \& {Krane}}{{Taranu} et~al.}{2014}]{Taranu:2014}
{Taranu} D.~S.,  {Hudson} M.~J.,  {Balogh} M.~L.,  {Smith} R.~J.,  {Power} C.,
  {Oman} K.~A.,   {Krane} B.,  2014, \mn@doi [\mnras] {10.1093/mnras/stu389},
  \href {https://ui.adsabs.harvard.edu/abs/2014MNRAS.440.1934T} {440, 1934}

\bibitem[\protect\citeauthoryear{{Taylor} et~al.,}{{Taylor}
  et~al.}{2011}]{Taylor:2011}
{Taylor} E.~N.,  et~al., 2011, \mn@doi [\mnras]
  {10.1111/j.1365-2966.2011.19536.x}, \href
  {https://ui.adsabs.harvard.edu/abs/2011MNRAS.418.1587T} {418, 1587}

\bibitem[\protect\citeauthoryear{{Tiley} et~al.,}{{Tiley}
  et~al.}{2016}]{Tiley:2016}
{Tiley} A.~L.,  et~al., 2016, \mn@doi [\mnras] {10.1093/mnras/stw936}, \href
  {http://adsabs.harvard.edu/abs/2016MNRAS.460..103T} {460, 103}

\bibitem[\protect\citeauthoryear{{Tiley} et~al.,}{{Tiley}
  et~al.}{2019}]{Tiley:2019a}
{Tiley} A.~L.,  et~al., 2019, \mn@doi [\mnras] {10.1093/mnras/sty2794}, \href
  {https://ui.adsabs.harvard.edu/abs/2019MNRAS.482.2166T} {482, 2166}

\bibitem[\protect\citeauthoryear{{Tully} \& {Fisher}}{{Tully} \&
  {Fisher}}{1977}]{Tully:1977aa}
{Tully} R.~B.,  {Fisher} J.~R.,  1977, \aap, \href
  {http://adsabs.harvard.edu/abs/1977A%26A....54..661T} {54, 661}

\bibitem[\protect\citeauthoryear{{Turner} et~al.,}{{Turner}
  et~al.}{2017}]{Turner:2017}
{Turner} O.~J.,  et~al., 2017, \mn@doi [\mnras] {10.1093/mnras/stx1366}, \href
  {http://adsabs.harvard.edu/abs/2017MNRAS.471.1280T} {471, 1280}

\bibitem[\protect\citeauthoryear{{{\"U}bler} et~al.,}{{{\"U}bler}
  et~al.}{2017}]{Ubler:2017}
{{\"U}bler} H.,  et~al., 2017, \mn@doi [\apj] {10.3847/1538-4357/aa7558}, \href
  {http://adsabs.harvard.edu/abs/2017ApJ...842..121U} {842, 121}

\bibitem[\protect\citeauthoryear{{Umetsu} et~al.,}{{Umetsu}
  et~al.}{2014}]{Umetsu:2014}
{Umetsu} K.,  et~al., 2014, \mn@doi [\apj] {10.1088/0004-637X/795/2/163}, \href
  {http://adsabs.harvard.edu/abs/2014ApJ...795..163U} {795, 163}

\bibitem[\protect\citeauthoryear{{Vaughan} et~al.,}{{Vaughan}
  et~al.}{2020}]{vaughan:2020}
{Vaughan} S.~P.,  et~al., 2020, MNRAS, submitted

\bibitem[\protect\citeauthoryear{{Wang}, {Liu}, {Qiu}, {Bai}, {Yang}, {Guo}  \&
  {Zhang}}{{Wang} et~al.}{2016}]{Wang:2016}
{Wang} S.,  {Liu} J.,  {Qiu} Y.,  {Bai} Y.,  {Yang} H.,  {Guo} J.,   {Zhang}
  P.,  2016, \mn@doi [ApJS] {10.3847/0067-0049/224/2/40}, \href
  {https://ui.adsabs.harvard.edu/\#abs/2016ApJS..224...40W} {224, 40}

\bibitem[\protect\citeauthoryear{{Wetzel}, {Tinker}, {Conroy}  \& {van den
  Bosch}}{{Wetzel} et~al.}{2013}]{Wetzel:2013}
{Wetzel} A.~R.,  {Tinker} J.~L.,  {Conroy} C.,   {van den Bosch} F.~C.,  2013,
  \mn@doi [\mnras] {10.1093/mnras/stt469}, \href
  {https://ui.adsabs.harvard.edu/abs/2013MNRAS.432..336W} {432, 336}

\bibitem[\protect\citeauthoryear{{White}}{{White}}{2001}]{White:2001}
{White} M.,  2001, \mn@doi [\aap] {10.1051/0004-6361:20000357}, \href
  {https://ui.adsabs.harvard.edu/abs/2001A&A...367...27W} {367, 27}

\bibitem[\protect\citeauthoryear{{Wisnioski} et~al.,}{{Wisnioski}
  et~al.}{2015}]{Wisnioski:2015}
{Wisnioski} E.,  et~al., 2015, \mn@doi [\apj] {10.1088/0004-637X/799/2/209},
  \href {http://adsabs.harvard.edu/abs/2015ApJ...799..209W} {799, 209}

\bibitem[\protect\citeauthoryear{{Wisnioski} et~al.,}{{Wisnioski}
  et~al.}{2018}]{Wisnioski:2018}
{Wisnioski} E.,  et~al., 2018, \mn@doi [\apj] {10.3847/1538-4357/aab097}, \href
  {https://ui.adsabs.harvard.edu/\#abs/2018ApJ...855...97W} {855, 97}

\bibitem[\protect\citeauthoryear{{Wright} et~al.,}{{Wright}
  et~al.}{2010}]{Wright:2010}
{Wright} E.~L.,  et~al., 2010, \mn@doi [\aj] {10.1088/0004-6256/140/6/1868},
  \href {http://adsabs.harvard.edu/abs/2010AJ....140.1868W} {140, 1868}

\bibitem[\protect\citeauthoryear{{Wu}, {Xue}  \& {Fang}}{{Wu}
  et~al.}{1999}]{Wu:1999}
{Wu} X.-P.,  {Xue} Y.-J.,   {Fang} L.-Z.,  1999, \mn@doi [\apj]
  {10.1086/307791}, \href
  {https://ui.adsabs.harvard.edu/abs/1999ApJ...524...22W} {524, 22}

\bibitem[\protect\citeauthoryear{{Wuyts} et~al.,}{{Wuyts}
  et~al.}{2013}]{Wuyts:2013}
{Wuyts} S.,  et~al., 2013, \mn@doi [\apj] {10.1088/0004-637X/779/2/135}, \href
  {https://ui.adsabs.harvard.edu/abs/2013ApJ...779..135W} {779, 135}

\bibitem[\protect\citeauthoryear{{Yee}, {Ellingson}  \& {Carlberg}}{{Yee}
  et~al.}{1996}]{Yee:1996}
{Yee} H.~K.~C.,  {Ellingson} E.,   {Carlberg} R.~G.,  1996, \mn@doi [\apjs]
  {10.1086/192259}, \href
  {https://ui.adsabs.harvard.edu/abs/1996ApJS..102..269Y} {102, 269}

\bibitem[\protect\citeauthoryear{{Yee} et~al.,}{{Yee} et~al.}{2000}]{Yee:2000}
{Yee} H.~K.~C.,  et~al., 2000, \mn@doi [\apjs] {10.1086/313426}, \href
  {https://ui.adsabs.harvard.edu/abs/2000ApJS..129..475Y} {129, 475}

\bibitem[\protect\citeauthoryear{{York} et~al.,}{{York}
  et~al.}{2000}]{York:2000}
{York} D.~G.,  et~al., 2000, \mn@doi [\aj] {10.1086/301513}, \href
  {http://adsabs.harvard.edu/abs/2000AJ....120.1579Y} {120, 1579}

\bibitem[\protect\citeauthoryear{{Zitrin} et~al.,}{{Zitrin}
  et~al.}{2009}]{Zitrin:2009}
{Zitrin} A.,  et~al., 2009, \mn@doi [\mnras]
  {10.1111/j.1365-2966.2009.14899.x}, \href
  {https://ui.adsabs.harvard.edu/abs/2009MNRAS.396.1985Z} {396, 1985}

\bibitem[\protect\citeauthoryear{{Zitrin} et~al.,}{{Zitrin}
  et~al.}{2013}]{Zitrin:2013}
{Zitrin} A.,  et~al., 2013, \mn@doi [\apjl] {10.1088/2041-8205/762/2/L30},
  \href {https://ui.adsabs.harvard.edu/abs/2013ApJ...762L..30Z} {762, L30}

\bibitem[\protect\citeauthoryear{{Zitrin} et~al.,}{{Zitrin}
  et~al.}{2015}]{Zitrin:2015}
{Zitrin} A.,  et~al., 2015, \mn@doi [\apj] {10.1088/0004-637X/801/1/44}, \href
  {http://adsabs.harvard.edu/abs/2015ApJ...801...44Z} {801, 44}

\bibitem[\protect\citeauthoryear{{da Cunha}, {Charlot}  \& {Elbaz}}{{da Cunha}
  et~al.}{2008}]{Cunha:2008}
{da Cunha} E.,  {Charlot} S.,   {Elbaz} D.,  2008, \mn@doi [\mnras]
  {10.1111/j.1365-2966.2008.13535.x}, \href
  {http://adsabs.harvard.edu/abs/2008MNRAS.388.1595D} {388, 1595}

\bibitem[\protect\citeauthoryear{{van Gorkom}}{{van
  Gorkom}}{2004}]{vanGorkom:2004}
{van Gorkom} J.~H.,  2004, in {Mulchaey} J.~S.,  {Dressler} A.,   {Oemler} A.,
  eds, Clusters of Galaxies: Probes of Cosmological Structure and Galaxy
  Evolution. p.~305 (\mn@eprint {arXiv} {astro-ph/0308209})

\bibitem[\protect\citeauthoryear{{van der Wel} et~al.,}{{van der Wel}
  et~al.}{2012}]{vanderWel:2012}
{van der Wel} A.,  et~al., 2012, \mn@doi [\apjs] {10.1088/0067-0049/203/2/24},
  \href {http://adsabs.harvard.edu/abs/2012ApJS..203...24V} {203, 24}

\makeatother
\end{thebibliography}




\appendix

\section{Map Masking}
\label{sec:mapmasking}

After we construct KMOS maps for the K-CLASH galaxies, we identify and iteratively mask any ``bad'' pixels that are the result of, for example, erroneous fits to residual sky line emission, edge effects in the outer spaxels of the cube, and/or ``hot" spectral pixels. Indeed, despite our efforts to mitigate some of these effects, for example by down-weighting spectral ranges corresponding to the positions of sky lines in the fitting process, all of them can act to skew the best-fitting triplet model (\S~\ref{subsec:measuringlinefluxes}), resulting in erroneous best-fitting parameters for some pixels in the KMOS maps (despite all fits having $\rm{S/N}_{\rm{H}\alpha} \geq 5$ by construction). We thus remove such pixels or groups of pixels from the KMOS maps in a series of steps. 

First, for each galaxy, we remove unresolved features from the maps, defined as contiguous regions with areas less than that of a resolution element (defined as 1.1 times the area enclosed within the FWHM of the PSF). Here we include a 10 per cent buffer to ensure we only consider features that are robustly spatially-resolved in the maps and disregard marginal cases that are more likely to be unresolved. 

Second, we apply a ``cleaning'' algorithm, whereby for each pixel in the $v_{\rm{obs}}$ map we consider the neighbouring pixels in a $9 \times 9$  $0\farcs1$ pixel box centred on the pixel in question. We calculate the median of the pixels in this box ($v_{\rm{box,med}}$) and then mask the central pixel if the difference between its velocity ($v_{\rm{pix}}$) and the median of the box is larger (in absolute terms) than a chosen limit, i.e. we mask the central pixel if $|v_{\rm{pix}} - v_{\rm{box,med}}| > \rm{limit}$. The limit here is taken as the 68$^{\rm{th}}$ percentile of the absolute value of the velocities in the $v_{\rm{obs}}$ map. In this way, we only mask pixels that significantly differ from their neighbouring pixels, whilst taking into account the total velocity gradient across the map (which can differ significantly between galaxies). The box size is chosen to be large enough to calculate a robust median velocity, but small enough to avoid a large spread of velocities within the box due to the intrinsic line-of-sight velocity gradient across the galaxy. This process is repeated iteratively three times, applying the new masking after each iteration. We then apply the same cleaning algorithm to the $\sigma_{\rm{obs}}$ map. Finally, for each pixel masked in a galaxy's $v_{\rm{obs}}$ or $\sigma_{\rm{obs}}$ map, we also mask the corresponding pixel in {\it all} of its KMOS maps. 

Following this we also remove very noisy pixels towards the outer-edges of each galaxy's KMOS maps, that likely correspond to spaxels at the edges of the KMOS image slicing mirrors, by inspecting the map of the root-mean-square (rms) noise for each cube and noting the spatial positions of large discontinuities in the rms values of spaxels. The rms of spectra corresponding to more central spaxels are typically lower than those of spaxels around the edges of the data cube, with a sharp difference between the two clearly visible even by eye. For any given cube, we thus define a bounding square containing the more central, lower rms, spaxels (making up the majority of the spaxels of the cube) and excluding spaxels at the very edges of the cube with much higher rms. We correspondingly mask the pixels in the maps that are outside of this box.  

As a final step, we once again check for spatially-unresolved regions and mask them.

\section{Table of Values}
\label{sec:appendixtable}

In Table~\ref{tab:kclashvals} we tabulate the key properties of the K-CLASH sample galaxies discussed in this work (see \S~\ref{sec:sampleoverview} and \S~\ref{sec:galaxyproperties}). Upon publication, this table will be made publicly available online in full, in machine-readable format.  

\begin{sidewaystable}
\vspace{8cm}
\centering
\caption{\normalsize Key properties of the K-CLASH sample galaxies discussed in this work.}\label{tab:kclashvals}
\small
\begin{tabular}{ lllllllllcccl}
\hline
ID  & R.A. & Dec & Redshift & H$\alpha$- & {\it cluster} & {\it field} & AGN & $\rm{R}_{50}$ & $\log_{10}\frac{\rm{M}_{*}}{\rm{M}_{\odot}}$ & $\log_{10}\frac{\rm{SFR}_{\sc ProSpect}}{\rm{M}_{\odot} \rm{yr}^{-1}}$ & $\log_{10}\frac{\rm{SFR}_{\rm{H}\alpha}}{\rm{M}_{\odot} \rm{yr}^{-1}}$ & NED/SIMBAD \\
  & (deg) & (deg) &  & detected  &  &  &  & (kpc) & (dex) & (dex) & (dex) & Name \\
(1) & (2)  & (3) & (4) & (5) & (6) & (7) & (8) & (9) & (10) & (11) & (12) & (13) \\
\hline
1311\_47439 & 197.710745 & -3.233176 & 0.493 & True & True & False & False & 3.4 & 10.4 & 0.7 & 0.6 & SDSS\phantom{W} J131050...  \\
1311\_40938 & 197.814597 & -3.261198 & 0.584 & True & False & True & False & 4.3 & 10.6 & 0.8 & 0.5  & WISEA J131115... \\
1931\_66185 & 292.996007 & -26.524327 & 0.416 & True & False & False & True & 3.4 & 10.6 & 1.1 & 1.3  & WISEA J193159... 
\\
\hline
\multicolumn{13}{l}{%
\begin{minipage}{22cm}%
\medskip
\normalsize Notes. {\bf (1)} Object ID.  {\bf (2)} Right ascension. {\bf (3)} Declination. {\bf (4)} Redshift (spectroscopic if detected in H$\alpha$, {\sc bpz} if not). {\bf (5)} H$\alpha$ detection flag. {\bf (6)} {\it cluster} sub-sample membership flag. {\bf (7)} {\it field} sub-sample membership flag. {\bf (8)} Candidate AGN host flag. {\bf (9)} Stellar light effective radius. {\bf (10)} {\sc ProSpect}-derived stellar mass. {\bf (11)} {\sc ProSpect}-derived total star-formation rate. {\bf (12)} H$\alpha$-derived total (i.e.\ aperture- and extinction-corrected) star-formation rate.  {\bf (13)} Object name(s) of any corresponding entry (within a $1''$ radius of the K-CLASH galaxy position) in the NASA/IPAC Extragalactic Database (NED) and/or the Set of Identifications, Measurements and Bibliography for Astronomical Data (SIMBAD) database. We note that candidate AGN hosts and H$\alpha$ non-detections are excluded from the {\it cluster} and {\it field} sub-samples, as discussed in \S~\ref{subsec:clustermembership}.  %
\end{minipage}%
}\\
\end{tabular}
\end{sidewaystable}

\vspace{-0.5cm}

\section{SED-Fitting Comparisons}
\label{sec:SEDcodecheck}

\begin{figure*}
\begin{minipage}[]{.95\textwidth}
\includegraphics[width=.33\textwidth,trim= 0 0 0 0,clip=True]{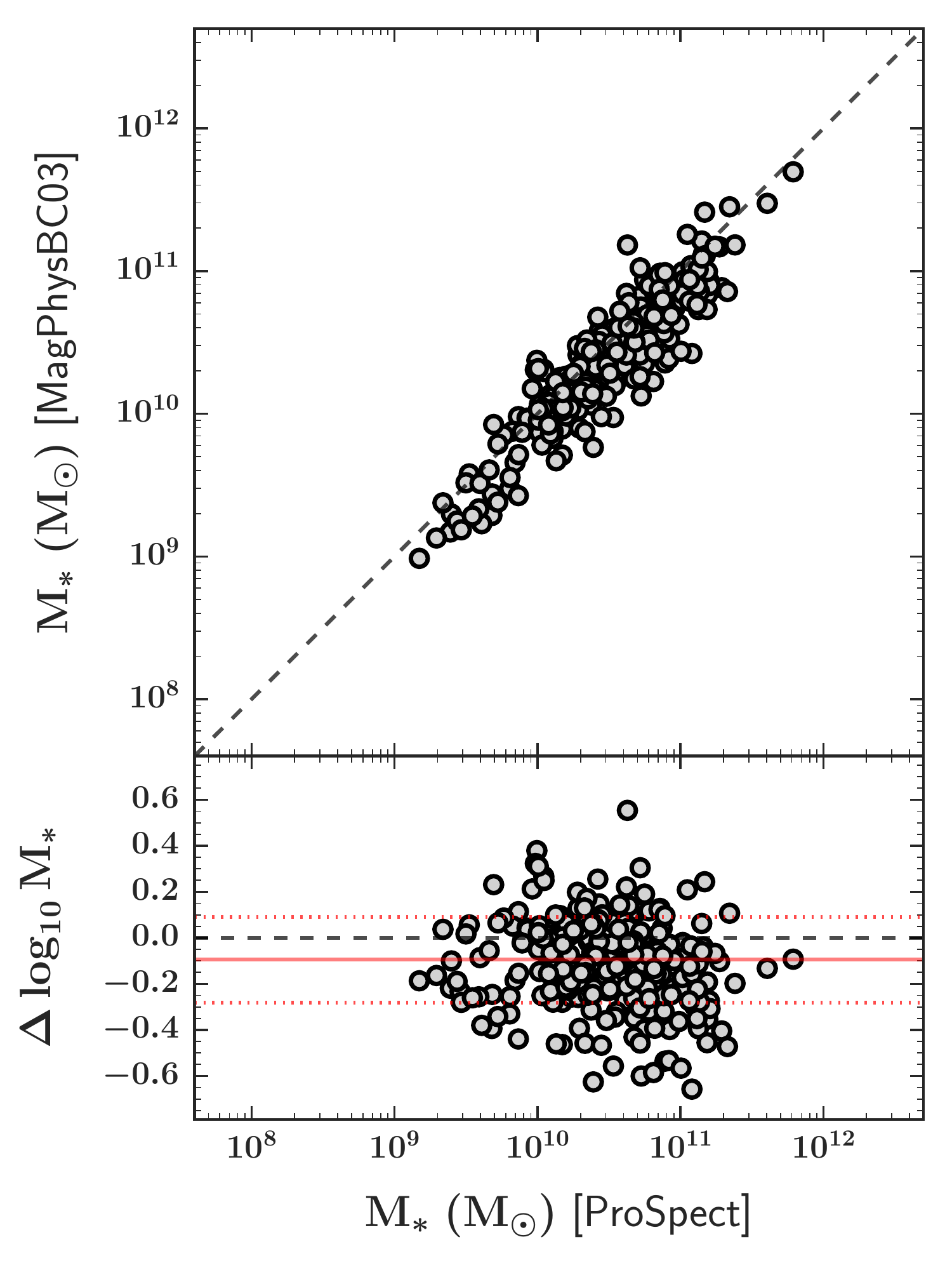}\includegraphics[width=.33\textwidth,trim= 0 0 0 0,clip=True]{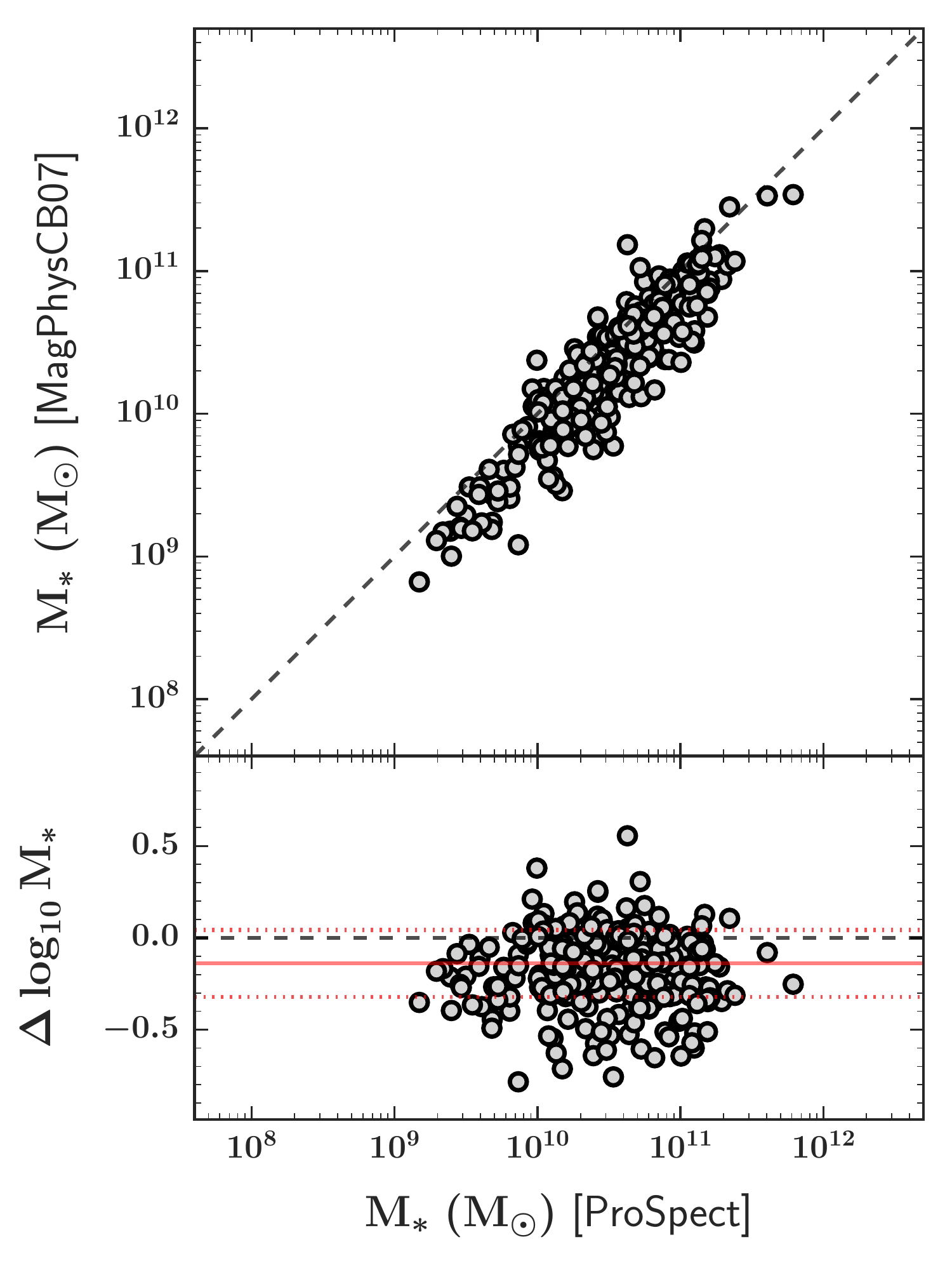}\includegraphics[width=.33\textwidth,trim= 0 0 0 0,clip=True]{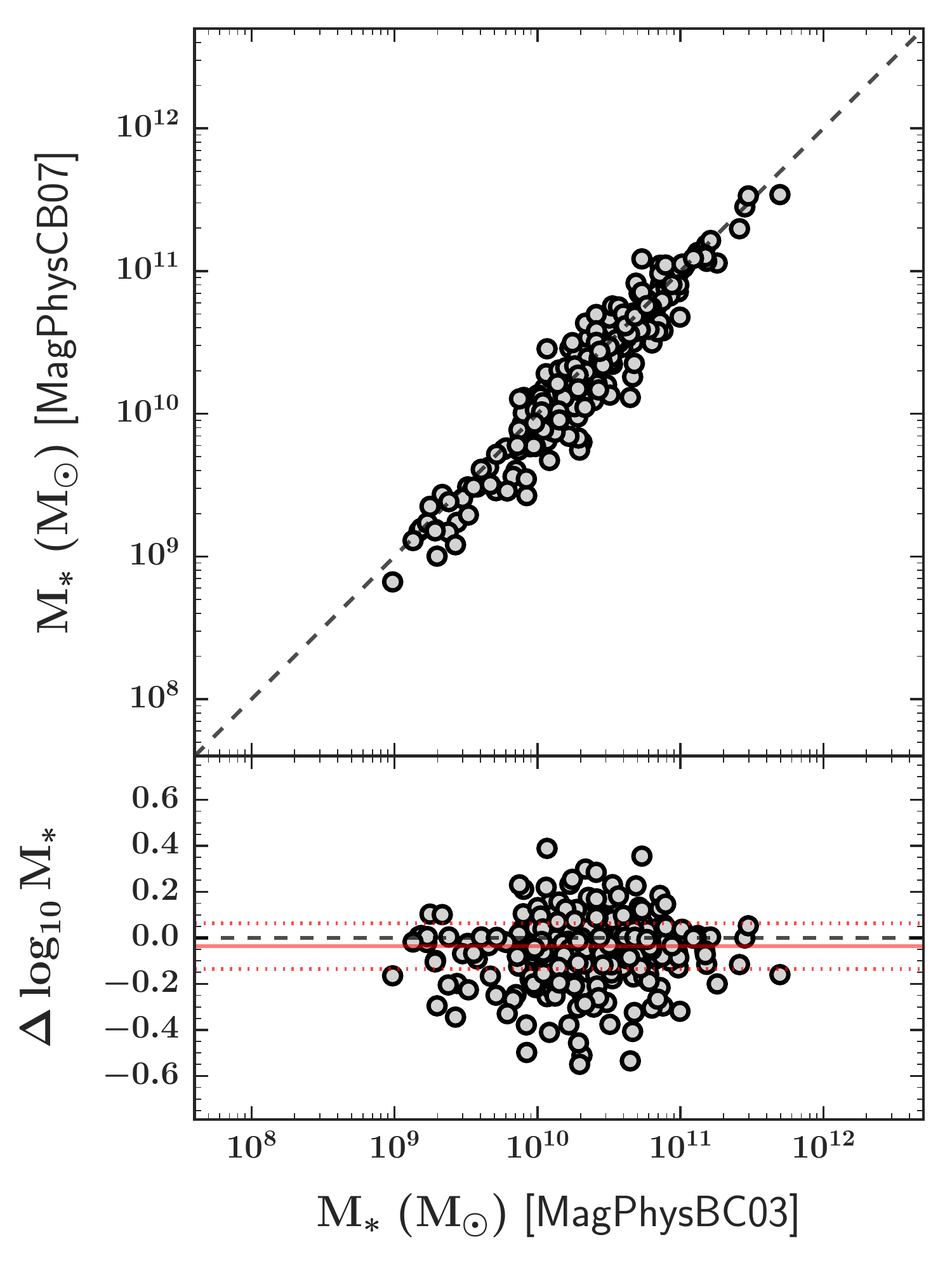}
\includegraphics[width=.33\textwidth,trim= 0 0 0 0,clip=True]{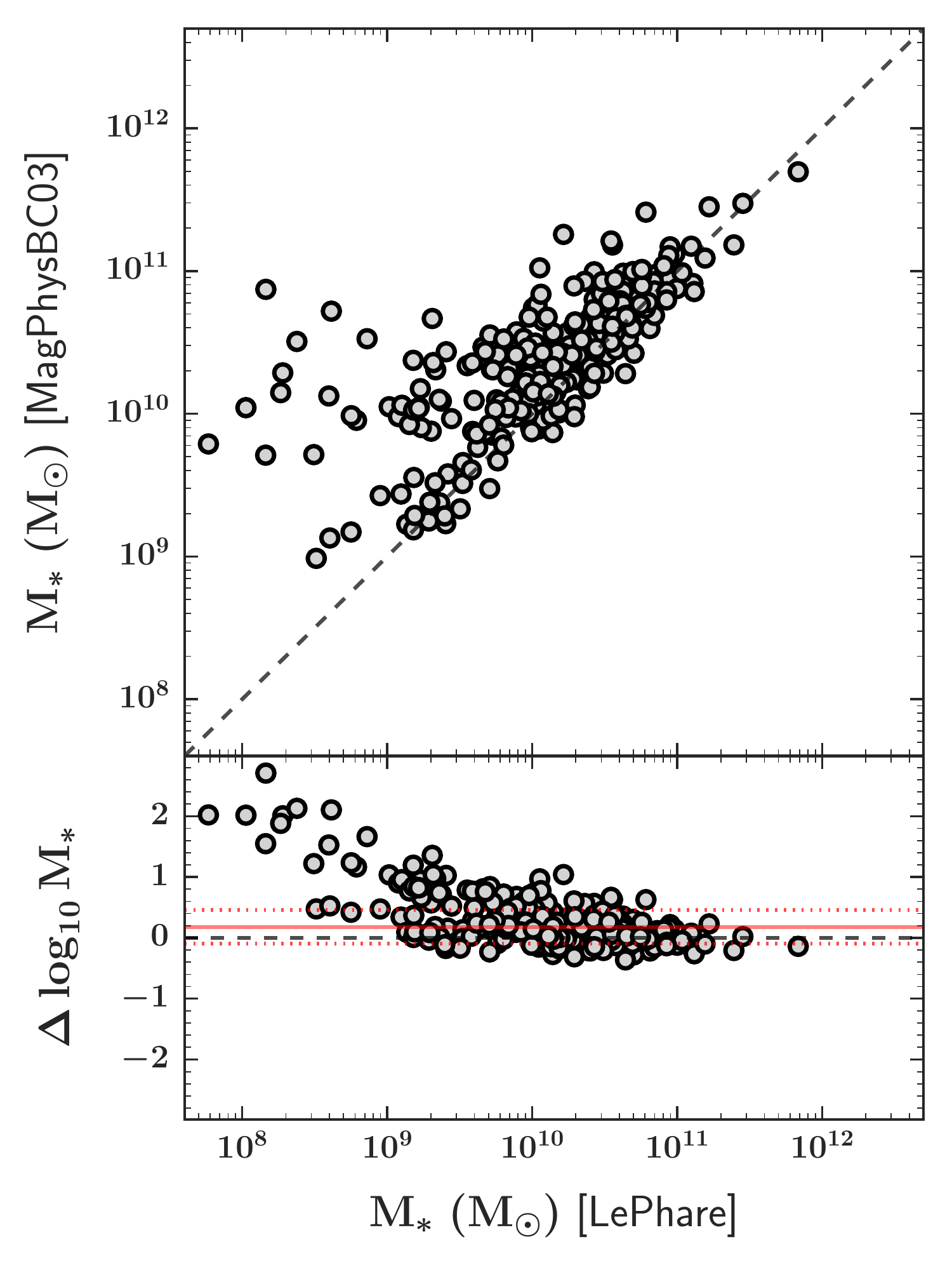}\includegraphics[width=.33\textwidth,trim= 0 0 0 0,clip=True]{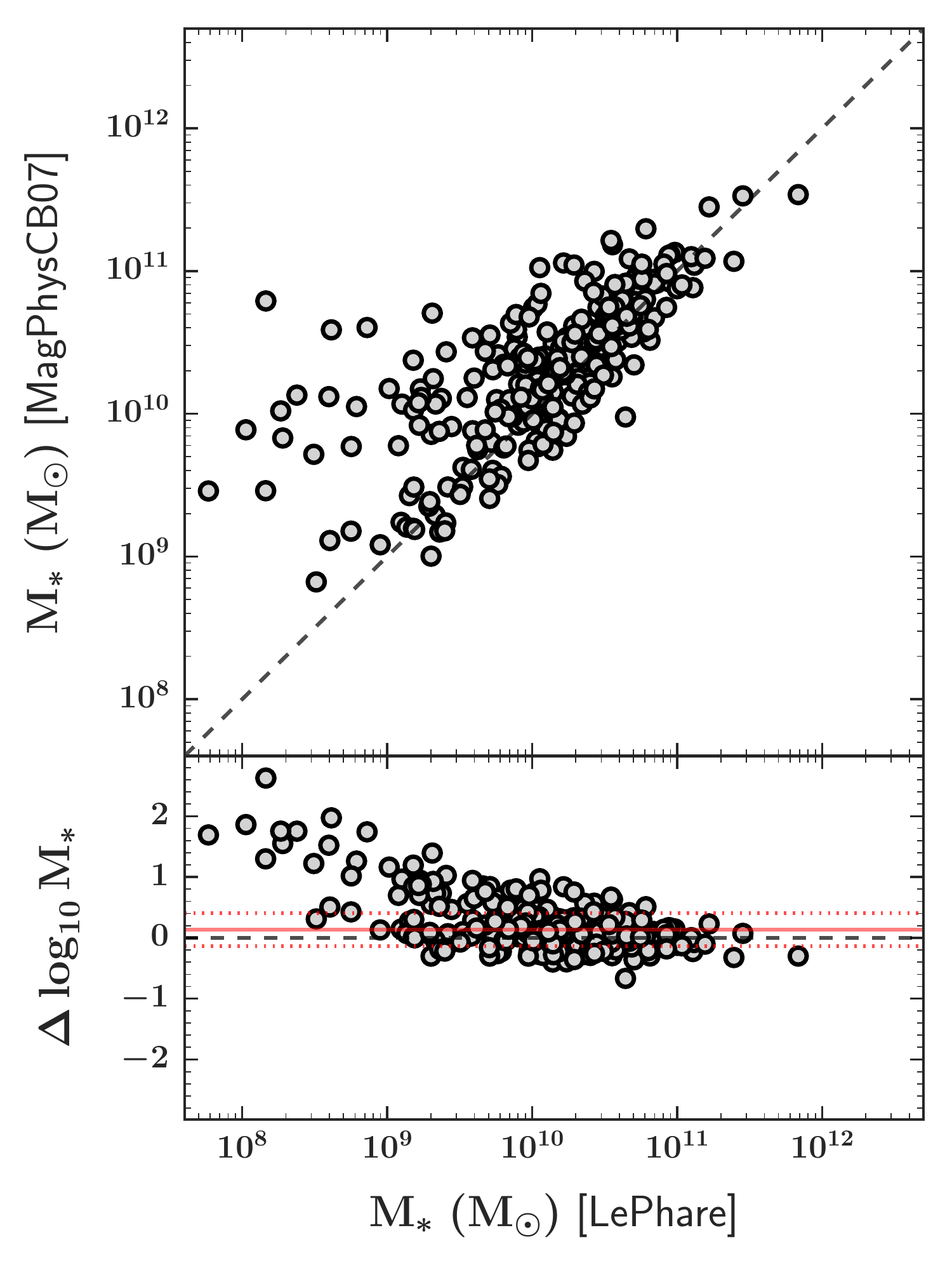}\includegraphics[width=.33\textwidth,trim= 0 0 0 0,clip=True]{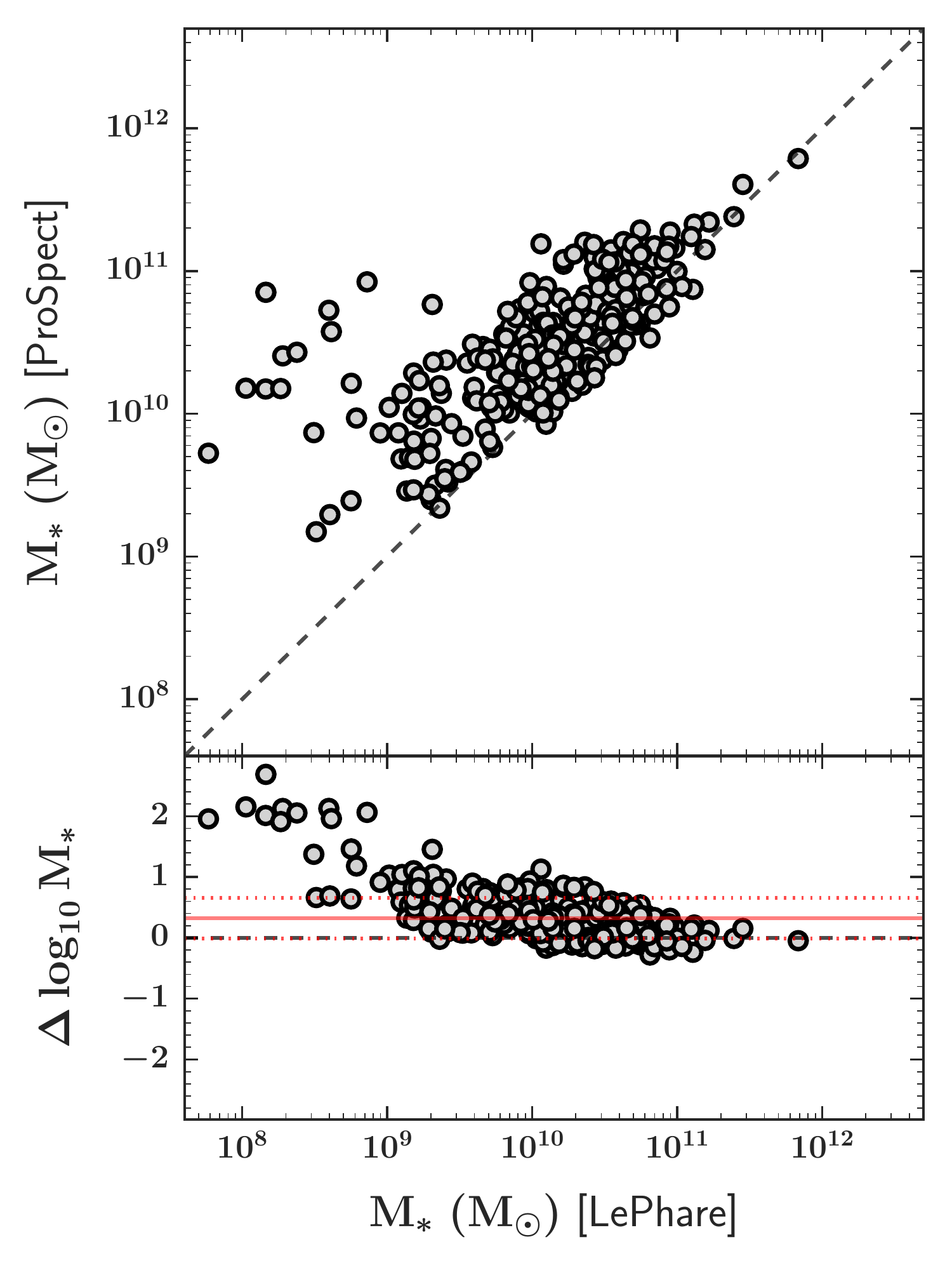}
\end{minipage}
\caption{%
Comparisons of the stellar mass estimates for K-CLASH galaxies derived using the {\sc ProSpect}, {\sc magphys} (BC03 and CB07) and {\sc LePhare} SED-fitting routines on photometry spanning the optical to the near-infrared. For each comparison between the routines, we also show the residuals between the two sets of measurements (ordinate $-$ abscissa). The 1:1 relation is indicated by a grey dashed line in both panels. The red solid and dashed lines in each lower panel are respectively the median and $\pm 1\sigma_{\rm{MAD}}$ scatter of the residuals. The {\sc ProSpect} and {\sc magphys} stellar mass estimates are in close agreement with one another, whilst the {\sc LePhare} estimates differ more substantially (although the differences are still only moderate for most galaxies).  %
     }%
\label{fig:SEDmasscomp}
\end{figure*}

\begin{figure*}
\centering
\begin{minipage}[]{.95\textwidth}
\centering
\includegraphics[width=.33\textwidth,trim= 0 0 0 0,clip=True]{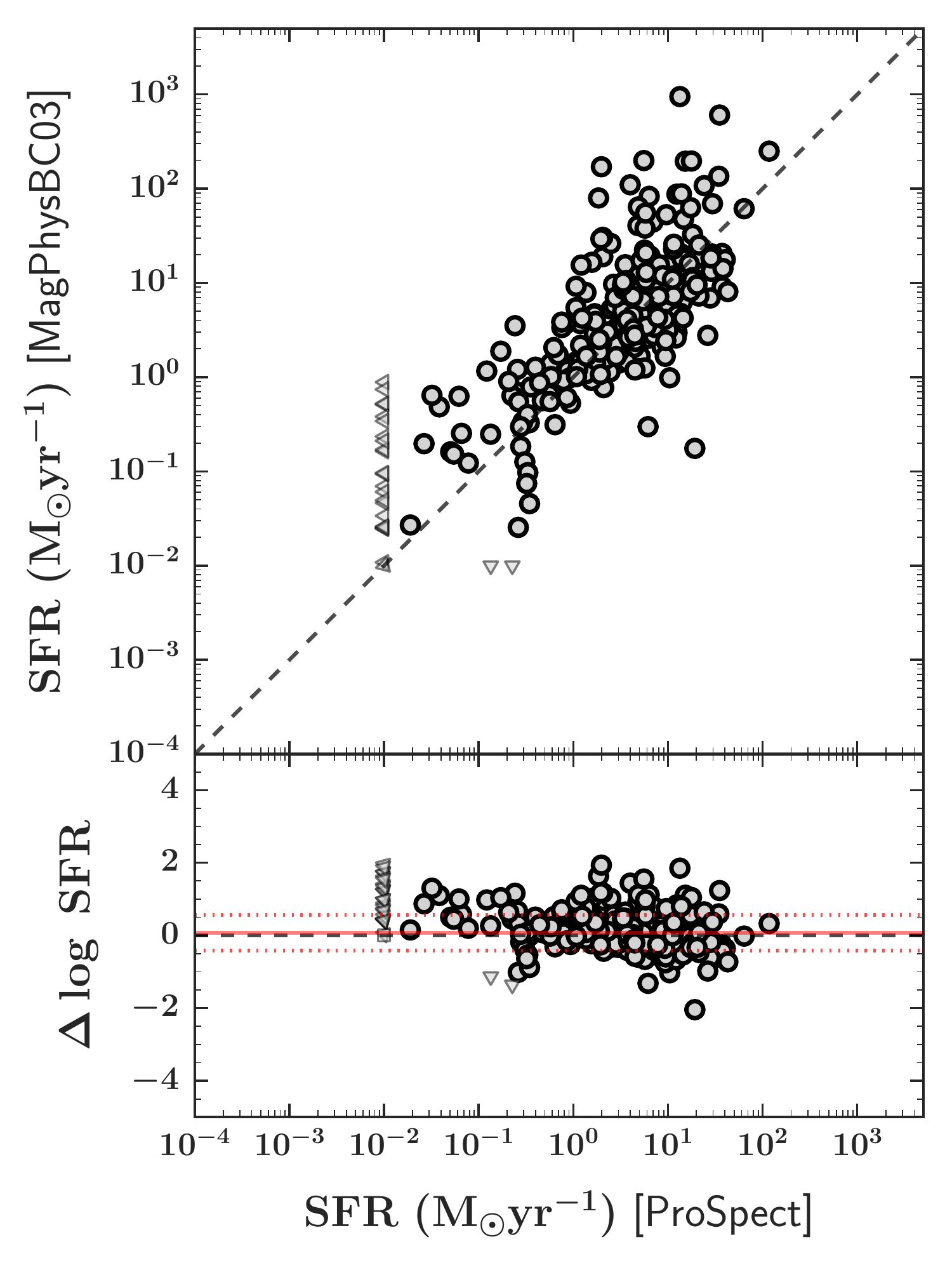}\includegraphics[width=.33\textwidth,trim= 0 0 0 0,clip=True]{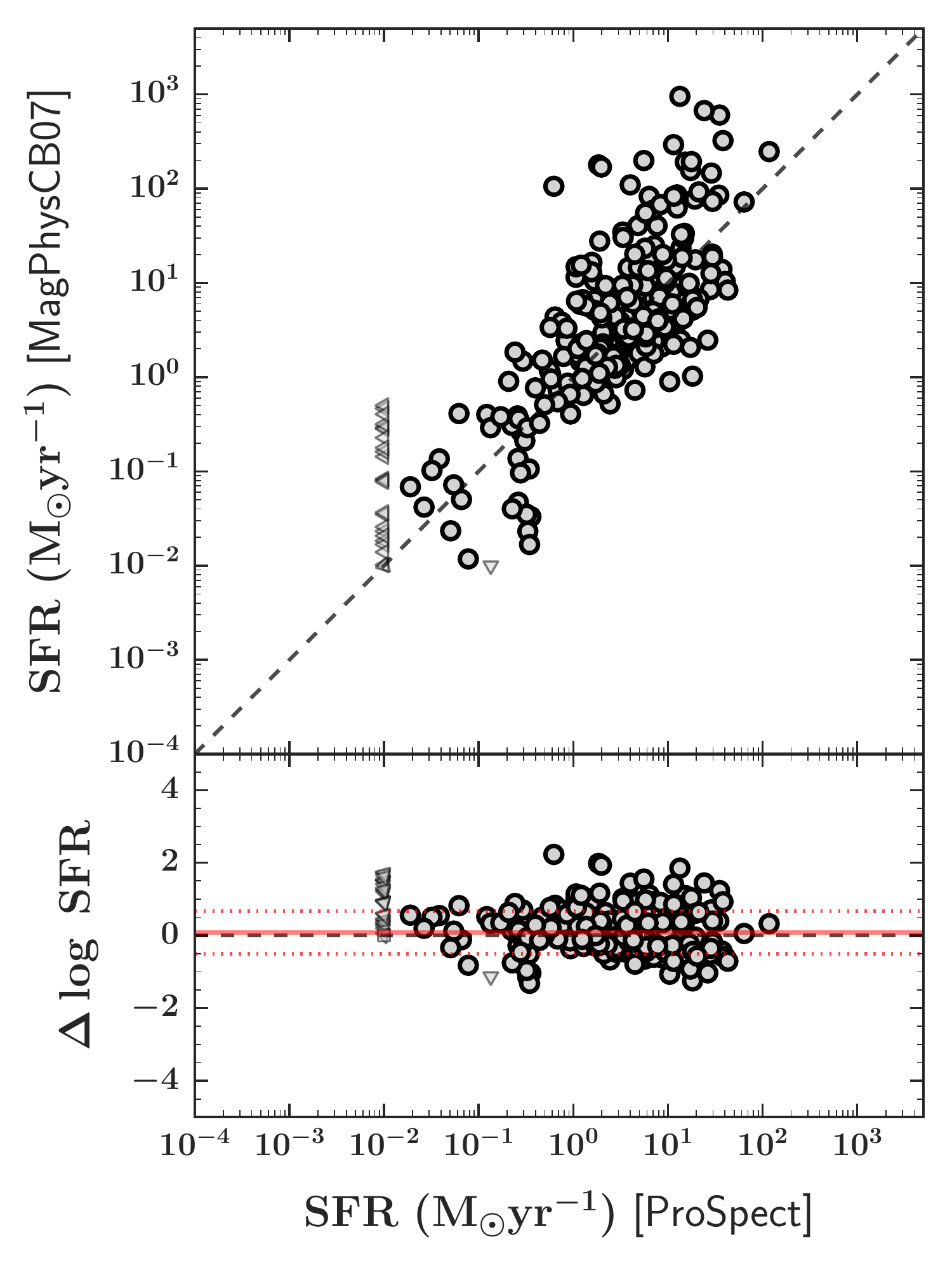}\includegraphics[width=.33\textwidth,trim= 0 0 0 0,clip=True]{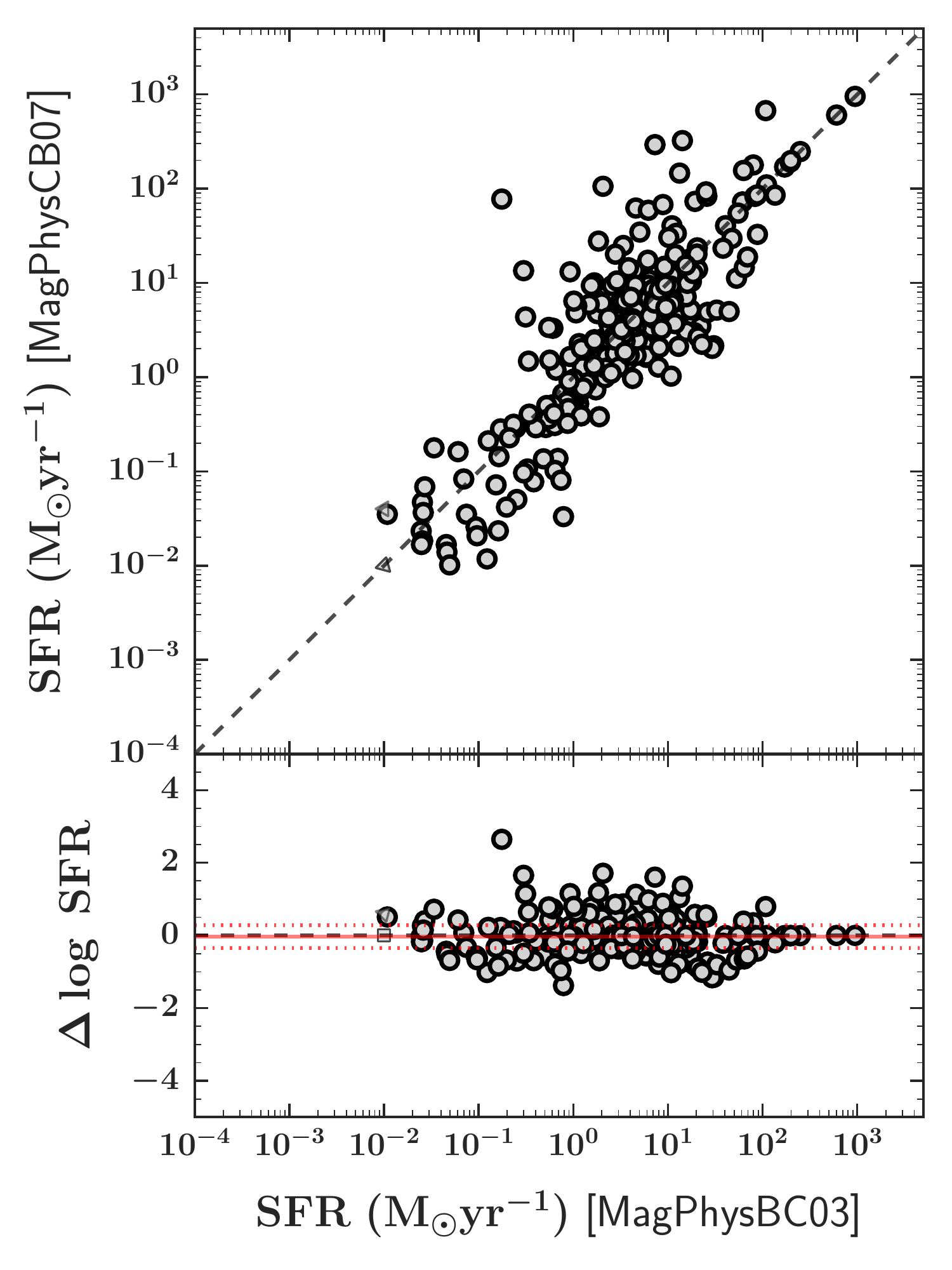}
\includegraphics[width=.33\textwidth,trim= 0 0 0 0,clip=True]{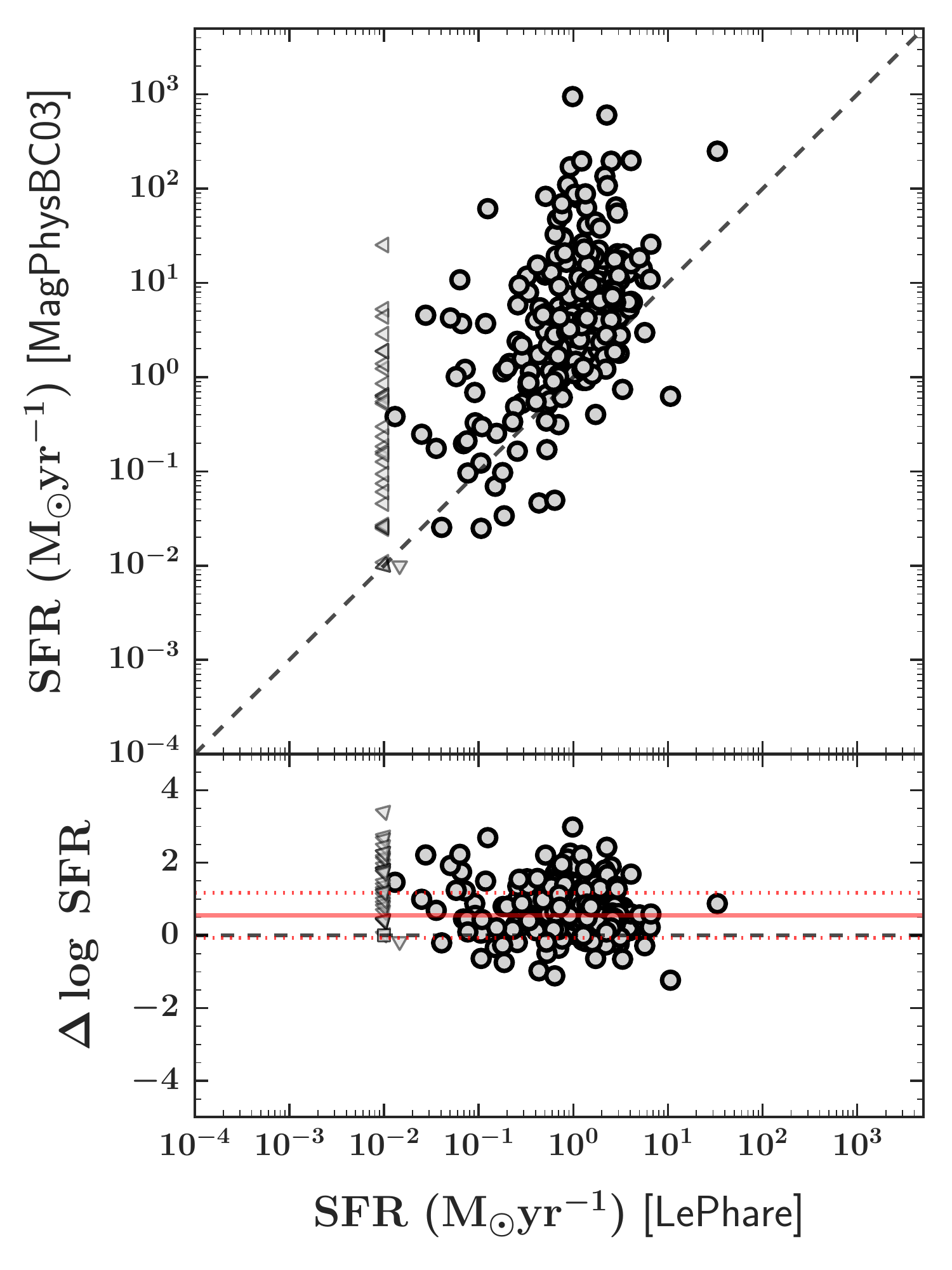}\includegraphics[width=.33\textwidth,trim= 0 0 0 0,clip=True]{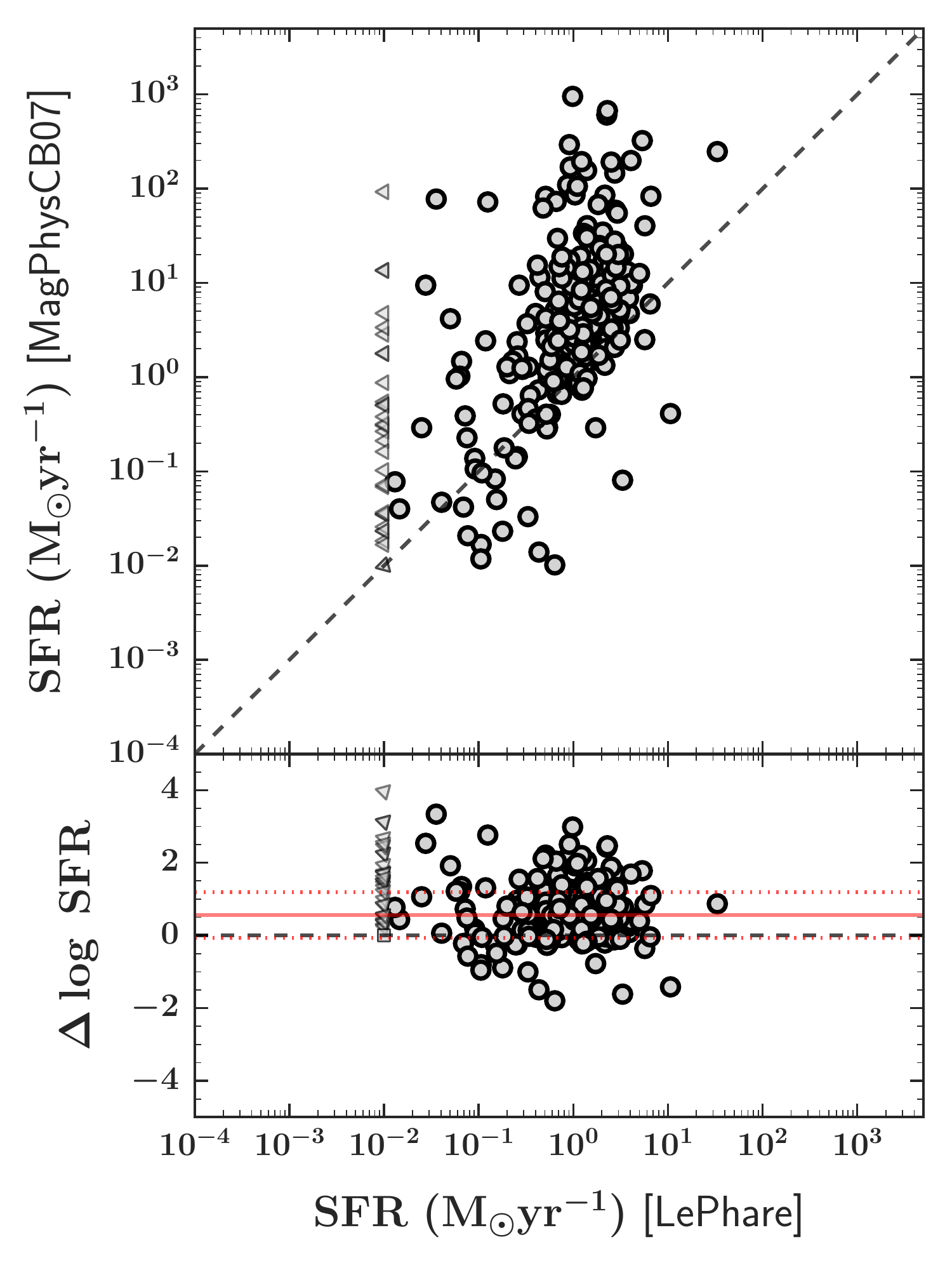}\includegraphics[width=.33\textwidth,trim= 0 0 0 0,clip=True]{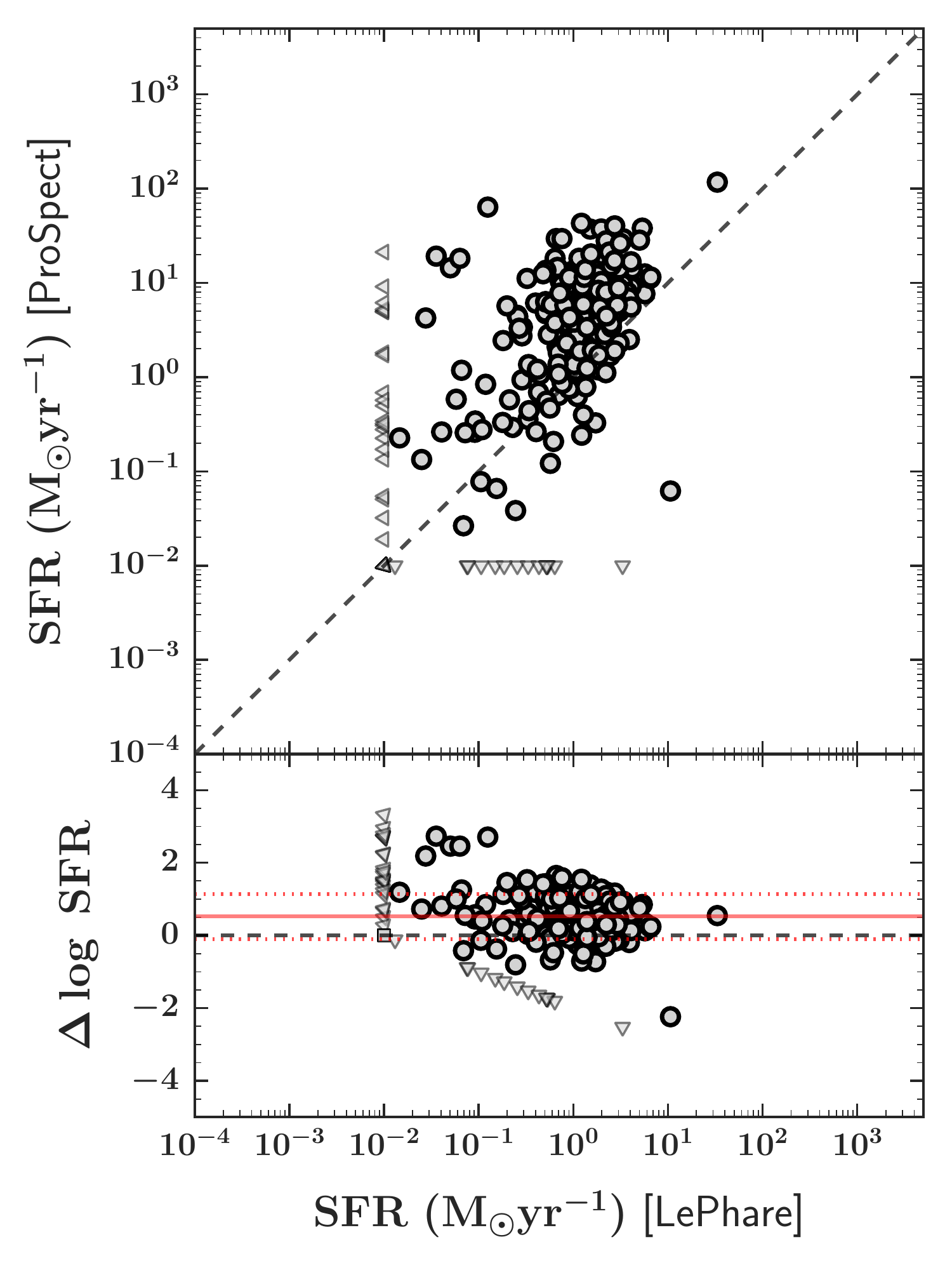}
\end{minipage}
\caption{%
Comparisons of the SFR estimates for K-CLASH galaxies derived using the {\sc ProSpect}, {\sc magphys} (BC03 and CB07) and {\sc LePhare} SED fitting routines on photometry spanning the optical to the near-infrared. For each comparison between the routines, we also show the residuals between the two sets of measurements (ordinate $-$ abscissa). The 1:1 relation is indicated by a grey dashed line in both panels. The red solid and dashed lines in each lower panel are respectively the median and $\pm 1\sigma_{\rm{MAD}}$ scatter of the residuals. Since SED-fitting solutions with very low SFRs are typically degenerate (e.g. a model SED with $\rm{SFR} = 0.01 \rm{M}_{\odot}\rm{yr}^{-1}$ can give as good a fit to the observed SED as one with $\rm{SFR} \ll 0.01 \rm{M}_{\odot}\rm{yr}^{-1}$, all else being the same), we replace $\rm{SFRs} < 0.01 \rm{M}_{\odot}\rm{yr}^{-1}$ with upper limits set at $0.01 \rm{M}_{\odot}\rm{yr}^{-1}$. Upper limits are indicated by grey arrows. The {\sc ProSpect} and {\sc magphys} SFR estimates are in close agreement with eachother, whilst the {\sc LePhare} estimates strongly disagree with those of the other two routines.%
     }%
\label{fig:SEDSFRcomp}
\end{figure*}

In \S~\ref{subsec:stellarmasses}, we discussed the stellar masses derived from the K-CLASH galaxies' optical to near-infrared photometry using the {\sc ProSpect}, {\sc magphys}, and {\sc LePhare} SED-fitting routines. Here we present a more detailed comparison of the results from the three codes.

\begin{figure*}
\centering
\begin{minipage}[]{1.\textwidth}
\centering
\includegraphics[width=.73\textwidth,trim= 0 0 0 0,clip=True]{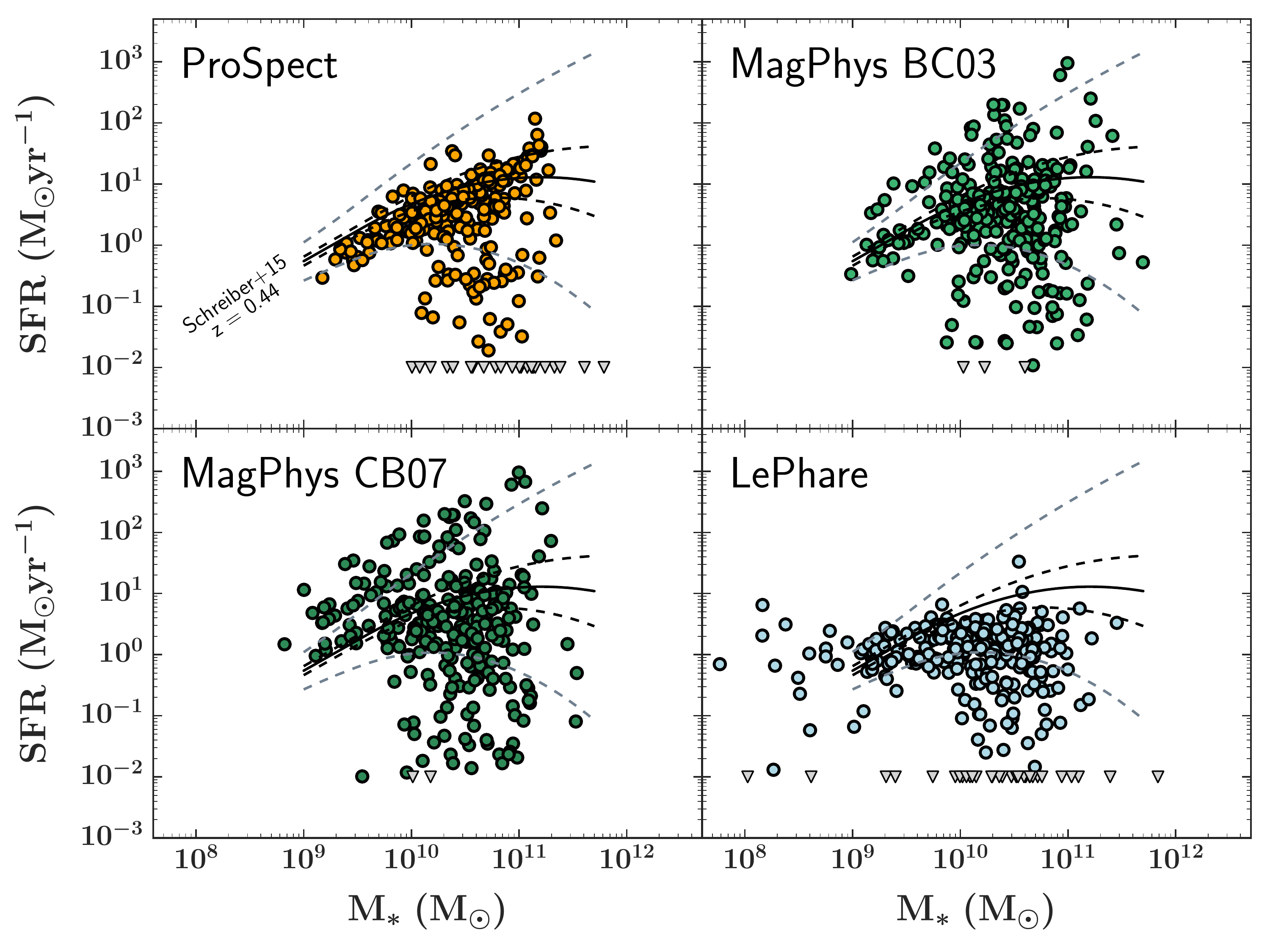}
\end{minipage}
\caption{%
Positions of the K-CLASH galaxies in the SFR--stellar mass plane, one for each of the {\sc ProSpect}, {\sc magphys} (BC03 and CB07), and {\sc LePhare} SED fits to available photometry spanning the optical and near-infrared. For reference we include a black solid line indicating the position of the main sequence of star formation for galaxies at the median K-CLASH galaxy redshift, according to \citet{Schreiber:2015}. The corresponding $1\sigma$ and $3\sigma$ uncertainty envelopes are indicated by respectively black dashed lines and grey dashed lines in each panel. The {\sc ProSpect} routine yields the best results, agreeing well with the expectation from \citet{Schreiber:2015} and having the least scatter. %
     }%
\label{fig:MSpanelplot}
\end{figure*}

In Figure~\ref{fig:SEDmasscomp}, we present a direct comparison of the stellar masses derived for the K-CLASH galaxies using the three routines. We find good agreement between the {\sc ProSpect} and {\sc magphys} masses, with a median difference of $-0.1 \pm 0.2$ dex (for either BC03 or CB07 in {\sc magphys}), where the uncertainty is the spread of the distribution ($\sigma_{\rm{MAD}}$). The {\sc LePhare} stellar masses are also in reasonably good agreement with those of the other routines, with the median differences approximately within that expected between different SED-fitting routines \citep[typically $\pm0.2$ dex;][]{Mobasher:2015}; the median difference ranges from $0.14 \pm 0.26$ dex (compared to {\sc magphys} CB07) to $0.3 \pm 0.3$ dex (compared to {\sc ProSpect}). However, we do find that the {\sc LePhare} masses become increasingly, and systematically, low compared to both the {\sc magphys} and {\sc ProSpect} estimates with decreasing stellar mass, i.e\ the differences between the {\sc LePhare} and {\sc magphys} or {\sc ProSpect} stellar mass estimates increase as the latter decrease. Nevertheless, the stellar masses derived from all three routines are generally consistent with each other within acceptable limits (although the {\sc ProSpect} and {\sc LePhare} masses differ the most).

We directly compare the SFRs derived using the three fitting routines in Figure~\ref{fig:SEDSFRcomp}.  Similar to the stellar masses, we find good agreement between the {\sc magphys} and {\sc ProSpect} SFRs, with median differences of $-0.1 \pm 0.5$ dex and $0.1 \pm 0.6$ dex when comparing respectively {\sc magphys} BC03 and {\sc magphys} CB07 to {\sc ProSpect}. However, we find the {\sc LePhare}-derived star-formation rates are much lower than those of {\sc magphys} and {\sc ProSpect}, with an average systematic difference ranging from $-0.5 \pm 0.6$ dex (compared to {\sc ProSpect}) to $-0.6 \pm 0.7$ dex (compared to {\sc magphys} BC03). 

It is clear from Figures~\ref{fig:SEDmasscomp} and \ref{fig:SEDSFRcomp} that the {\sc ProSpect} and {\sc magphys} routines are in good average agreement, whilst the {\sc LePhare} routine in comparison deviates in its outputs, particularly for SFR estimates. This does not neccessarily mean that the {\sc ProSpect} and {\sc magphys} routines produce more {\it accurate} estimates than those of {\sc LePhare} for these key galaxy properties (or vice versa), only that they differ. Nevertheless, we must select one set of results, out of the three, to use for our analysis. As a final step to aid in this decision, in Figure~\ref{fig:MSpanelplot} we plot the positions of the K-CLASH galaxies in the SFR--stellar mass plane, one plot for each of the routines, and compare to the expected position of ``main sequence'' star-forming galaxies at the same (median) redshift as the K-CLASH sample, according to \citet{Schreiber:2015}. From their colour-magnitude selection (blue and bright), we expect the K-CLASH sample to mainly comprise star-forming systems. We therefore expect the positions of the K-CLASH galaxies in the SFR--stellar mass plane  to be in closest agreement with the  measurements of \citet{Schreiber:2015} when the most accurate SED fitting estimates are used. Furthermore, given the homogenous selection criteria of the K-CLASH sample galaxies, the scatter of the K-CLASH data points in this plane should also at least be an indicator of how self-consistent the outputs of the code are across the sample.

From Figure~\ref{fig:MSpanelplot}, the {\sc LePhare} routine appears to perform worst, producing a flat distribution of SFRs as a function of the stellar masses, with large scatter. The {\sc LePhare} estimates also place the K-CLASH galaxies significantly and systematically below the  \cite{Schreiber:2015} main sequence of star formation for galaxies at the median K-CLASH redshift. Both the {\sc ProSpect} and {\sc magphys} estimates agree well, on average, with the expectation from \cite{Schreiber:2015}, but the {\sc ProSpect} data points have considerably smaller scatter than the {\sc magphys} data points. We therefore adopt the {\sc ProSpect} results for our analysis in this work.

\section{Estimating Total H$\alpha$ Star-formation Rates}
\label{sec:halphasfr}

In this section we describe the corrections we apply to the K-CLASH galaxies' H$\alpha$ fluxes measured within circular apertures, to estimate the total SFR of each galaxy. In \S~\ref{subsec:aperturecorrection} we describe the aperture correction applied to the measured H$\alpha$ flux of each galaxy to account for the finite size of the circular aperture. In \S~\ref{subsec:checkextinctioncorr} we discuss the application of an extinction correction to account for the attenuating affect of dust within each K-CLASH galaxy. We note that in \S~\ref{subsubsec:halphaSFRs}, we discussed a direct comparison between the {\sc ProSpect}- and total H$\alpha$-derived SFRs of galaxies in the {\it cluster} and {\it field} sub-samples (\S~\ref{subsec:clustermembership}), finding the two measures to be in good agreement. 

\subsection{Aperture correction}
\label{subsec:aperturecorrection}

In this work we measure H$\alpha$ fluxes for the K-CLASH galaxies within three different circular apertures, with diameters of respectively $\rm{D} = 0\farcs6$, $1\farcs2$, and $2\farcs4$. For most of the K-CLASH galaxies, these apertures do not encompass the total extent of the galaxy, and thus presumably the total H$\alpha$ flux. Since the FOV of each KMOS IFU is only $2\farcs8 \times 2\farcs8$, we in fact cannot measure the flux within an aperture large enough to encompass the total spatial extent of the H$\alpha$ emission of most K-CLASH galaxies. Instead, we calculate and apply aperture corrections to the measured H$\alpha$ fluxes, to account for the finite aperture sizes.  

\begin{figure}
\centering
\begin{minipage}[]{.5\textwidth}
\includegraphics[width=.93\textwidth,trim= 10 30 -10 0,clip=True]{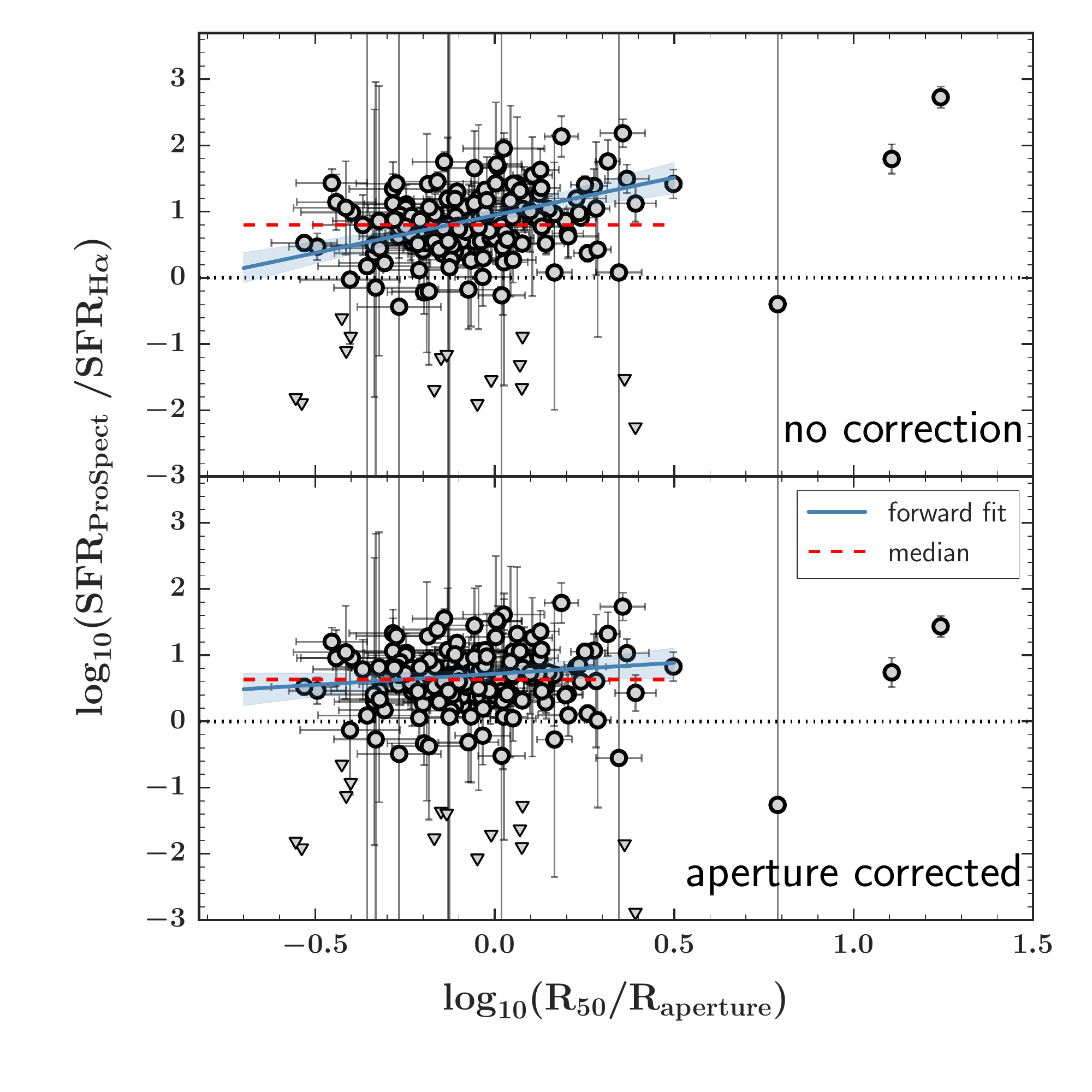}
\end{minipage}
\caption{%
{\bf Upper panel: } The ratio of the total {\sc ProSpect}-derived SFR and the H$\alpha$-derived SFR (with no correction for aperture or extinction effects) as a function of the ratio of the galaxy size and aperture size. The best bisector linear fit and the corresponding $1\sigma$ uncertainty are indicated by the blue solid line and light-blue shaded region, respectively. For reference, the median ratio of the {\sc ProSpect} and H$\alpha$-derived SFRs is indicated by a red horizontal dashed line. The 1:1 relation is indicated by a black dotted line. There is a positive correlation between the two ratios, indicating that an increasing fraction of the H$\alpha$ flux of each galaxy is missed by our fixed size circular apertures with increasing galaxy size. {\bf Lower panel:} As for the upper panel, but with our aperture correction applied to the H$\alpha$-derived SFRs. The best linear fit is now consistent with a horizontal line fixed at the median along the ordinate, indicating that our aperture correction is effective. %
     }%
\label{fig:checkapcorr}
\end{figure}

To calculate the aperture correction factors, we assume that the H$\alpha$ emission distribution exactly follows that of the stellar light. For each galaxy, we thus measure the curve-of-growth from the intrinsic stellar light model best-fitting the observed $R_{\rm{C}}$-band image (see \S~\ref{subsec:sizes}), where the model image is here convolved with a two-dimensional Gaussian with the same width as that of the KMOS PSF. From each galaxy's curve-of-growth, we then measure the fractions of the total model stellar light contained within the three apertures. We then calculate the required aperture corrections as the inverse of these fractions. To minimise extrapolation uncertainty, we take the total H$\alpha$ flux of each galaxy as the aperture corrected flux measured from the largest aperture in which we detect H$\alpha$. 

In Figure~\ref{fig:checkapcorr}, we verify that our aperture corrections are accurate by considering the ratio of the {\sc ProSpect}-derived total SFR and the H$\alpha$-derived SFR before and after applying the aperture correction, and how this ratio relates to the size of the galaxy ($\rm{R}_{50}$) compared to the size of the aperture ($\rm{R}_{\rm{aperture}}$). Before the aperture correction is applied, we find a positive trend between the SFR ratios and $\rm{R}_{50}/\rm{R}_{\rm{aperture}}$. In other words, the larger the galaxy compared to the aperture, the smaller the H$\alpha$-derived SFR is compared the total SFR measured from SED fitting. This is expected, since the smaller the fraction of the total H$\alpha$ emission enclosed within the aperture, the more we underestimate the total SFR calculated from the H$\alpha$. 

The median and $\sigma_{\rm{MAD}}$ of the aperture corrections we make are respectively $1.42 \pm 0.03$ and $0.36 \pm 0.04$. After applying the corrections, we find a flat relation between the ratios of {\sc ProSpect}-derived and H$\alpha$-derived SFRs and the ratios of galaxy and aperture sizes. This implies that our aperture correction is functioning correctly, removing any dependence of the total H$\alpha$-derived SFR on the aperture size. Similarly, the scatter (in the ordinate) about the best-fitting line in the lower panel of Figure~\ref{fig:checkapcorr} is consistent with that in the upper panel ($\sigma_{\rm{MAD}} = 0.37 \pm 0.03$ versus $\sigma_{\rm{MAD}} = 0.37 \pm 0.04$), implying the scatter introduced into the H$\alpha$-derived SFR via application of the aperture correction is very small (and thus that the correction is, on average, accurate). 

Lastly, we note that even after application of our aperture correction, we still find that the H$\alpha$-derived SFRs are a factor of $\approx4$ smaller than the total SFRs derived from {\sc ProSpect} SED fitting. This is accounted for by the application of an extinction correction, discussed in \S~\ref{subsubsec:halphaSFRs} and Appendix~\ref{subsec:checkextinctioncorr}.

\vspace{-0.2cm}

\subsection{Extinction correction}
\label{subsec:checkextinctioncorr}

\begin{figure}
\centering
\begin{minipage}[]{.5\textwidth}
\includegraphics[width=.77\textwidth,trim= -30 10 15 10,clip=True]{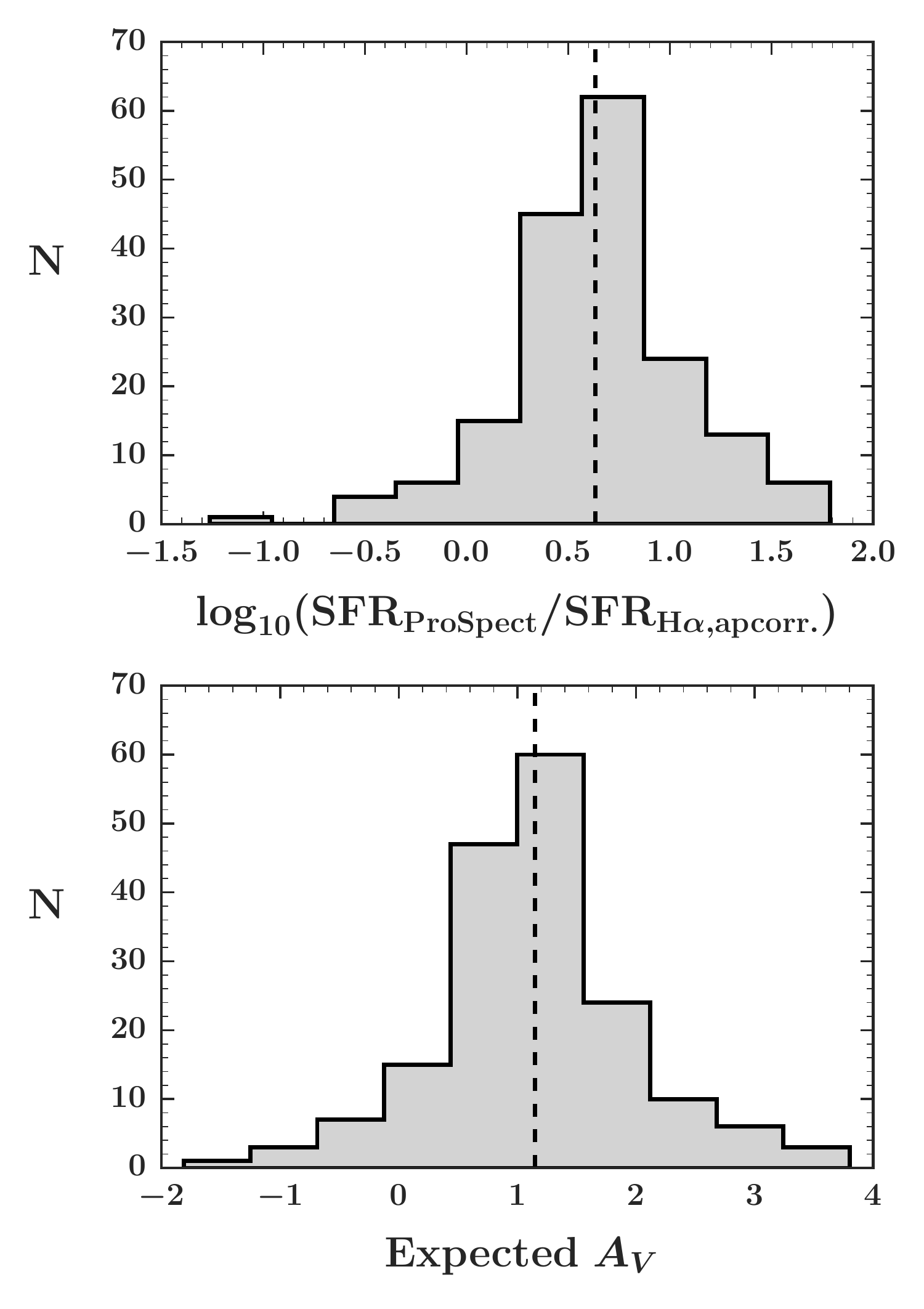}
\end{minipage}
\caption{%
{\bf Upper panel:} The distribution of the multiplicative correction factors required for the aperture-corrected H$\alpha$-derived SFRs to match the {\sc ProSpect}-derived SFRs exactly. {\bf Lower panel:} The corresponding distribution of expected $A_{V}$ required for the aperture-corrected, extinction-corrected H$\alpha$-derived SFRs to match the {\sc ProSpect}-derived SFRs exactly. In each panel, the black vertical dashed line indicates the mean.%
     }%
\label{fig:extinctioncorrcheck}
\end{figure}

In \S~\ref{subsubsec:halphaSFRs} we described how we calculate an inferred $A_{V}$ for each K-CLASH galaxy by comparison with those of mass-matched, main sequence galaxies in the UDS field (derived in D20 via SED fitting of photometry covering a much more extensive wavelength range for each target than is available for the K-CLASH sample galaxies).  The mean and standard deviation of these inferred $A_{V}$ are respectively $\overline{A_{V}} = 1.26 \pm 0.01$ mag and $\sigma_{A_{V}} = 0.19 \pm 0.01$ mag. 

As a check, we also calculate here the distribution of {\it expected} $A_{V}$ values, assuming the {\sc ProSpect}-derived total SFRs are, on average, representative of the real SFRs. We consider only those K-CLASH galaxies detected in H$\alpha$ (see \S~\ref{subsec:measuringlinefluxes} and \S~\ref{subsec:detectionstats}) and deemed to be star-forming by {\sc ProSpect} (SFR $> 0.01\ \rm{M}_{\odot}\rm{yr}^{-1}$). For each galaxy, we then calculate the H$\alpha$-derived SFR, applying the aperture correction described in \S~\ref{subsec:measuringlinefluxes} and Appendix~\ref{subsec:aperturecorrection}, but excluding any correction for the effects of dust extinction. We then calculate the multiplicative factor required for each galaxy to make the two SFR measures ({\sc ProSpect}- and H$\alpha$-derived) agree ($\rm{SFR}_{\rm{ProSpect}}/\rm{SFR}_{\rm{H}\alpha,\rm{apcorr.}} = 1$). The distribution of these multiplicative factors is shown in the upper panel of Figure~\ref{fig:extinctioncorrcheck}. 

We then simply adopt these multiplicative factors as $A_{\rm{H}\alpha,\rm{gas}}$ of the K-CLASH galaxies, and work backwards using Equation~\ref{eq:AstarToAgas} and the \citet{Calzetti:1994} extinction law to find the $A_{V}$ required such that $\rm{SFR}_{\rm{ProSpect}}/\rm{SFR}_{\rm{H}\alpha,\rm{apcorr.}} = 1$ for each galaxy. The distribution of these expected $A_{V}$ is shown in the lower panel of Figure~\ref{fig:extinctioncorrcheck}. The mean of this distribution is $\overline{A_{V}} = 1.15  \pm 0.07$ mag, consistent with that of the inferred $A_{V}$ from the D20 catalogue. The standard deviation is $\sigma_{A_{V}} = 0.85 \pm 0.06$ mag, much larger than that of the inferred D20 values. However, we refrain from a direct interpretation of this, since there will inevitably be scatter introduced into both the {\sc ProSpect}-derived and H$\alpha$-derived SFRs due to measurement uncertainties (also explaining why the distribution of $A_{V}$ in the lower panel of Figure~\ref{fig:extinctioncorrcheck} extends to negative values).

\bsp	
\label{lastpage}
\end{document}